\documentclass[traditabstract]{aa}

\usepackage{graphicx}
\usepackage{txfonts}
\usepackage{natbib}
\usepackage{bm}
\usepackage{xspace}
\usepackage{color}
\bibpunct{(}{)}{;}{a}{}{,}
\usepackage[colorlinks=true,linkcolor=blue,citecolor=blue]{hyperref}%
 
\newcommand{\ncase}{\ensuremath{n_{\mathrm{case}}}\xspace}
\newcommand{\nrea}{\ensuremath{n_{\mathrm{rea}}}\xspace}

\newcommand{\euclid}{\textit{Euclid}\xspace}
\newcommand{\galsim}{\texttt{GalSim}\xspace}
\newcommand{\snr}{\ensuremath{\mathrm{S/N}}\xspace}
\newcommand{\sersic}{S\'ersic\xspace}

\begin{document}

\title{Weak-lensing shear measurement with machine learning}
\subtitle{Teaching artificial neural networks about feature noise}
\titlerunning{Shear measurement with machine learning}

\author{
M. Tewes\inst{\ref{bonn}} \and
T. Kuntzer\inst{\ref{epfl}} \and
R. Nakajima\inst{\ref{bonn}} \and
F. Courbin\inst{\ref{epfl}} \and 
H. Hildebrandt\inst{\ref{bonn}} \and
T. Schrabback\inst{\ref{bonn}}
}

\institute{
Argelander-Institut f\"ur Astronomie, Auf dem H\"ugel 71, D-53121 Bonn, Germany \label{bonn}
\and
Institute of Physics, Laboratory of Astrophysics, Ecole Polytechnique F\'ed\'erale de Lausanne (EPFL), Observatoire de Sauverny, CH-1290 Versoix, Switzerland  \label{epfl}
}

\date{\today}

\abstract{Cosmic shear, that is weak gravitational lensing by the large-scale matter structure of the Universe, is a primary cosmological probe for several present and upcoming surveys investigating dark matter and dark energy, such as \euclid or WFIRST.
The probe requires an extremely accurate measurement of the shapes of millions of galaxies based on imaging data.
Crucially, the shear measurement must address and compensate for a range of interwoven nuisance effects related to the instrument optics and detector, noise in the images, unknown galaxy morphologies, colors, blending of sources, and selection effects.
This paper explores the use of supervised machine learning as a tool to solve this inverse problem.
We present a simple architecture that learns to regress shear point estimates and weights via shallow artificial neural networks.
The networks are trained on simulations of the forward observing process, and take combinations of moments of the galaxy images as inputs.
A challenging peculiarity of the shear measurement task, in terms of machine learning applications, is the combination of the noisiness of the input features and the requirements on the statistical accuracy of the inverse regression.
To address this issue, the proposed training algorithm minimizes bias over multiple realizations of individual source galaxies, reducing the sensitivity to properties of the overall sample of source galaxies.
Importantly, an observational selection function of these source galaxies can be straightforwardly taken into account via the weights.
We first introduce key aspects of our approach using toy-model simulations, and then demonstrate its potential on images mimicking \euclid data.
Finally, we analyze images from the GREAT3 challenge, obtaining competitively low multiplicative and additive shear biases despite the use of a simple training set.
We conclude that the further development of suited machine learning approaches is of high interest to meet the stringent requirements on the shear measurement in current and future surveys.
We make a demonstration implementation of our technique publicly available.
}

\maketitle

\section{Introduction}
\label{sec:intro}

Images of distant galaxies appear slightly distorted, typically at the percent level, as light bundles reaching the observer are differentially deflected owing to gravitational lensing by massive structures along the line of sight.
Since galaxies come in a variety of intrinsic shapes, inclinations, and orientations, these weak distortions are not identifiable on individual sources.
In this sense, galaxies give us only a very noisy view of the distortion field.
However, despite this intrinsic ``shape noise'', the weak lensing (WL) effect imprints spatial correlations on the apparent galaxy shapes.
Observing these spatial correlations, ideally as a function of redshift, allows us to infer properties of the large-scale matter structure of the Universe, and how this structure has grown over time.

This probe, known as cosmic shear, is one of the main scientific drivers for surveys poised to explore dark matter and dark energy, such as KiDS\footnote{\url{http://kids.strw.leidenuniv.nl/}} \citep{deJong:2015jl}, the Dark Energy Survey \citep[DES,][]{TheDarkEnergySurveyCollaboration:2016co}, the ESA \euclid\footnote{\url{http://www.euclid-ec.org/}} mission \citep{Laureijs:2011wi}, and NASA's Wide Field InfraRed Survey Telescope WFIRST\footnote{\url{http://wfirst.gsfc.nasa.gov}}.
\citet{Kilbinger:2014tb} and \citet{Mandelbaum:2018ij} provide recent reviews on the field, with a particular focus on the analysis methods to interpret the data from wide field surveys.

The statistical uncertainty of cosmic shear measurements, which is related to the finite number of galaxies probing the shear field, decreases with the increasing sky coverage and depth of the surveys.
To make full use of large surveys, the accuracy of the data analysis methods must therefore be high enough to avoid that systematic errors dominate the error-budget of the cosmological parameter inference \citep{Refregier:2003jl}.
For \euclid, surveying 15\,000 square degrees of extra-galactic sky, the resulting accuracy requirements are unprecedented.
These requirements flow down, on the observational side, to (1) the determination of redshifts and (2) the measurement of shear.
The cosmology community is working intensively on both aspects and on the required algorithmic improvements, often addressing effects that could previously be neglected due to the limited survey size.

Regarding the problem of photometric redshift determination, ``empirical'' and machine learning methods are now considered as at least equivalent to traditional template-fitting methods in terms of precision and accuracy.
They are also complementary, as they are based on fundamentally different principles and assumptions.
Several applications of artificial neural networks (NN) yield highly competitive results, especially when predicting redshift probability distributions \citep[e.g., ][]{Bilicki:2018iy, Bonnett:2015ku}. Furthermore, \citet{DIsanto:2018jd} demonstrate how deep convolutional NNs can infer redshifts by directly processing multiband image data at the pixel level, as compared to using fluxes measured in apertures.

The shear measurement problem has not yet seen a similar evolution toward machine learning methods.
The problem of shear measurement is also referred to as ``shape measurement'' in the literature, as the shape (more precisely the ellipticity) of galaxies yields an estimator for the lensing shear.
There are two traditional categories of shear measurement techniques: (1) methods based on the measurement of weighted quadrupole moments of the observed light distribution and (2) methods that forward-fit a model.
Mathematically, these categories share strong similarities \citep{Simon:2017kj}, and both have to tackle the same sources of biases in order to serve as accurate shear estimators.

The most prominent observational issues are the deformation of the sheared galaxy light distribution by the telescope optics and atmospheric seeing (often seen as convolution by a point-spread function PSF), the pixellation of the image by the detector, and the pixel noise.
Information about the original galaxy shape and the shear is lost or compromised by each of these effects.
Even for space-based instruments, the shape of the PSF varies over the field of view and in time, and the PSF for each galaxy must therefore first be reconstructed with high fidelity before shear can be estimated.
In addition, the low signal-to-noise ratio (\snr) of the galaxy images leads to biases that have to be accounted for, as shear estimators are not linear functions of the image pixel values \citep[see, e.g.,][]{Refregier:2012jt}.
A large variety of shear measurement methods have been developed to deal with these effects, notably in context of the public Shear Testing Program (STEP) and the GRavitational lEnsing Accuracy Testing (GREAT) challenges \citep{Heymans:2006ix, Massey:2007jy, Bridle:2010bq, Kitching:2012hy, Mandelbaum:2015gc}.
Today's state-of-the-art shape measurement methods involve various forms of simulation-based calibration to account for different biases \citep[e.g.,][]{FenechConti:2017cq, Huff:2017ww, Jee:2016eq}, yet without embracing a full machine learning approach.
The computational cost of the shape measurement process is also of importance, with \euclid set to observe about 1.5 billion galaxies.
Rigorously testing a method will typically imply applying it to simulations larger than the survey itself, underlining the need for fast algorithms.

In this paper, we use supervised machine learning (ML) to address the problem of shear measurement, building upon the few previous applications of ML to this specific problem (\citealt{Gruen:2010ii}; \citealt{Tewes:2012gy}; \citealt{Graff:2014ij}, see also\footnote{Between submission and acceptance of the present paper, these authors released an analysis of a deep learning approach to the shear estimation problem, in the context of galaxy cluster lensing.} \citealt{Springer:2018tv}).
Specifically, we simulate noisy and PSF-convolved galaxy images with known shear, and train NNs to regress shear estimates based on features of these images, so to minimize shear prediction biases rather than shear errors.
While we participated in the GREAT3 challenge \citep{Mandelbaum:2015gc} with an early attempt of this approach under the name \emph{\mbox{MegaLUT}}, the present paper describes a fundamentally revised methodology.
The development of a machine learning approach is motivated by
\begin{enumerate}
\item the low CPU cost of ML predictions, as compared for example to iterative forward-fitting methods (either frequentist or Bayesian),
\item the unavoidable need for some form of shear calibration via image simulations for any state-of-the-art technique, due to practical effects such as galaxy blends and CCD charge transfer inefficiency,
\item the potential of simulation-driven methods to easily embrace further complex bias sources not identified at the moment, without affecting the initial formalism,
\item the possibility to control and penalize the tradeoff between sensitivity of the method to parameters affecting bias, such as prior knowledge of galaxies, and the bias itself.
\end{enumerate}

A distinctive aspect of this ML application is the noisiness of the data.
For the low-\snr galaxies of interest to cosmic shear studies, the measurement uncertainty on the shape of individual sources is larger than the WL distortion we wish to recover accurately.
The cost function of the training algorithm and the structure of the training data must therefore be adapted so that the NNs can learn to correct for biases resulting from the propagation of noisy inputs through them.

To ease the analysis and comparison with other methods, the present work is limited to the prediction of point estimates and weights for each component of the shear. This was also the format adopted by GREAT3 \citep{Mandelbaum:2014dq}.
The large number of source galaxies in weak lensing surveys led the community to (so far) favor these over shear probability distributions.
Furthermore, traditional shape measurement methods only produce point estimates, and they are also easier to analyze, for example when computing correlation functions.
The situation is however changing, with several current methods adopting some more descriptive probabilistic formalisms \citep[e.g.,][]{Bernstein:2014jf}.
We see the implementation presented in this paper as a stepping stone toward a probabilistic machine learning approach.

This article is organized as follows:
we introduce the required WL-formalism in Sect.~\ref{sec:wl}.
In Sect.~\ref{nn}, we describe how NNs can be trained to achieve accurate predictions in the presence of noise in their inputs. 
The input features measured on galaxy images are presented in Sect.~\ref{sec:InputFeatures}.
We then detail how we connect these steps to form a shear measurement method in Sect.~\ref{sec:shearestimation}.
The method is demonstrated on simple simulations in Sect.~\ref{sec:fiducial}, and with a variable PSF in Sect.~\ref{sec:fiducial:varpsf-field}. 
We apply it to more realistic \euclid-like simulations with source selection in Sect.~\ref{sec:euclid}, and on GREAT3 data in Sect.~\ref{sec:great3}.
Lastly, we offer perspectives for the practical integration in a shear measurement pipeline in Sect.~\ref{sec:outlooks}, and summarize in Sect.~\ref{sec:conclusion}.

\section{Formalism of weak gravitational lensing} \label{sec:wl}

In the following we give minimal definitions of the formalism of weak lensing and its estimation. Recent reviews include \citet{Kilbinger:2014tb} and \citet{Bartelmann:2017eh}, and a comprehensive introduction can be found in \citet{Schneider:2006vd}.

\subsection{Shear and ellipticity}
\label{sec:ellip}

The weak-lensing distortion seen in a given field of view can locally be approximated as a linear transformation between the ``true'' unlensed coordinates and the observed coordinates, expressed by a Jacobian matrix. In coordinates centered respectively on the unlensed and observed source, this local transformation is often written as

\begin{equation}
\label{jacobi}
\left( \begin{array}{c}
x^{\mathrm{\,true}}\\
y^{\mathrm{\,true}}\\
\end{array} \right) =
(1- \kappa)
\left( \begin{array}{cc}
1-g_1 & -g_2 \\
-g_2 & 1 + g_1 \\
\end{array} \right)
\left( \begin{array}{c}
x^{\mathrm{\,obs}}\\
y^{\mathrm{\,obs}}\\
\end{array} \right),
\end{equation}
where $g_1$ and $g_2$ are the two components of the reduced\footnote{For the rest of the paper, we designate this quantity $g$ simply as shear.} shear, causing a change in the ellipticity of observed galaxies, and where $\kappa$ is the convergence, describing the change in their apparent size. It is often convenient to write the shear as a complex number $g = g_1 + g_2 \mathrm{i}$.

Most traditional methods measuring the lensing shear deal with expressions for the ellipticity of a galaxy, as for example with KSB \citep{Kaiser:1995bi, Hoekstra:1998cl}.
In contrast, our proposed method constructs a direct estimator of the shear signal $g$ as defined above. Inevitably, this estimator will be noisy, due to shape noise (see Sect.~\ref{sec:intro}). But it does not require a formal description of the ellipticity at any stage of the process.
Indeed, the notion of the ellipticity of a galaxy is not trivial, as real galaxies have complex morphologies without simple elliptical isophotes, not to mention pixellation and noise.
For idealized galaxies with elliptical isophotes, we do however define an ellipticity in the following, which we use in the analysis of the sensitivity of our method and for some experiments.
For such a galaxy, with semi-major axis $a$ and semi-minor axis $b$, we follow the notation of the GREAT3 challenge \citep{Mandelbaum:2014dq} and define the ellipticity $\varepsilon$ as a complex number of modulus $|\varepsilon| = (1-b/a) / (1+b/a)$ and a phase determined by the position angle $\phi$ of the major axis such that $\varepsilon_1 = |\varepsilon| \cos(2 \phi)$ and $\varepsilon_2 = |\varepsilon| \sin(2 \phi)$.
With this definition, and considering only weak shear $|g| \ll 1$, the observed ellipticity $\varepsilon^{\mathrm{obs}}$ of an idealized lensed galaxy is related to its true intrinsic ellipticity, $\varepsilon^{\mathrm{true}}$, by $\varepsilon^{\mathrm{obs}} \approx \varepsilon^{\mathrm{true}} + g$.
The ellipticity $\varepsilon^{\mathrm{obs}}$ of each galaxy subject to some shear is a noisy but unbiased estimator of this shear, $\langle\varepsilon^{\mathrm{obs}}\rangle \approx g$, under the assumption that the source galaxies are intrinsically randomly oriented, that is $\langle\varepsilon^{\mathrm{true}}\rangle = 0$.

\subsection{Biases and sensitivity of shear estimation}
\label{sec:biases}

Biases of a shear estimator $\hat{g}$ are commonly quantified using a linear bias model following \citet{Heymans:2006ix}, decomposing the bias into a multiplicative part, $\mu$, and an additive part, $c$, for each component,
\begin{equation} \label{eq:mc}
\hat{g}_i-g^\mathrm{true}_i = \mu_i \cdot g^\mathrm{true}_i + c_i + \mathrm{noise}.
\end{equation}
Given shear measurements on simulations with known true shears, estimates of these biases $\mu$ and $c$ are obtained by fitting a line to the shear estimation residuals $\hat{g}_i-g^\mathrm{true}_i$ against the true shear value\footnote{\citet{Pujol:2018ud} propose an alternative method to estimate these bias parameters efficiently, which might be of interest for future developments of ML shear measurement methods.}.
The commonly used components $i=\{1,2\}$ of the shear and the biases $\mu_i$ and $c_i$ are defined by the coordinate grid used in Equation \ref{jacobi}, usually the image pixel grid. The first (second) component describes deformations along the axes (along the diagonals) of this grid.
In addition, following GREAT3 conventions \citep{Mandelbaum:2015gc} and in line with  \citet{FenechConti:2017cq}, we use the indices $i=\{+,\times\}$ to relate to components in a frame rotated to be aligned with the anisotropy of the PSF. The estimation of these PSF-oriented biases is done on simulations with variable orientation of the PSF.
More precisely, to estimate those biases, one first rotates the components 1 and 2 of $\hat{g}$ and $g^\mathrm{true}$ by $-2 \theta$, where $\theta$ is the position angle of the PSF anisotropy, and then performs the linear regressions on these rotated components.
We stress that we focus in this paper on probing the bias of the shear measurement method only, assuming for instance that the PSF is perfectly known.

In line with the above linear bias model, the numerous sources for bias are also often categorized into ``multiplicative'' and ``additive'' \citep{Mandelbaum:2018ij}. 
For example, the size of the PSF and the noise in the images are sources for multiplicative bias, as both effects tend to make galaxies look rounder and therefore less sheared \citep[see, e.g.,][]{Melchior:2012fb}. 
A typical source for an additive bias is the imperfect correction for an anisotropic PSF, leading to a net shift in the measured galaxy ellipticity. 
We refer to \citet{Massey:2013hr} for a more comprehensive list of biases and studies of their propagation into cosmic shear results. 
These authors also establish that Stage IV weak-lensing experiments \citep{Albrecht:2006tg} require multiplicative (additive) biases and the uncertainty on these biases to be on the order of $|\mu|\lesssim2 \cdot 10^{-3}$ ($|c|\lesssim2 \cdot 10^{-4}$). 

In this paper, we will perform evaluations of $\mu$ and $c$ in different bins of ``true'' parameters potentially affecting the bias, such as the intrinsic size of the  galaxies. This is made possible by carrying out numerical experiments using simulated data. It is crucial to be aware that any binning or selection according to some noisy ``observed'' parameters might lead to shear estimation biases due to selection effects. For example, the estimate of the size or the signal-to-noise ratio of a galaxy can in practice depend on the orientation and magnitude of the shear. For a discussion of selection biases, see, e.g., \citet{FenechConti:2017cq}.
An illustration of the intricate dependencies of biases on true and observed parameters can be found in \citet{Pujol:2017wy}. 

As mentioned by \citet{Hoekstra:2017hg}, an important goal for a shear measurement method should be to minimize the sensitivity $|\partial \mu / \partial p|$ to any parameter $p$ potentially affecting the multiplicative bias $\mu$ of a measurement.
A tradeoff between this sensitivity and the overall bias will have to be made.
Let us consider some extreme examples.
Suppose that a method shows a strong multiplicative bias on a given set of simulations.
Applying a plain multiplicative scaling to all its shear estimates will apparently remove this overall bias.
However, the sensitivity of this method to the galaxy population and simulation parameters might be increased by this rough rescaling.
The rescaled method would therefore show a low bias on these particular simulations, but a potentially unacceptable sensitivity to the actual galaxy profiles.
As another example, we can consider a method strongly driven by a prior on the galaxy shapes, but failing to use some of the available information from the observed data.
While this method might show a low sensitivity to some parameters of the observed galaxy population, it would certainly have a biased overall response to shear, and a strong sensitivity to its prior assumptions.
When designing a shear measurement method, both sensitivity and integrated biases should therefore be kept under control simultaneously.

\section{Accurate regressions from artificial neural networks in presence of feature noise}
\label{nn}

In this section, we describe how we train neural networks to perform accurate regressions despite noisy input features, building upon ideas 
from \citet{Gruen:2010ii}.
We keep this part generic to any inverse problem, and will introduce the particular application to weak-lensing shear measurement in Sect.~\ref{sec:shearestimation}.

\subsection{The inverse regression problem}
\label{invregprob}
A standard feedforward neural network (NN) with $N$ input nodes and one output node can be seen as a ``free-form'' fitting function of $\varmathbb{R}^N \to \varmathbb{R}$ \citep[see, e.g., ][for an introduction to NNs and applications to astronomy]{Tagliaferri:2003df}.
As such, the property of a NN to be nonlinear in its inputs (also called features) is explicitly desired, to allow for flexibility of the fitting function.
A natural consequence of this nonlinearity is that if noisy realizations of input data are to be propagated through the NN, the resulting distribution of outputs might well differ from the noise distribution of the inputs.
In particular, the expectation value of the output can be offset from the output which would be obtained from noise-free or less noisy inputs, leading to a net noise bias. 
We note that this holds for any nonlinear estimator.

Let us consider a NN of sufficient capacity, that is flexibility to adapt to the data, for a given regression problem.
The NN regression is parametrized by all the weights and biases of the network's nodes.
For a fixed network architecture, the shape of this regression is then entirely determined by the training of the network.
This training consists in optimizing the network parameters so as to minimize a cost function which compares network predictions to some known truth or ``target'' values.
A simple and common choice for such a cost function is the mean square error (MSE) between the network predictions and the target values, in analogy to an ordinary least squares or maximum likelihood method.
When fitting a model to noisy observations that depend on noiseless explanatory variables, the MSE does lead to the usually desired fitting curve (or hypersurface, in case of several input nodes). The latter traces, in the limit of many observations, the average values of the observed variable in bins of the explanatory variables.

In this work, our use of NNs is however ``inverse''.
We want to regress estimates for the explanatory variable (the NN target) based on noisy observations of the dependent variables (the NN inputs), a problem known in statistics as an inverse regression or calibration.
We provide a illustrated example of an inverse regression and the terminology in Appendix \ref{appendix:invreg}.

As mentioned by \citet{Gruen:2010ii}, it is counterproductive in such a situation to train a NN to minimize a MSE expressed between targets and individual predictions of the explanatory variable.
We can however formulate other cost functions which explicitly favor accuracy in the predictions of the explanatory variable, when facing noise in the observed dependent variables.
For this, the training data has to be structured so that the neural network can experience several realizations of the noise in the dependent variables for each value of the target explanatory variable.

\subsection{Training with realizations and cases}
\label{sec:nomenclature}

To structure our training data, we introduce the distinction between ``realizations'' and ``cases'':
\begin{itemize}
\item A training realization is a single observation of the noisy dependent variables, for a particular (known) value of the explanatory variable. Measurements of the dependent variables resulting from a physical process give us such realizations, except that the value for the explanatory variable is usually not known.
\item A training case is an ensemble of realizations obtained for the same value of the explanatory variable.
In other words, for our application of NNs, it is an ensemble of (input, target) pairs all sharing the same target value.
When the training data is entirely simulated, cases can easily be generated to contain as many realizations as desired.
\end{itemize}
The training data therefore consists of an ensemble of cases, each containing an ensemble of realizations.
Cost functions can now take advantage of this structure.
We define the mean square bias (MSB) cost function, which penalizes the estimated prediction bias over the realizations in each case, as\footnote{This expression is comparable to Eq.~7 of \citet{Gruen:2010ii}. We note that for the later application to shear measurement, these authors average the NN output over heterogeneous galaxy populations, while we average over rotated realizations of single galaxies.}
\begin{equation}
\label{eq:msb}
\mathrm{MSB}(\bm{p}, \bm{D}, \bm{t}) \doteq \frac{1}{\ncase} \sum_{k=1}^{\ncase} \left[\frac{1}{\nrea} \sum_{i=1}^{\nrea}  o(\bm{p}, \bm{D}_{i, k}) - t_{k} \right]^2,
\end{equation}
where $\bm{p}$ groups all the parameters (weights and biases) of the NN, $\bm{D}$ represents the training inputs containing the input vector $\bm{D}_{i, k}$ for each of the $\nrea$ realizations in each of the $\ncase$ cases, $o(\bm{p}, \bm{D}_{i, k})$ is the NN output for each realization, and $\bm{t}$ represents the training targets (with the target $t_k$ of each case).
With the same notations, the classical MSE cost function making no distinction between realization and cases would be written
\begin{equation}
\label{eq:mse}
\mathrm{MSE}(\bm{p}, \bm{D}, \bm{t}) \doteq \frac{1}{\ncase} \sum_{k=1}^{\ncase} \frac{1}{\nrea} \sum_{i=1}^{\nrea} \left[ o(\bm{p}, \bm{D}_{i, k}) - t_{k} \right]^2.
\end{equation}
The apparently small difference between MSB and MSE is therefore that the MSB averages the NN outputs over the realizations in each case before comparing them to the target values.
For both cost functions, the NN still learns how to predict one output for each realization.
Let us note some consequences of the MSB cost function, which plays an important role in this paper.

First, in the limit of a sufficiently large number of realizations per case, the MSB does not penalize scatter in the predictions.
A network trained to minimize MSB will, as desired, trade precision for accuracy, but it could potentially go beyond the optimal use of information and introduce additional unnecessary noise in its predictions. 
In practice, one can control this behavior, as well as potential overfitting to the training data, by limiting the capacity of the NN (i.e., limiting the number of its nodes and/or layers).

Second, one has to acknowledge that an inverse regression problem might simply not have an ``accurate'' solution, in the sense of a solution with vanishing MSB.
If the observed dependent variables (the NN input features) do not carry information about the explanatory variable (the NN target) the corresponding target values will not be accurately estimated.
And even if this information is still there, given a finite number of realizations and cases, a sufficiently strong noise in the input features will lead to biased predictions.
We note that this might affect some ``difficult'' realizations only, while other regions of parameter space allow for sufficient accuracy.

More generally, not only the accuracy but also the achievable precision of the predictions might vary from one realization to another.
In situations where the noisiness of an observed realization can be estimated from the observation itself, we can therefore further mitigate the effect of noise and extract more information by going beyond the prediction of point estimates.
In this paper, we explore the simplest extension to the prediction of point estimates, by including the prediction of weights.

\subsection{Predicting weights}
\label{subsec:weights}

Ideally, the prediction of weights and point estimates should be learned simultaneously.
To maximize insight, we propose in the scope of this work the use of a separate NN for the weight prediction, in addition to the NN predicting the point estimates.
The two networks are trained successively.
In the first step, the NN yielding point estimates is trained using the MSB cost function. Then, the second NN is trained to predict an optimal weight for each realization, in order to increase the accuracy of each case.
For this second NN, with parameters $\bm{p}_{\mathrm{W}}$ and exclusively positive outputs $w$, we define the mean square weighted bias (MSWB) cost function
\begin{equation}
\label{eq:mswb}
\mathrm{MSWB}(\bm{p}_{\mathrm{W}}, \bm{O}, \bm{D}, \bm{t}) \doteq \\
\frac{1}{\ncase} \sum_{k=1}^{\ncase}
\left[
\frac{\sum_{i=1}^{\nrea} o_{i, k} \cdot w(\bm{p}_{\mathrm{W}}, \bm{D}_{i, k})}
{\sum_{i=1}^{\nrea} w(\bm{p}_{\mathrm{W}}, \bm{D}_{i, k})}
- t_{k} 
\right]^2,
\end{equation}
where $\bm{O}$ contains the predicted point estimates $o_{i, k} = o(\bm{p}, \bm{D}_{i, k})$ obtained through the first NN.
A pecularity of this cost function is that no explicit target values for the weights is given. Furthermore, by construction, the weights $w$ minimizing the MSWB might have an arbitrary scale.
We impose both the positivity and an upper bound to the weights by using an activation function $\varmathbb{R} \to (0, 1)$ for the output layer of this second NN.

The training data $(\bm{D}, \bm{t})$ for the weight training can have a different structure of realizations and cases than the training data for the point estimates.
It is always possible to obtain the point estimate predictions $\bm{O}$ from the first NN by running it on the training data of the second NN.
We can thus make use of two training datasets, each optimized for its purpose.

We will further discuss the properties and behavior of NNs trained with the MSB and MSWB cost functions and the importance of the distributions of cases and realizations in Sect.~\ref{sec:shearestimation}, in the context of the practical application to weak lensing shear estimation.

\subsection{Neural network implementation, training optimization algorithm, and committees}
\label{subsec:nndetails}

All results of this paper are obtained using an experimental custom NN library implemented in \texttt{python}, which we make publicly available (see Appendix \ref{Code}). In the following, we briefly summarize details and default settings of the NNs.
If not stated otherwise through this paper, these configurations were chosen based on previous experience or trial-and-error attempts.
We do not claim that these choices are optimal, and expect many other configurations to yield equivalent or better results.

For both types of networks (point estimates and weights), we use small fully-connected NNs with typically two hidden layers of five nodes each. All input and hidden nodes use the hyperbolic tangent $f(x) = \tanh(x)$ activation function.
For the output layer, we use an identity activation function for the prediction of point estimates, and a variant of a sigmoid, $f(x) = 1/(1+\exp(-4 x))$ for the weight-predicting networks.
We follow the conventional practice to deal with highly heterogeneous feature scales, and prepend a normalization (or whitening) of the input data vectors to our networks \citep{Graff:2014ij}.
This normalization independently scales and shifts the features seen by each node of the input layer, so that, for the training data, all inputs cover the interval $[-1, 1]$.

Instead of using the conventional back-propagation \citep{Rumelhart:1986er}, we train our networks with a Broyden-Fletcher-Goldfarb-Shanno (BFGS) iterative optimization algorithm \citep[][and references therein]{Nocedal:2006uv} in its \texttt{scipy} implementation\footnote{\url{https://www.scipy.org}}.
The use of an algorithm that is agnostic of the network details, and therefore computes all required gradients numerically, allows for easy experimentation with cost functions and also with unconventional nodes, such as product units \citep{Durbin:1989kc, Schmitt:2002kk}. 
To increase the efficiency of the training, we implement a caching mechanism for the results computed by each layer of the network. 
We also use so-called mini-batch optimization \citep[see, e.g.,][]{Nielsen:2015ul}, that is we randomly select a ``batch'' of typically $25\%$ of the training cases, perform several (typically 30) optimization iterations on this batch, and iteratively pursue with the next randomly selected batch.

We start the training iterations from a randomized initial parameter state, with network weights and biases drawn from a centered normal distribution with a standard deviation of $0.1$.
Owing to this random initialization as well as the mini-batch optimization, networks trained on exactly the same data yield different estimators.
We exploit this stochastic behavior to increase the robustness of our training procedure, by systematically using so-called committees of typically eight NNs in place of single networks.
After the parallel training of such a committee, and a repeated evaluation of the performance of each member on an independent validation dataset during the training, we retain the best half of the members to form our final estimator.
This allows in particular to reject badly converged optimizations, and to verify the overall stability of the training procedure \citep[see also][]{Zhou:2002jm}.
We take averages of the predictions made by the retained committee members as output of a committee.
For the weight-predicting NNs described in Sect.~\ref{subsec:weights}, the unconstrained scale of the predicted weights could potentially require a prior normalization.
In practice, we observe however that the use of the sigmoid output activation function results in members predicting weights of very similar scales.

Finally, we note that our implementation allows to individually mask realizations of each case, which is important to handle failures of the input feature measurements, discussed in the next section.

\section{Feature measurement on galaxy images} \label{sec:InputFeatures}

The raw data of a weak-lensing study consists of survey images.
In this section we describe how we measure a small set of features based on moments of the observed galaxy light profiles from which the shear is to be inferred.
Those features will serve as input to the machine learning algorithm, potentially together with information from a PSF-model, multiband photometry, or other relevant parameters.
For this exploratory work we deliberately opt for a small number of selected features describing the galaxy images, to ease experimentation, efficiency, and also to set a benchmark.
Deep-learning approaches with convolutional NNs, which directly learn filters to extract optimal galaxy features from image pixels are an obvious alternative \citep{Tuccillo:2018dj, DIsanto:2018jd}.
However, we expect that few simple ``hand-crafted'' features\footnote{In machine learning, ``hand-crafted'' features are statistics designed and selected by an expert, versus ``learned'' features which are developed by an algorithm based on training data.} are sufficient to capture a very large fraction of the shear information from the noisy galaxy images of interest to a weak lensing analysis, especially on simple simulations.

\subsection{Adaptive weighted moments} \label{sec:adaptativemoments}

To describe the galaxy shapes we use statistics based on moments computed with an adaptive elliptical Gaussian weight function (in contrast to the circular weight function used in \citet{Tewes:2012gy}, which we observe to yield less precise results).
We employ the well-tested and efficient implementation offered by the \texttt{HSM} module of the \texttt{GalSim} software package \citep[][and references therein]{Bernstein:2002gq,Hirata:2003ji,Mandelbaum:2012gj,Rowe:2015ema}.
The same or very similar moment computations are used in other shape measurement techniques, such as DEIMOS \citep{Melchior:2011je} and the methods directly implemented within \texttt{GalSim}.

To stress the computational nature of these features and connect them with the \texttt{HSM} implementation, we denote them in a fixed-width font.
We define the following moment-based features:

\begin{description}
\item[\texttt{adamom\_flux}] corresponds to the total source flux of the best-fit elliptical Gaussian profile (\texttt{ShapeData.moments\_amp} in \texttt{GalSim}), expressed in ADU. This is a biased estimate of the flux of any realistic (i.e., non-Gaussian) galaxy profile, but such biases have no direct consequences for ML input features.
\item[\texttt{adamom\_g1} and \texttt{adamom\_g2}] are components of the observed ellipticity (\texttt{ShapeData.observed\_shape.g1}/\texttt{2} in \texttt{GalSim}), which would correspond, for a simple elliptical Gaussian profile and without PSF, noise, and pixellation, to the ellipticity defined in Sect.~\ref{sec:wl} as an estimator for shear.
\item[\texttt{adamom\_sigma}] gives a measurement of the radial extension of the profile, in units of pixels (\texttt{ShapeData.moments\_sigma}).
In the case of a circular Gaussian profile, it would estimate its standard deviation.
\item[\texttt{adamom\_rho4}] gives a weighted radial fourth moment of the image, measuring the concentration, i.e., a kurtosis, of the light profile (\texttt{ShapeData.moments\_rho4} in \texttt{GalSim}).
\end{description}

\subsection{Noise measurement and signal-to-noise ratio}

The signal-to-noise ratio (\snr) of galaxy images is a key quantity when assessing the quality of a shear measurement.
A scientific analysis of a shear catalog will tend to include galaxies with a \snr as low as tolerable, for a given shear measurement technique.
Unfortunately, \snr measurements mentioned across the literature are often difficult to compare, as the observational definition of a \snr is not trivial and not always fully described. In the following, we present the simple observational \snr that we use to evaluate our method.

First, we quantify the background pixel noise for each target galaxy using a rescaled median absolute deviate (MAD) to estimate the standard deviation \citep[see, e.g.][]{Rousseeuw:1993jp}
\begin{equation}
\label{equ:skymad}
\sigma_{\textrm{sky}} \doteq 1.4826 \cdot \textrm{median}( | \xi_i - \textrm{median}(\xi_i) | ) ,
\end{equation}
where $\xi_i$ are the pixel values (in ADU) along the edge of a ``stamp'' of sufficient size centered on the target galaxy. Generalizations of this procedure, for better precision, are easily conceivable if required. The robust MAD statistic has the advantage, over a plain standard deviation, that potential field stars, galaxies, or image artifacts on the stamp edge have a reduced impact.

In the second step, we combine this background pixel noise measurement with the results from the adaptive moment measurements described above to obtain a \snr. Naturally, our definition of \snr follows from the CCD equation \citep[see, e.g.,][]{Chromey:2010vv}, and we choose a circular aperture with a radius of three times the measured half-light radius of the source as effective area for the background noise contribution. More precisely,
\begin{equation}
\label{equ:sn}
\snr \doteq \frac{G \cdot \texttt{adamom\_flux}}{\sqrt{G \cdot \texttt{adamom\_flux} + A_{\textrm{eff}} \cdot (G \cdot \sigma_{\textrm{sky}})^2}},
\end{equation}
where
\begin{equation}
\label{equ:aeff}
A_{\textrm{eff}} \doteq \pi \cdot \left(3 \cdot \texttt{adamom\_sigma} \cdot \sqrt{2 \cdot \ln(2)}\right)^2,
\end{equation}
and $G$ is the gain in electrons per ADU.
For a Gaussian profile, the numerical factor $\sqrt{2 \cdot \ln(2)} \approx 1.1774$  would rescale the standard deviation into the desired half-light radius.
This choice of effective aperture $A_{\textrm{eff}}$ has a strong influence on the \snr, and might seem arbitrary as galaxy light profiles are not Gaussian.
We observe however that this definition gives results within a few percent of the ratio $\texttt{FLUX\_AUTO} / \texttt{FLUXERR\_AUTO}$ given by the \texttt{SExtractor} software \citep{sextractor, Bertin:2010wt}, for all simulations considered in this paper.
The advantage of defining our own measure of \snr based on the described simple input features is to ease reproducibility and to avoid introducing the dependency on an additional software.
 
\begin{figure}[tbp]
\begin{center}
\includegraphics[width=0.7\linewidth]{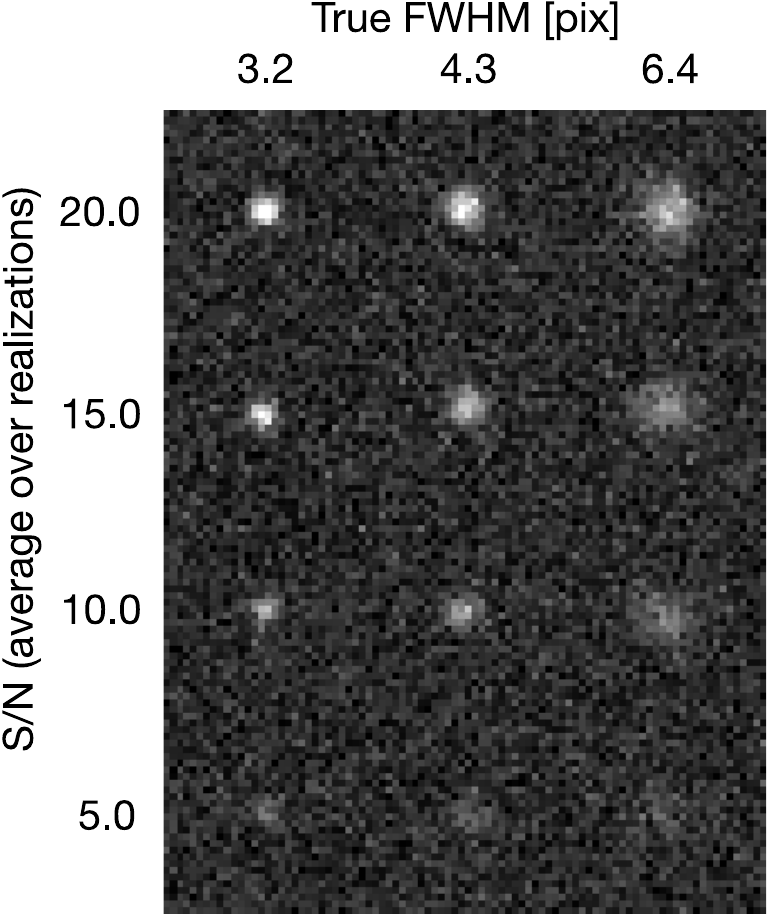} 
\caption{\label{fig:snr}
Illustration of the \snr on simple Gaussian profiles with Gaussian pixel noise. The fluxes are chosen so that the average \snr, measured on many realizations of each source, matches the scale given on the left.}
\end{center}
\end{figure}

To mimic ``sky-limited'' observations, simulated images are sometimes drawn purely with a stationary Gaussian noise. In this approximation, Eq.~\ref{equ:sn} simplifies to 
 \begin{equation}
\label{equ:sngaussian}
\mathrm{\snr_\mathrm{Gaussian}} \doteq \frac{\texttt{adamom\_flux}}{\sqrt{A_{\textrm{eff}} \cdot \sigma_{\textrm{sky}}^2}}.
\end{equation}
We show some simulated sources with Gaussian profiles for different \snr and sizes in Fig.~\ref{fig:snr} (see also Fig.\ref{fig:snrgems} for an illustration with PSF-convolved elliptical S\'ersic profiles).

We note that for ML shear measurement, a measured \snr is potentially an interesting input feature of each galaxy, especially if the number of features needs to be small \citep{Tewes:2012gy}.
However, in the following, we will not use the \snr as input feature, but provide instead separately the more fundamental flux and size measurements to the ML algorithm, complemented by a sky noise measurement if required. 
This use of flux instead of \snr allows, in particular, for testing a single training on test sets with different noise levels, or for training on data with a lower noise than the actual observations. Nevertheless, we will extensively use the observed $\mathrm{S/N}$ defined above in the analysis of shear estimation biases.

\section{Machine learning shear estimation}
\label{sec:shearestimation}

We now describe how we use and train NNs to predict an estimate for the shear of each galaxy, using the NN cost functions and the input features introduced in the previous sections.
We focus on the core principles of the ML approach, and defer for now the numerous complications that a full shear measurement pipeline has to face.

Recall that we consider here the prediction of point estimates of the shear components $\hat{g}_i$, $i \in \{1, 2\}$, and associated weights which we denote $w_i$.
We will predict these point estimates and weights with independent NNs that are trained with different cost functions.
For the sake of simplicity, we also distribute the predictions related to the two components to independent networks, instead of considering networks with multiple output nodes.
We therefore train four scalar estimators, each consisting of a committee of several NNs for increased robustness.

Depending on the conditions in which the shear estimation method is to be applied, such as ground- or space-based data, variability of the sky background, instrumental effects in the data, selection of the source galaxies, accuracy to be achieved, different ways to setup and train these estimators can be considered.
In the following, we present and motivate one simple fiducial approach in generic terms, using two different ``training sets'', that is forward-simulations of observed galaxies with known shear.

\subsection{Step I: shear point estimates with low conditional bias} \label{sec:predict_shear}
\label{subsec:shearest}

We start by training the two shear point estimators $\hat{g}_i$.
A simple toy-model choice of the input features could be measures of the ellipticity components, the flux, the size of the observed galaxy image, the noise of the sky background, and the ellipticity and size of the PSF model at the location of the considered galaxy.
These eight input values summarize key information required by the shear estimator to account for the PSF shape and noise bias.
We note that a different approach to inform the ML about the variability of a space-telescope PSF is discussed in Sect.~\ref{sec:fiducial:varpsf-field}.

We use the MSB cost function (Eq.~\ref{eq:msb}), which, muting the explicit dependency on the training data, takes the form
\begin{equation}\label{eq:shearcost}
\mathrm{MSB}(\bm{p}) = \frac{1}{\ncase} \sum_{k=1}^{\ncase}
\left[
\frac{1}{\nrea} \sum_{j=1}^{\nrea} \hat{g}_{jk}(\bm{p}) - g^{\mathrm{true}}_k
\right]^2
\end{equation}
with $\bm{p}$ designating the parameters of the estimator.
Recall that the networks for the two components of $\hat{g}_i$ are entirely independent.
For each component, $\hat{g}_{jk}(\bm{p})$ designates the predicted value for the realization $j$ of the case $k$.

The structure of the training set, that is its composition of cases and realizations, is the next most important choice.
To train the estimator to be both accurate and as insensitive as possible to the distribution of true galaxy properties, we aim at penalizing its ``conditional'' bias, that is its bias in any subregion of this true parameter space.
In other words, we aim at a potential estimator which would be accurate for any PSF, and any true galaxy size, elongation, flux, etc.

\begin{figure}[tbp]
\begin{center}
\includegraphics[width=1.0\linewidth]{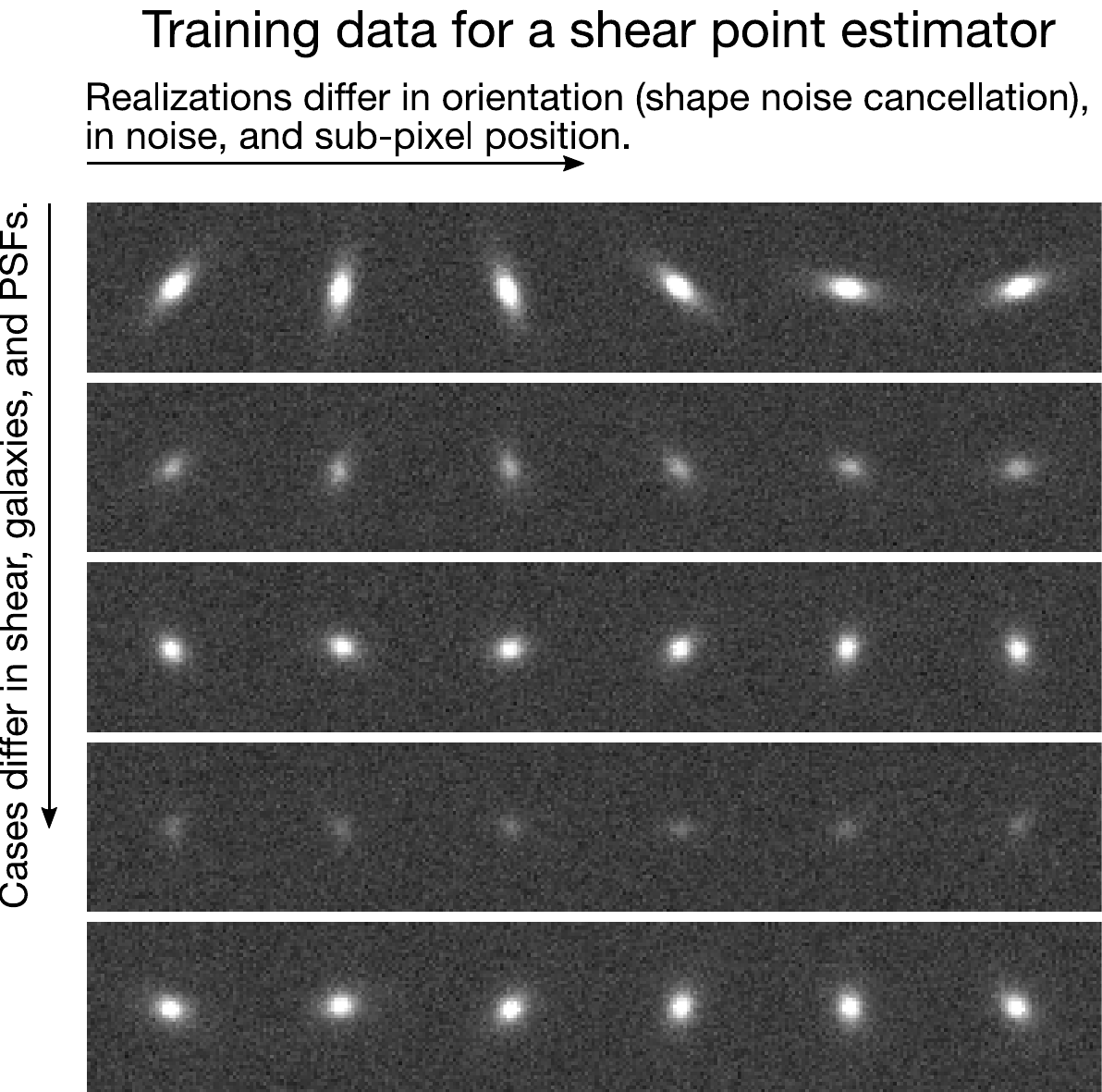} 
\caption{\label{fig:shear}
Illustration of the structure of a training set to train a shear estimator $\hat{g_i}$ with an MSB cost function.
The horizontal frames correspond to different ``cases'', each containing different ``realizations'' of a galaxy.
All galaxies of a case are simulated with the same true shear, and the same PSF.
Despite the circular symmetry of the PSFs used in this illustration, the typical cosmic shear is too small to be noticed by eye.
}
\end{center}
\end{figure}

We generate a training set as illustrated in Fig. \ref{fig:shear}.
Within each case $k$, the realizations share the same true shear $g^{\mathrm{true}}_k$ (the target value for the training), but also the same value for other explanatory variables that we can request the estimator to attempt to become insensitive to, given the information it obtains from its input features.
Consequently, each case contains only one true galaxy combined with one particular PSF, always seen under the same shear.
While other aspects of the data, such as the position angle of the galaxy, its exact position on the pixel grid, and the realization of the pixel noise do have a direct influence on the shear estimate, they have to be dealt with statistically.
Indeed, a shear estimator cannot be insensitive to the intrinsic orientation of a galaxy, which is degenerate with the shear.
This orientation acts as a form of unavoidable shape noise for the shear measurement.
Therefore, within each case, we draw $\nrea$ realizations of these noise sources, and train the estimator to yield unbiased predictions despite this noise and pixellation.

The required value $\nrea$ to sufficiently average-out the noise effects with respect to a significant bias can be reduced by noise cancellation techniques.
With such techniques, a controlled ensemble of compensating samples is taken, to improve the precision on the bias of a case beyond what would be achieved by randomly drawing the realizations.
In Fig. \ref{fig:shear}, the intrinsic orientations of the galaxies are rotated in regular intervals on a ring in the ($\varepsilon_1$, $\varepsilon_2$)-plane, so that the average intrinsic ellipticity within each case exactly vanishes \citep[following ][]{Nakajima:2007gv}. 
Such techniques have become known as shape noise cancellation \citep[see, e.g., ][and references therein]{Mandelbaum:2014dq}.

Let us consider again the cost function.
If a hypothetical estimator would achieve a zero MSB cost, for an infinite amount of realizations per case, this estimator could be said to be fully insensitive to the distribution of galaxy and PSFs it is presented with, among the population it was trained on.
It is important to acknowledge that this is not possible in practice for all regions of this ``true'' parameter space: consider the example of an intrinsically small, unresolved galaxy, whose observed shape will not carry shear information.
The PSF, the noise, and the pixellation lead to a loss of information which cannot be compensated for by the point estimator.

This limitation has important consequences.
The presence of ``difficult'' cases in the training set, such as unresolved galaxies without useable shear information, or cases for which the precision is insufficient, can affect the performance of the estimator even in ``easier'' regions of parameter space.
Indeed, the cases are connected by the one scalar MSB cost function summarizing the whole training set.
If the features do not allow the NN to differentiate well enough between these ``difficult'' and ``easy'' cases, or if ML capacity is insufficient to exploit the features, the ML might have to settle with bad performance on easy cases in order to avoid excessively bad scores on the difficult ones. 
It's a price of the simple sequential point estimate and weight training explored in this paper.

The role of the second step is to build a function that downweights galaxies from which an unbiased estimate cannot be obtained.

\subsection{Step II: weight prediction}
\label{subsec:shearweights}

\begin{figure}[tbp]
\begin{center}
\includegraphics[width=1.0\linewidth]{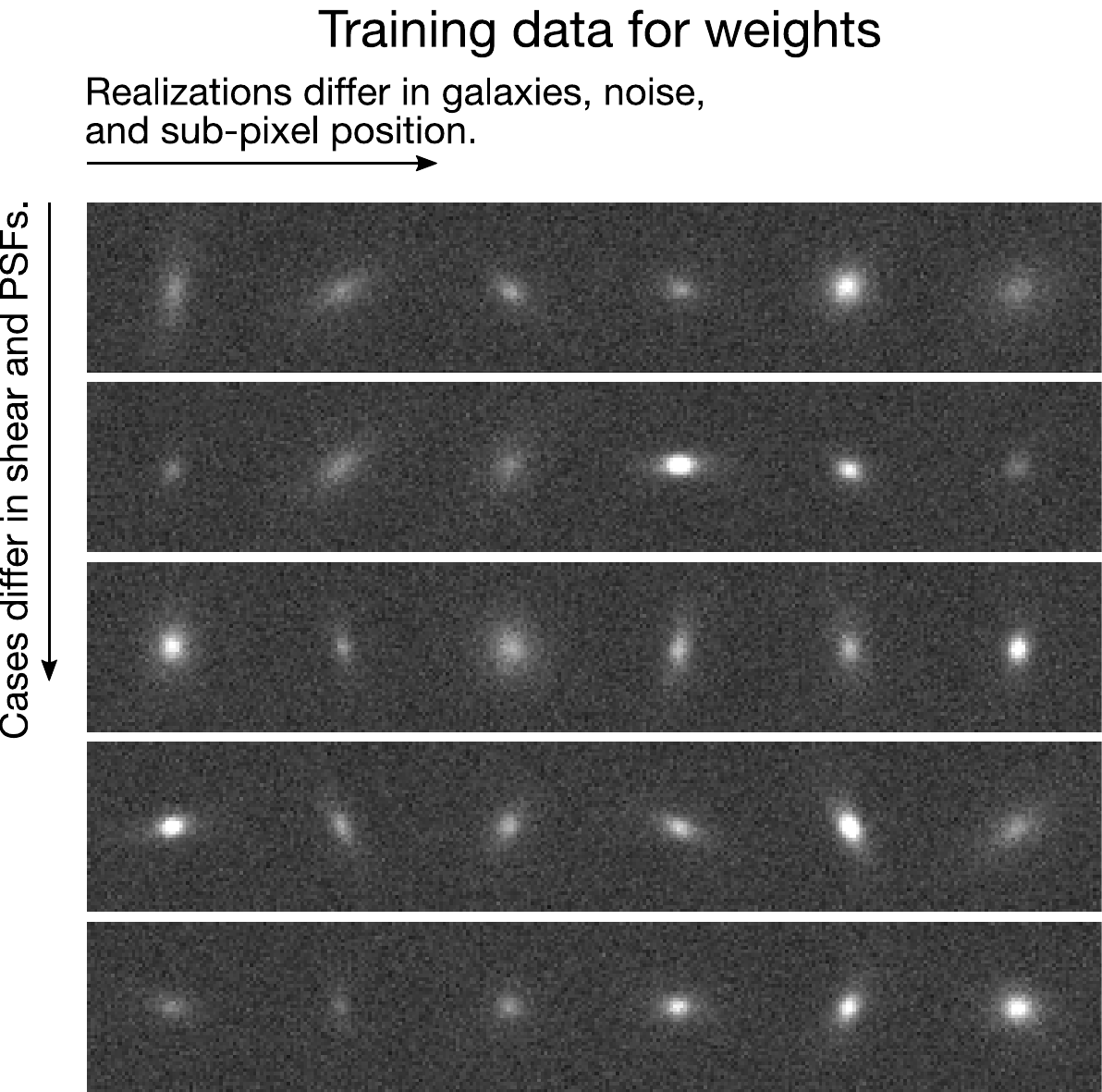} 
\caption{\label{fig:shearw}
Structure of a training set to train a weight estimator, $w_i$, with a MSWB cost function.
Within each case, this training informs the method about approximated distributions of properties of the source galaxies and selection functions.
}
\end{center}
\end{figure}

Given the estimators $\hat{g}_i$, $i \in \{1, 2\}$, we now train independent NNs to predict the associated weights $w_i$.
We use an MSWB cost function from Eq. (\ref{eq:mswb}), which can be written, separately for each component of the shear and its estimators, as
\begin{equation}\label{eq:shearwcost}
\mathrm{MSWB}(\bm{p}_{\mathrm{W}}) = \frac{1}{\ncase} \sum_{k=1}^{\ncase}
\left[
\frac{ \sum_{j=1}^{\nrea}  \hat{g}_{jk} \cdot w_{jk}(\bm{p}_{\mathrm{W}}) }{\sum_{j=1}^{\nrea} w_{jk}(\bm{p}_{\mathrm{W}})} - g^{\mathrm{true}}_k 
\right]^2.
\end{equation}
We recall that the $w$ are constrained to the interval $(0,1)$ by design of the NNs, and that the estimates $\hat{g}$ are to be computed ahead of the training of the weight-predictor, for each galaxy in the training data.

Again, putting aside technical details of the ML algorithm, we consider the choice of input features and the structure of the training data.
Regarding the input features, it could seem intuitive that a small set of features, describing for example the observed size and flux, is sufficient to optimally down-weight low-\snr source galaxies for which an unbiased shear estimation cannot be achieved.
After all, if the $\hat{g}_i$ achieve low conditional biases, the act of removing intrinsically small and faint galaxies from the sample cannot introduce any additional biases.
This reasoning is however wrong, as we don't have access to any true galaxy parameters, which are uncorrelated with the shear.
Instead, the input features, including the measurement of size and flux, are based on the observed galaxy and will inevitably show dependencies on the shear, at some level.
Using such a small set of features would lead to a shear-dependent weighting, and thereby lead to biases even if the $\hat{g}_i$ itself is accurate.
Weighting acts in this regard exactly as any selection function, leading to selection biases \citep{Miller:2013bb, Kaiser:2000bw, Bernstein:2002gq}.

Two conclusions can be drawn from this observation.
First, it is justified to maintain the full set of features when training the weight-estimator, so that the weights can exploit the full information from each source to counter selection effects.
Second, if selection biases prior to the shape measurement are affecting the data, this step of training the weights is a natural place to inform the ML-alogrithm about the selection function.
Applications in Sects.~\ref{sec:fiducial} and~\ref{sec:euclid}  will illustrate this point.

We structure the training data for the weights as illustrated in Fig. \ref{fig:shearw}.
The realizations within a case still all share the same true shear and PSF, but now also sample ideally the full population of observed galaxies.
By this structure, we therefore aim at predicting weights so that, for any shear and any PSF, the overall shear prediction error (both statistical and bias) gets minimized.

We stress that the introduction of these weights, estimated on the noisy observations, to the shear estimation formalism will potentially re-introduce some small conditional biases that we attempted to minimize in Sect.~\ref{subsec:shearest}.
For example, the different realizations of a galaxy shown in Fig. \ref{fig:shear} will get slightly different weights.
Given the loss of information in the observation process, it is expected that a shear estimator cannot be fully insensitive to the true galaxy parameters.
The approach presented here attempts to minimize this sensitivity to the smallest achievable level.

Furthermore, we stress that the trained weight-estimator does depend on the distribution of galaxy properties in the training data.
To pick again an extreme example for the purpose of illustration, the prevalence of unresolved galaxies (or mis-identified stars) in the source population will influence how conservative the rejection of small observed galaxies needs to be in order to avoid biases.

Finally, we note that in the context of the weight training described above, shape-noise cancellation (SNC) should in general not be used in the simulated datasets.
As the shape-measurement precision increases with \snr, SNC is more efficient on high-\snr than on low-\snr galaxies.
On simulations with SNC, the training of $w_i$ would easily learn to exploit this, by yielding weights that exaggeratedly favor bright and large galaxies.
Weights trained in this fashion are closer to optimal weights for ellipticity measurement than to weights for shear measurement, and would have to be corrected before being applicable to real survey data.
For simplicity, we consider in this paper the direct estimation of weights for shear estimation, and therefore avoid when feasible the use of SNC in our training and validation sets.

\subsection{On the estimation of ellipticity, size, magnification or other parameters}
\label{sec:ellipshortcut}

Variants of the described approach can be used to estimate parameters other than shear, such as galaxy shape model parameters. 
More generally, any parameter defined by a measurement on an idealized source, as would be seen with an infinite-resolution and noise-free imaging system, can serve as target for the ML output.

\begin{figure}[tbp]
\begin{center}
\includegraphics[width=1.0\linewidth]{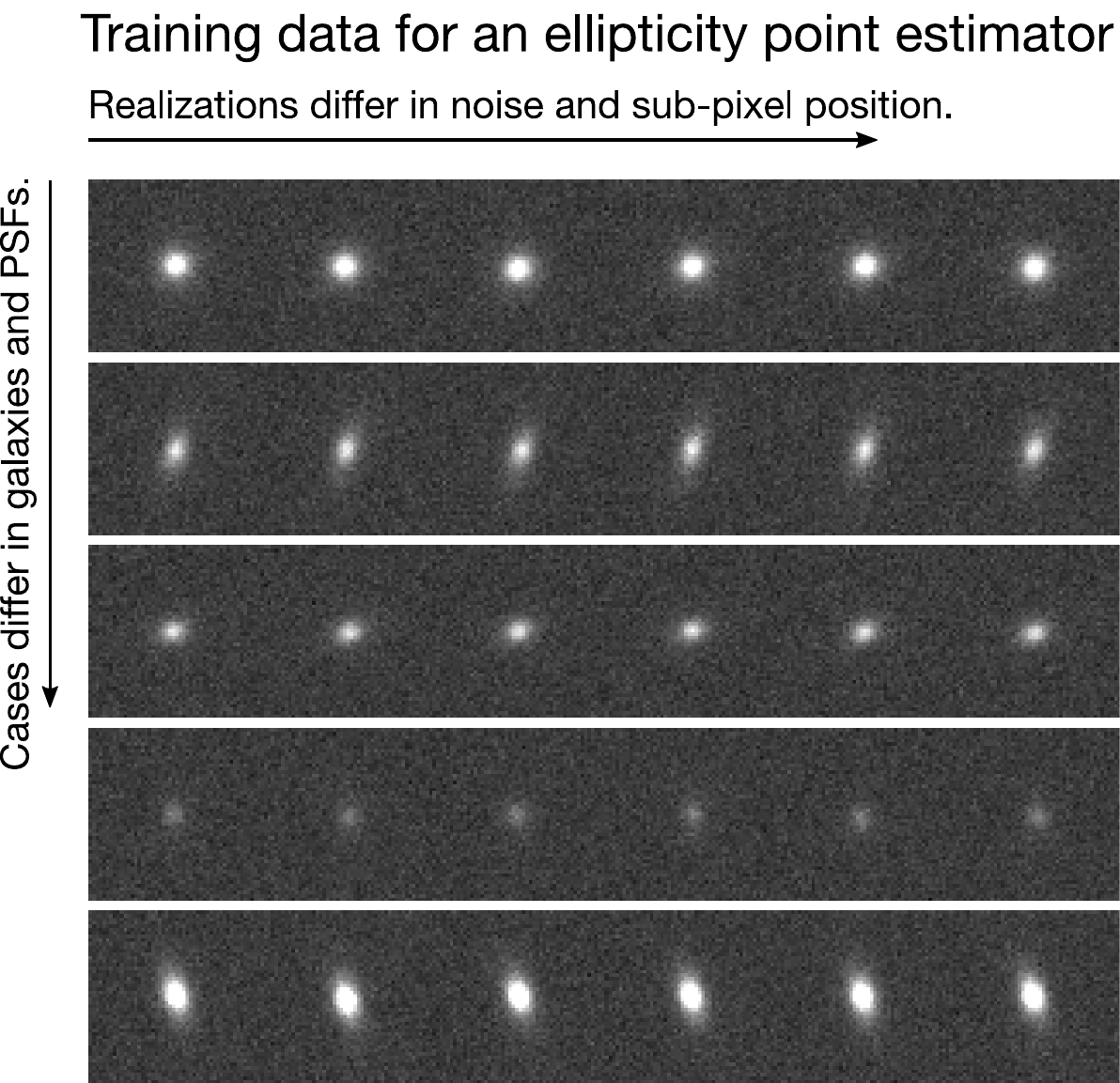} 
\caption{\label{fig:ellip}
Structure of a training set to train an ellipticity estimator, which can serve as a shortcut to the training of a shear estimator, when considering galaxies with simple elliptical profiles.
}
\end{center}
\end{figure}

Let us first consider the estimation of the ellipticity components of idealized galaxy images, as defined before PSF-convolution and noise (see Sect.~\ref{sec:ellip}).
Recall that real galaxy profiles have no simply-defined ellipticity.
However, if the use of galaxies with complex morphology is required, ellipticity measurements can be used as target values in an identical manner.
Figure \ref{fig:ellip} illustrates a simulation structure to train a point estimator for ellipticity.
Cases cover a variety of galaxies and PSFs, and the realizations within each case differ only in noise and their exact positioning on the pixel grid.
No shear is added to the training simulations.
The set of input features remains unchanged from the previous examples, and the NNs are trained with an MSB cost function.
Under the hypothesis of the idealized galaxy morphology, the resulting point estimate of the ellipticity can be seen as an estimate for the shear.
Associated weights for this use can then be trained exactly as done in Sect.~\ref{subsec:shearweights}.

Such a prediction of ellipticity is of particular interest, as it provides a computational shortcut for shear estimation, and we will later use it to pretrain NNs.
The advantage, with respect to training a shear point estimator, is that fewer realizations per case are needed to achieve comparable results, as the shape noise has been removed from the problem.

Another estimator of interest is the angular size of a galaxy, again before PSF-convolution and noise effects.
The availability of an accurate size measurement is mandatory for galaxy size-magnification studies \citep[][]{Schmidt:2011ju}, which suffer from the same instrumental bias sources as the ones affecting shear measurements.
An approach to directly predict a magnification estimator could also be explored, in analogy to the shear estimator presented in this paper.
Doing so, the ML algorithm could learn to exploit physical correlations between galaxy properties \citep[see, e.g.,][]{Huff:2013bc} while compensating for the observational correlations introduced by the measurement process on noisy images.
Discussing these estimators in more detail is, however, beyond the scope of this paper.

\subsection{Practical notes on the training convergence and data}
\label{sec:pracMLnotes}

The successful training of supervised ML alogrithms typically requires some experimentation with hyperparameters, such as the size of the NNs and the size of a training set.
In the following, we briefly list important observations and advices which ease the methodical optimization of the architecture and training of the neural networks.
While some of these suggestions might seem elementary to ML-practitioners, we detail them in the particular context of the presented galaxy shape measurement problem.
We assert that these principles are useful for any ML shear measurement approach.

\begin{enumerate}

\item \emph{Validation set}: arguably the most important idea is to always use a separate validation dataset to evaluate a training performed on some training dataset.
This validation set can be simulated in the similar way as the training set, but should otherwise be independent (i.e., contain different cases and realizations).
Monitoring the cost function value on both validation and training sets during the training allows the detection of training convergence and potential overfitting of the ML algorithm.
Overfitting could happen if the training set is too small (in terms of $\nrea$ and $\ncase$) and/or if the NN-capacity is unnecessarily high.
Validation sets with both structures shown in Fig. \ref{fig:shear} and \ref{fig:shearw} are useful. The first one can be used to test the achieved quality of the point shear estimator, and the second one to test the overall shear estimate including the predicted weights.

\item \emph{Start with small NNs}: the experimentation should start with few and simple features and a very low-capacity network, to obtain a benchmark solution.
Before adding features, or increasing the NN capacity by adding nodes, a validation set of sufficient size on which one can clearly visualize the limitation of the benchmark solution should be available.
We would like to point out that the shear estimation problem based on the input features described in this paper does not require a large capacity.
Indeed, the dependency of the shear estimate on the observed features is rather smooth, and no discontinuities are expected.

\item \emph{Training set adjustments}: it is often advantageous to use different source galaxy property distributions in the training set than in the data one wishes to process. 
In particular, when training the shear point estimator with MSB (Sect.~\ref{subsec:shearest} and Fig.~\ref{fig:shear}), cases from which no unbiased shear estimates can be expected may harm the training and should be avoided. 
A typical example is given by unresolved galaxies, with an intrinsic extension much smaller than the PSF.
Their observed features will carry no shear information. 
Even a small number of these cases can dominate the cost function value, and lead the NNs to overfit and yield biased estimates on much easier cases.
For the same type of training, it can also be beneficial to fill the true parameter space relatively uniformly, and to extend the range of true simulation parameters (such as shear, galaxy flux, and size) beyond what the real-sky data contains.
We stress that the training set for the weight prediction should mimic the real-sky data and therefore include all problematic cases in a representative way.
The same is true for an overall validation set.

\item \emph{Pretraining on ``simpler'' data}: the computational cost to train the point estimator with MSB can be reduced by starting the NN training on data from simulations with an artificially reduced noise.
The low noise allows for a smaller training set, and therefore a faster training, in many cases by at least an order of magnitude.
The motivation for such a pretraining is that the NNs can learn for instance how to perform the PSF correction in an efficient way.
Afterwards, the pretrained NNs are further optimized to correct for noise bias on the conventional training set, requiring far less iterations than if no pretraining was done.
If an input feature informs the NNs about the noise, it can be necessary to alter this feature during the pretraining, so that the values encountered by the NNs when training on low-noise data approximatively correspond to the values seen on the conventional simulations.

\item \emph{Amount of simulated data}: ideally, the size of a training set should be increased up to a point at which no further improvement of predictions made on an even larger validation sets is seen.
As a rule of thumb, the validation set size needed to probe biases to some desired accuracy gives a good indication of the required training set size.
For example, as the different shear ``cases'' of a constant-shear GREAT3 branch contain 10\,000 galaxies each (including shape noise cancellation), one needs a training set with about as many realizations to obtain satisfactory results.

\end{enumerate}

\section{Application to fiducial toy-model simulations} \label{sec:fiducial}

\begin{table}[tbp]
\caption{Galaxy parameters of the fiducial experiments.}
\label{table:fidgalparams}
\centering
\begin{tabular}{l l} 
\hline\hline
Parameter & Distribution  \\ 
\hline
Shear components $g_1$, $g_2$ & $\mathcal{U}(-0.1, 0.1)$ \\
Intrinsic galaxy ellipticity modulus $\varepsilon^{\mathrm{true}}$ & $\mathcal{R}(0.2)_{[0,\,0.6]}$ \\
S\'ersic index\tablefootmark{a} $n$ & $\mathcal{U}(1.0, 4.0)$  \\
Half-light radius $R$ [pix] & $\mathcal{U}(2.0, 8.0)$ \\
Surface brightness $S$ [pix$^{-2}$] & $\mathcal{U}(1.0, 15.0)$ \\
\hline
\end{tabular}
\tablefoot{
We use $\mathcal{U}(a, b)$ to denote the uniform distribution between $a$ and $b$, and $\mathcal{R}(\sigma)$ for a Rayleigh distribution with mode $\sigma$. Intervals in subscript denote the range to which we clip a distribution, so that no sample falls outside of the given interval.\\
\tablefoottext{a}{In practice, we grid the values for the S\'ersic index instead of drawing them from a continuous distribution. This significantly speeds up the galaxy stamp generation, as \texttt{GalSim} can reuse cached S\'ersic profiles.}
}
\end{table}

As a first proof of concept of the proposed machine learning approach, we demonstrate it on a simple set of easily reproducible simulations, which we call ``fiducial''.
We stress that the main purpose of these simulations is to allow qualitative examinations and comparisons.
We do not seek to optimize or explore every aspect of the algorithm at this stage, and focus instead on illustrating the core ideas introduced in the previous sections with small NNs.
We defer experiments based on more realistic simulations to Sect.~\ref{sec:euclid}.

\subsection{Fiducial image simulation parameters}
\label{sec:fiducial:data}

We use \texttt{GalSim} \citep{Rowe:2015ema} to generate training and validation data in the form of \sersic profiles \citep{Sersic:1963uw} convolved with a Gaussian PSF, on stamps of 64 by 64 pixels.
For these first experiments the PSF is circular with a standard deviation of $2.0$ pixels, while we introduce a non-stationary PSF in Sect.~\ref{sec:fiducial:varpsf-field}.
Table \ref{table:fidgalparams} lists the simple distributions of true galaxy parameters that we use through all these experiments.
For efficiency reasons, we couple the size and flux distributions by first drawing a surface-brightness parameter $S$ for each galaxy, and then computing the true flux $F = S \cdot \pi R^2$, where $R$ is the true half-light radius of the \sersic profile.
This choice allows us to reduce the generation of undetectable galaxies with large size and a low flux.
For each realization of a stamp, the true position of the galaxy is uniformly drawn within one pixel around the stamp center, to simulate random pixellation.
We mimic background-limited noise conditions by drawing stationary pixel noise from a normal distribution with zero mean and a standard deviation of 1.0 counts.

\begin{figure}[tbp]
\begin{center}
\includegraphics[width=0.60\linewidth]{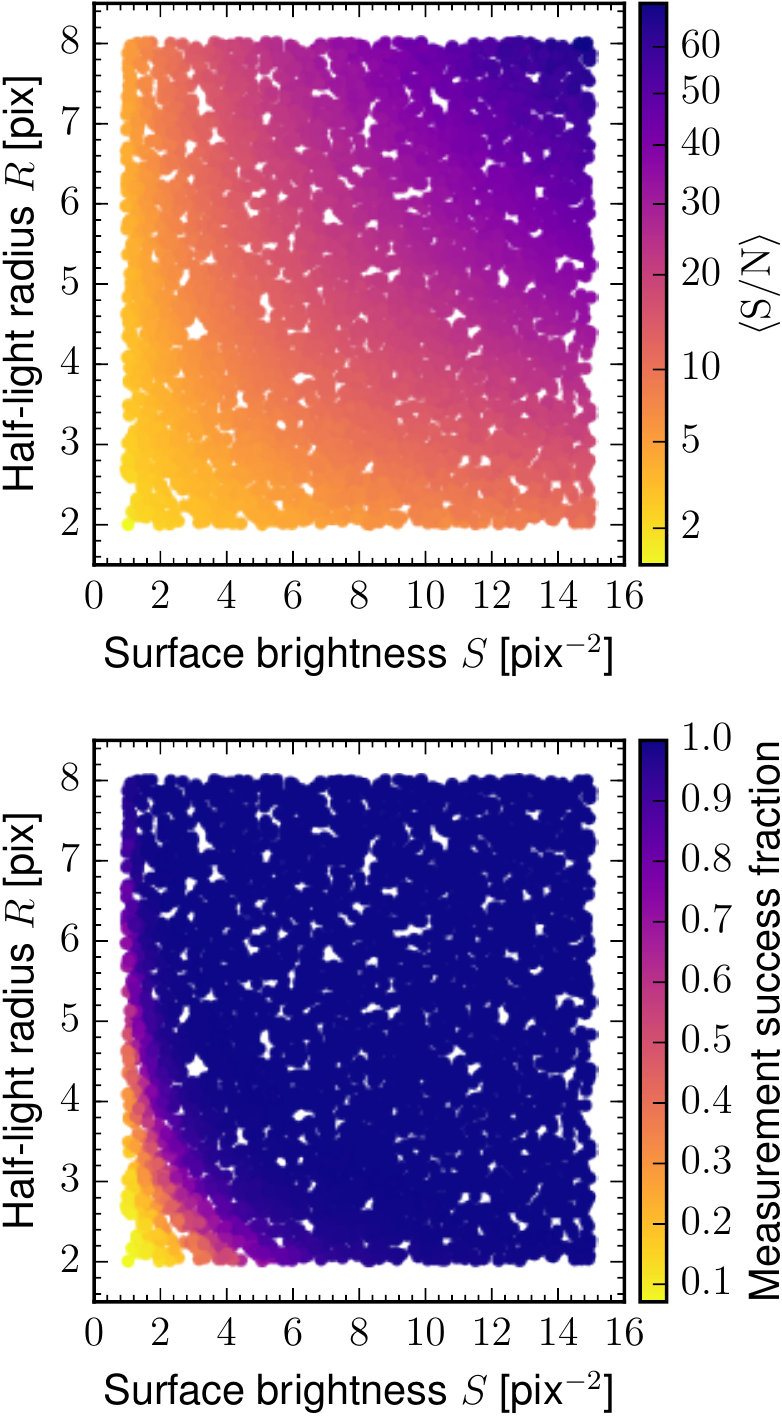} 
\caption{\label{fig:fid-snr}
Evolution of the \snr (\emph{top panel}) and of the selection function imposed by the feature measurement (\emph{bottom panel}) as function of the parameters $R$ and $S$ of the fiducial simulations. Each point is computed from 10\,000 rotated realizations of a galaxy, convolved by the stationary circular PSF used in Sect.~\ref{sec:fiducial:statpsf} (dataset {\tt VP}).
}
\end{center}
\end{figure}

Figure \ref{fig:fid-snr} shows the distribution of average observed \snr (as computed via Equation \ref{equ:sngaussian}) and of the relative frequency of feature measurement successes on these fiducial simulations.
The points in the two panels represent cases from a dataset as shown in Fig.~\ref{fig:shear}: both statistics are computed over many orientations and noise realizations of the same true galaxies.

Roughly 30\% of the galaxies drawn from the fiducial parameters have a \snr below 10.
We intentionally design the fiducial simulations to include those galaxies, and even to reach into regions of parameter space in which the feature measurement regularly fails due to the noise.
Our feature measurement via adaptive weighted moments imposes a selection on the galaxies, analogously to what the detection of sources in a real survey would do.
Any adaptive or non-trivial feature measurement will show a similar behavior.
With these fiducial simulations, we can therefore make a demonstration of handling a selection function with the ML-predicted weights.

\begin{figure}[tbp]
\begin{center}
\includegraphics[width=1.0\linewidth]{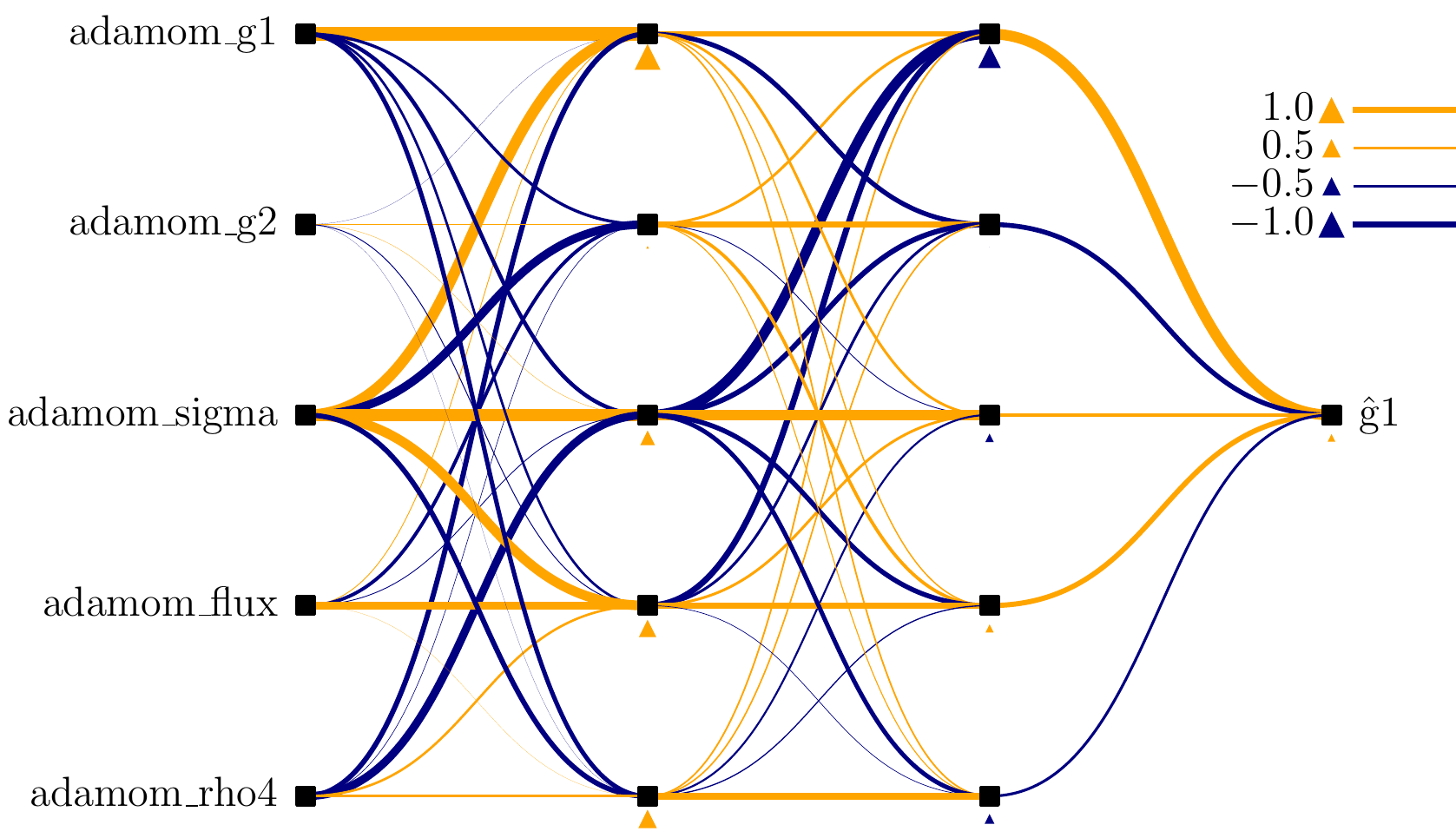} 
\caption{\label{fig:fid-netviz}
Visualization of one NN from the committee predicting a shear point estimate $\hat{g}_1$, trained for the fiducial experiment.
The NN nodes are represented by black squares, in a configuration with only two hidden layers of five nodes each.
The connections between these nodes depict the values of the NN weights, by their thickness and color.
The NN biases are visualized by the triangles below the nodes.
The legend gives the scale of these elements.
From the relative amplitude of the weights, one can observe, for example, that \texttt{adamom\_g2} has relatively low impact, while \texttt{adamom\_sigma} plays an important role in the prediciton of $\hat{g}_1$.
All input features are normalized prior to entering the NNs (see Sect.~\ref{subsec:nndetails} for details). 
}
\end{center}
\end{figure}

\begin{figure}[tbp]
\begin{center}
\includegraphics[width=1.0\linewidth]{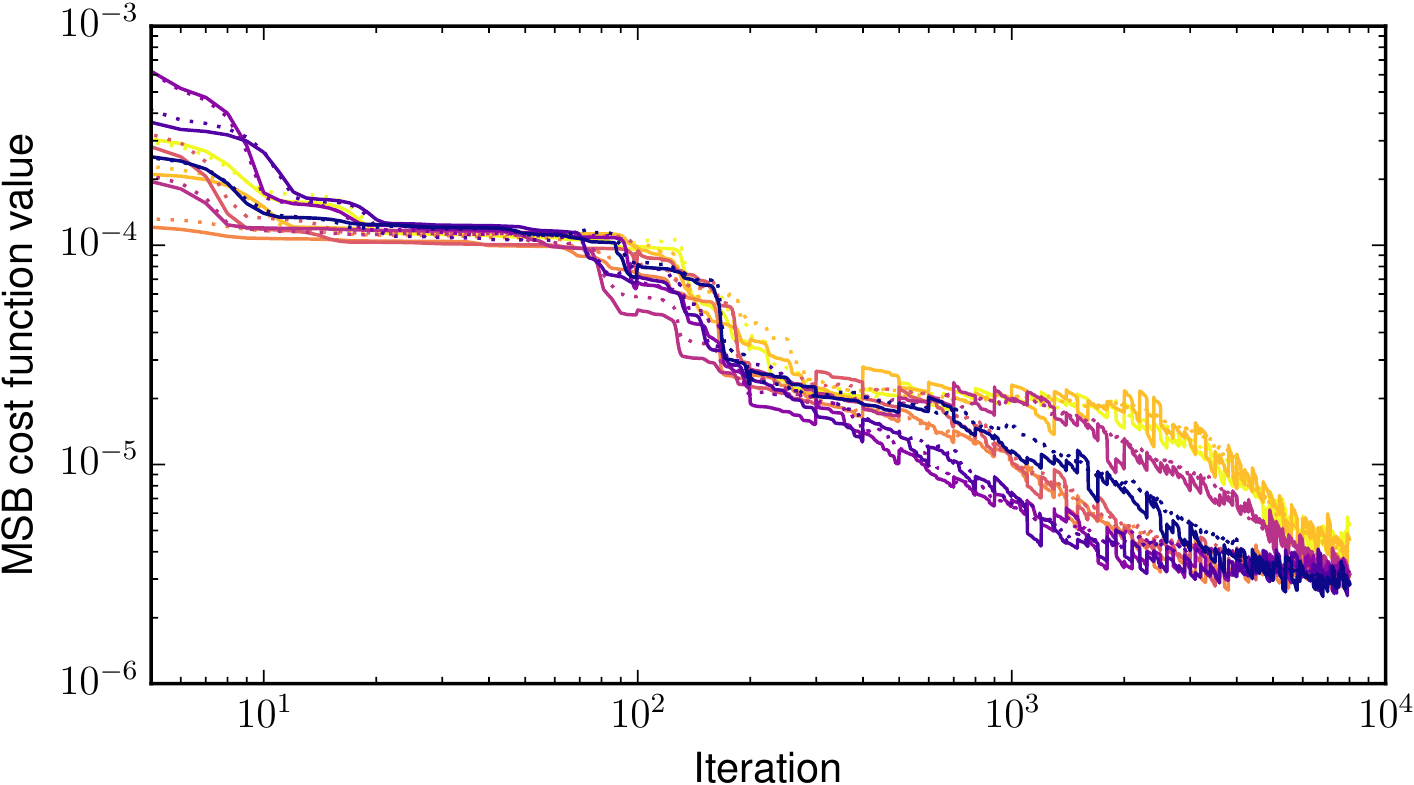} 
\caption{\label{fig:fid-msbevo}
Evolution of the MSB cost function value during the training of the shear point estimation for the fiducial experiment. Each color shows a different committee member. Evaluations on a validation set are shown with dotted lines.
Owing to the implemented mini-batch algorithm (Sect.~\ref{subsec:nndetails}), the cost function is not always monotonically decreasing.
}
\end{center}
\end{figure}

\begin{figure*}[tbp]
\begin{center}
\includegraphics[width=0.8\linewidth]{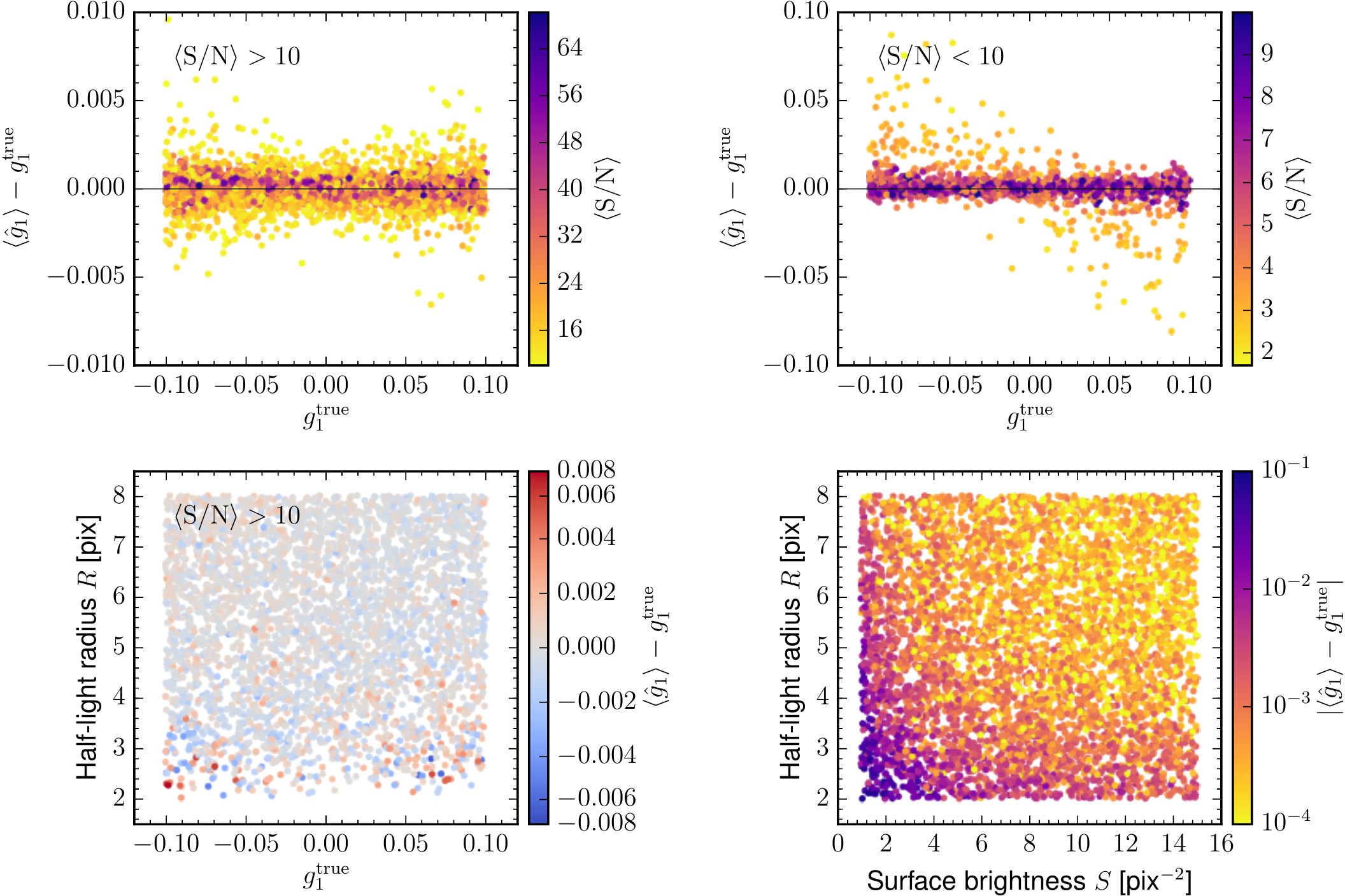} 
\caption{\label{fig:fid-resi}
Overview of the shear point estimation for the fiducial simulations.
Each point is a case of the {\tt VP} validation set.
\emph{Top panels}: residuals of the unweighted average $\hat{g}_1$ as function of the true shear $g_1^{\mathrm{true}}$ for the high-$\langle \snr \rangle$ cases (\emph{left}) and low-$\langle \snr \rangle$ cases (\emph{right}, on a wider scale).
The same residuals are shown as function of $g_1^{\mathrm{true}}$ and true half-light radius $R$ in the \emph{bottom left panel}, for the high-$\langle \snr \rangle$ cases.
\emph{Bottom right panel}: absolute value of the shear residuals as function of $R$ and the surface brightness $S$.
Results for $\hat{g}_2$ are highly similar.
}
\end{center}
\end{figure*}

\begin{figure*}[tbp]
\begin{center}
\includegraphics[width=0.75\linewidth]{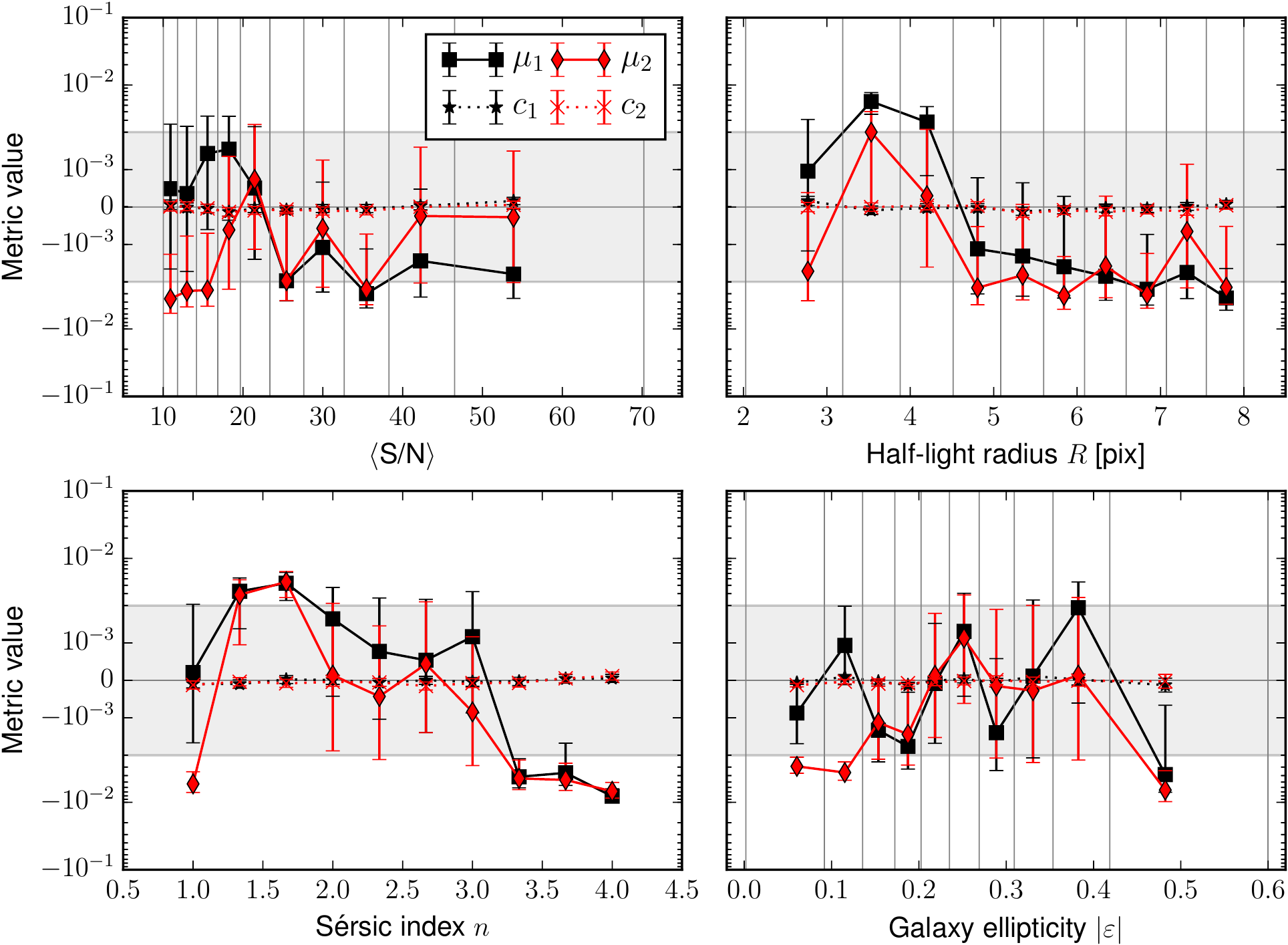} 
\caption{\label{fig:fid-condbias}
Multiplicative ($\mu_i$) and additive ($c_i$) bias parameters characterizing the $\hat{g}_i$ point estimates, for the fiducial experiment, as function of galaxy parameters.
The four panels visualize the dependence of these bias terms on the average \snr per case, the true half-light radius $R$, the \sersic index, and the intrinsic ellipticity $\varepsilon$ of the galaxies.
In each panel, the $\mu_i$ and $c_i$ are shown respectively with solid and dotted lines, in black for the first component, in red for the second (see legend).
The bin limits of source galaxy parameters divide the population into quantiles and are indicated with light vertical lines, except for the \sersic index, which follows a discrete distribution.
The parameters are shown on a linear scale within the shaded area, and using a logarithmic scale outside of this region.
Error bars depict $1\sigma$ uncertainties originating from the validation set. In addition, the training itself introduces some noise, so that details of these curves vary depending on the realization of the training set, without changing the qualitative observations discussed in the text.
}
\end{center}
\end{figure*}

\begin{figure}[tbp]
\begin{center}
\includegraphics[width=0.94\linewidth]{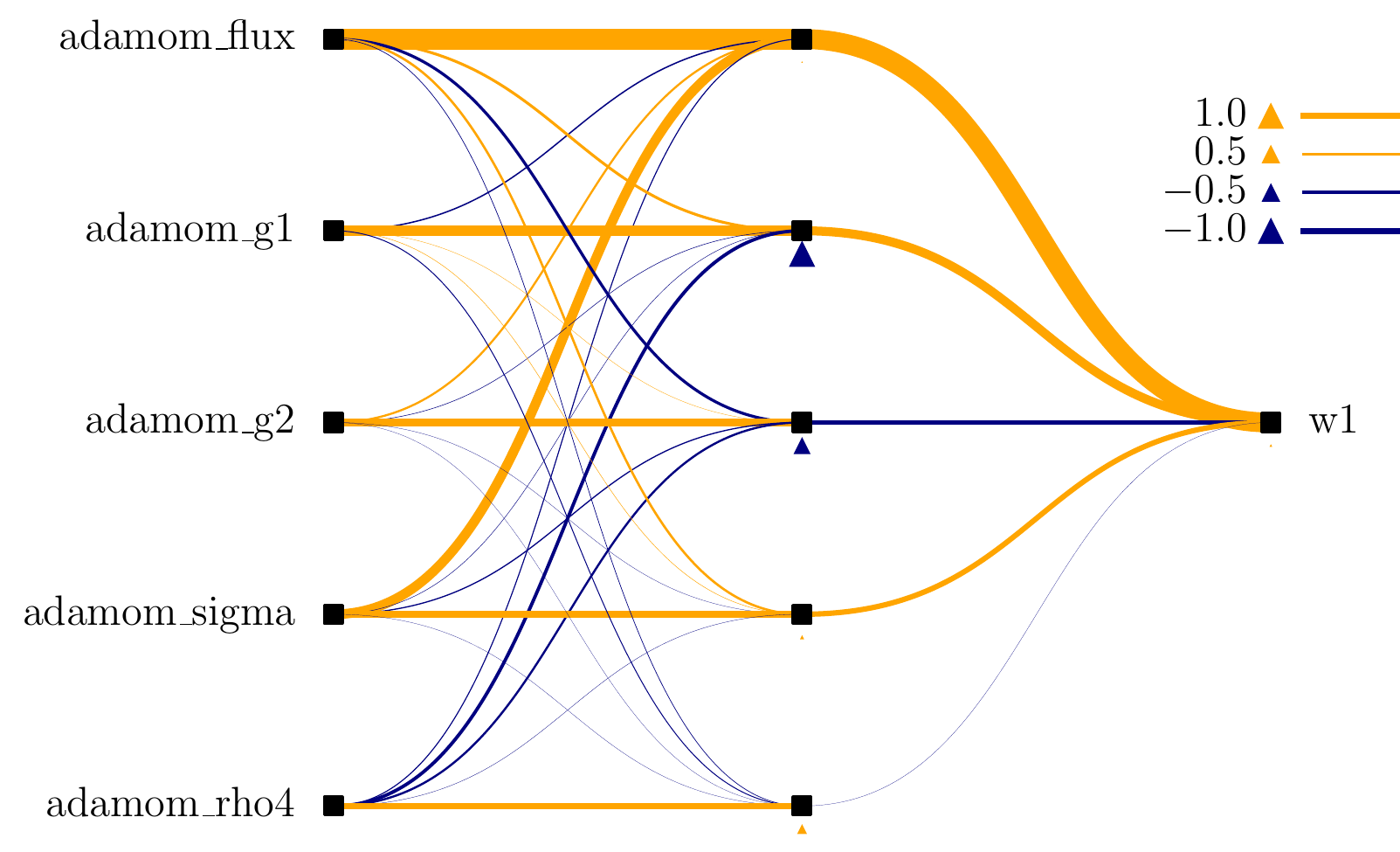} 
\caption{\label{fig:fid-netvizw}
Similar to Fig.~\ref{fig:fid-netviz}, but showing a network predicting a weight estimate $w_1$.
}
\end{center}
\end{figure}

\begin{figure*}[tbp]
\begin{center}
\includegraphics[width=1.0\linewidth]{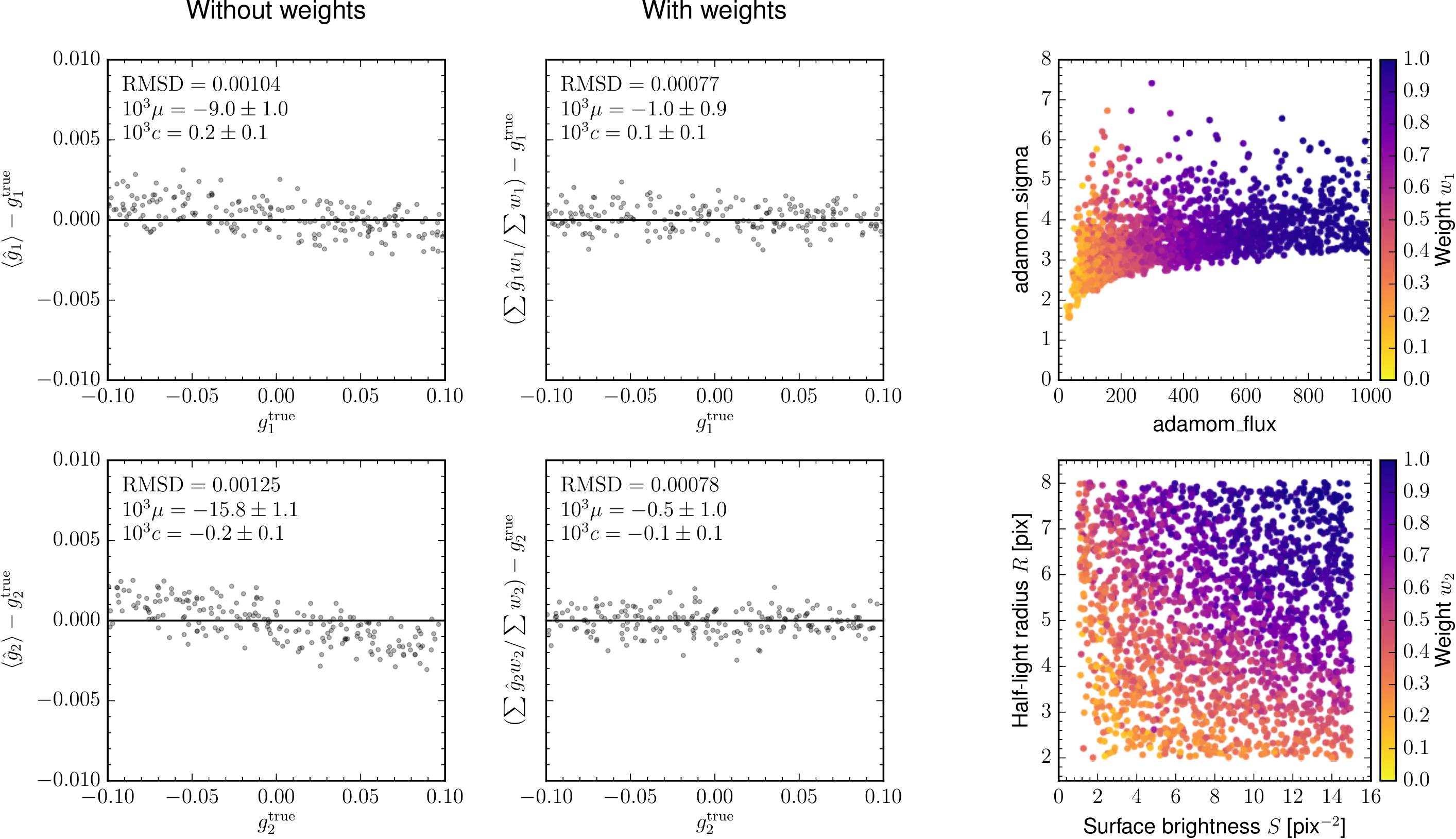} 
\caption{\label{fig:fid-wbias}
Overall validation of the shear estimation of the fiducial experiment with stationary PSF, demonstrating the achieved performance on galaxies selected only by the measurability of the input features (see Fig.~\ref{fig:fid-snr}).
The \emph{leftmost panels} show residuals of the unweighted average point estimates as function of the true shear component (one point per case of the {\tt VO} dataset), while the \emph{central panels} show residuals of the full shear estimates including the weights. Sums and averages are computed over the realizations within each case.
The bias parameters $\mu_i$ and $c_i$ obtained from linear regressions are indicated within the panels.
The \emph{rightmost panels} illustrate the weight values for a random subsample of galaxies, as function of the measured features {\tt adamom\_sigma} and {\tt adamom\_flux} (\emph{top}), and the true parameters $R$ and $S$ (\emph{bottom}).
The equivalent distributions of the respective other weight components are highly similar.
}
\end{center}
\end{figure*}

\begin{figure*}[tbp]
\begin{center}
\includegraphics[width=0.75\linewidth]{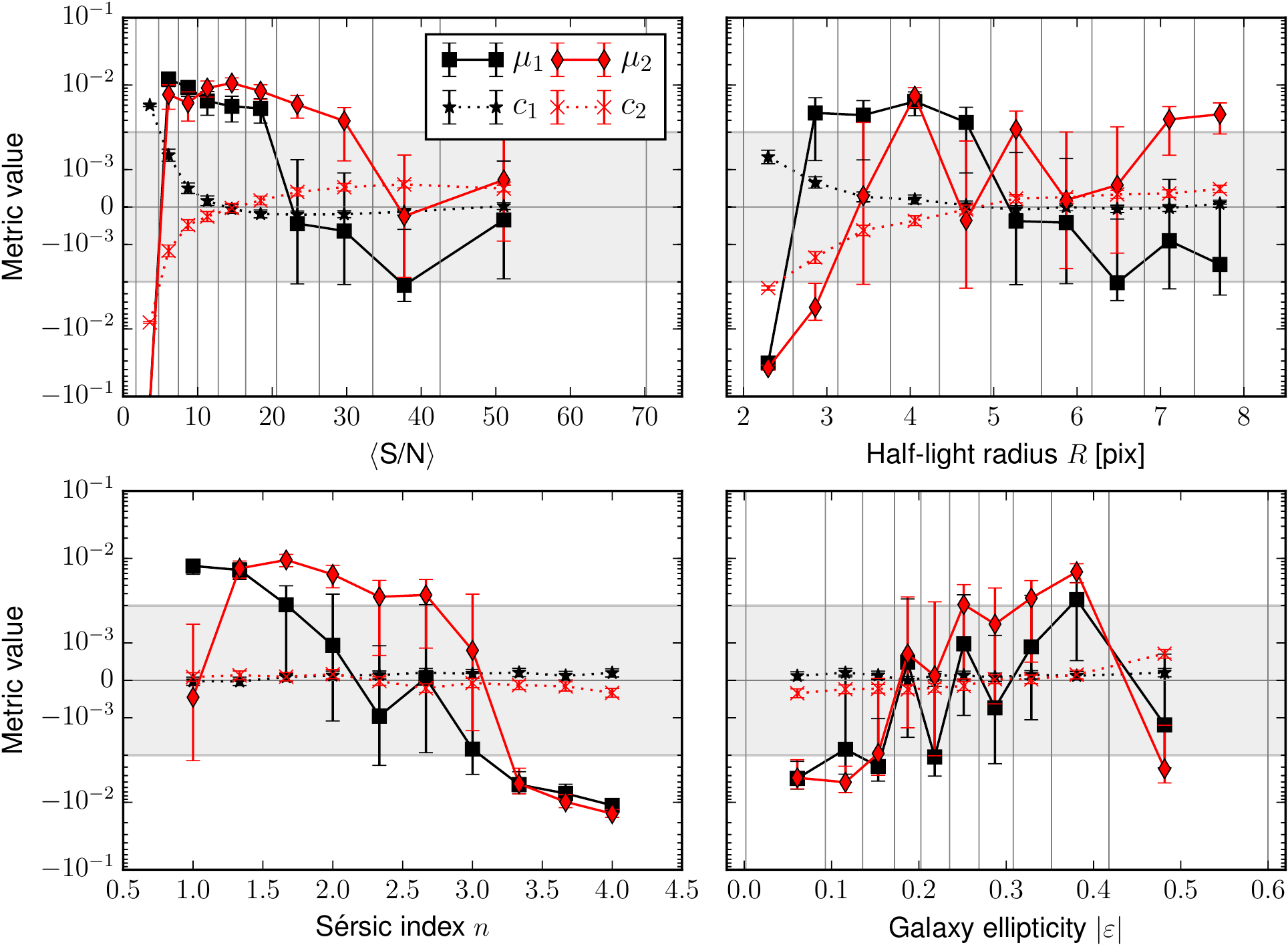} 
\caption{\label{fig:fid-condbias-weights}
Conditional biases of the full shear estimator (point estimates and weights) on the fiducial experiment.
As discussed in the text, these conditional biases give a very pessimistic view, as the population within each bin strongly differs from the assumed overall population with which the weights are trained.
The analysis uses the {\tt VP} dataset, and includes all galaxies for which features could be measured, without any cut in $\langle \snr \rangle$.
A weighted least squares regression is performed in each bin to obtain $\mu$ and $c$.
}
\end{center}
\end{figure*}

\subsection{Training and validation datasets}
\label{sec:fiducial:statpsf}
All training and validation data are drawn based on the fiducial parameter distributions introduced above.
The feature measurement on the simulated galaxy images directly follows their generation, and might not always succeed, due to the pixel noise.
Individual galaxy realizations for which this happens are simply masked out from the catalog.
We generate four different datasets, which we describe in the following.
\begin{description}
\item[{\tt TP}] designates the dataset used to train the shear point estimators, with a structure as illustrated in Fig.~\ref{fig:shear}.
We draw 5\,000 independent galaxies and shears (``cases''), and generate 2\,000 realizations of each galaxy, for a total of 10 million stamps.
Prior to the shearing, the PSF-convolution and the pixellation, the galaxy is rotated for each realization so that the position angle of the galaxy uniformly and regularly describes a full circle per case.
We then remove about $30\%$ of the cases that have $\langle \snr \rangle < 10$ from this dataset, as motivated below.

\item[{\tt VP}] serves as an intermediate validation dataset for the point estimators. It has the same structure and number of cases as the above {\tt TP}, but with 10\,000 realizations per case, amounting to 50 million stamps. This allows for a higher precision of the bias analysis.

\item[{\tt TW}] serves to train the weight estimators, with a structure as illustrated in Fig.~\ref{fig:shearw}. We draw 200 cases of different shears, and 100\,000 realizations of different galaxies per case (20 million stamps), without any shape noise cancellation scheme.

\item[{\tt VO}] is an overall validation dataset. It has the same structure and size as {\tt TW}, and we again opt for not using shape noise cancellation.

\end{description}
These datasets are drawn with different initializations of the random number generators.

To summarize, the training of the point estimates will use a dataset avoiding ``difficult'' cases of small and faint galaxies, while the training of the weights uses the full fiducial parameter space.
Clearly, some of the small and faint galaxies within the parameter space do not allow accurate $\hat{g}_i$ estimates, even with 2\,000 realizations.
Their presence in {\tt TP} would perturb the training process, as the NNs would attempt to fit these outliers instead of obtaining accurate $\hat{g}_i$ on the ``easy'' cases (see Sect.~\ref{sec:pracMLnotes}).
We stress that this selection based on the average \snr per case rejects entire cases, and not only some particular realizations.
It does not introduce the biases and complications that a selection on individual measurements would bring.
Furthermore, we note that the use of this particular threshold on $\langle \snr \rangle$ is somewhat arbitrary, and that more optimized selections based on true galaxy parameters could result in a better performance.

On average, and with the demonstration scripts which we make available, the generation of one 64-by-64-pixel stamp takes about 10 ms, and the feature measurement takes 3 ms on a contemporary CPU.

\subsection{Machine learning shear estimation}
\label{sec:fid:ml}

As input for all NNs, we use the five features \texttt{adamom\_g1}, \texttt{adamom\_g2}, \texttt{adamom\_sigma}, \texttt{adamom\_flux}, \texttt{adamom\_rho4} described in Sect.~\ref{sec:adaptativemoments}.
We do not inform the NNs about the PSF and the background noise, given that these are stationary in the present section.

In a first step, we train the two point estimators $\hat{g}_1$ and $\hat{g}_2$ on the dataset {\tt TP}.
For each component, we use a committee of eight individual NNs with two hidden layers of five nodes, and one output node.
Figure~\ref{fig:fid-netviz} illustrates one of these NNs.
All other parameters concerning the NN setup follow the description given in Sect.~\ref{subsec:nndetails}.

The evolution of the MSB cost function during the training is shown in Fig.~\ref{fig:fid-msbevo}.
After about $10\,000$ iterations no further improvement is seen for the best committee members in this particular setup, and we stop the training.

Before training the weight estimators, we inspect the achieved quality of this first step by applying the point estimators to the {\tt VP} validation set.
Figure~\ref{fig:fid-resi} presents a first quality check of $\hat{g}_1$, visualizing estimation biases per case as a function of source galaxy parameters.
Some panels show results for a subset of cases selected according to their $\langle \snr \rangle$.
As first observation, we note that the simple networks succeed in predicting accurate shear point estimates for almost all $\langle \snr \rangle > 10$ cases.
Remarkably, this performance even extrapolates to some low-$\langle \snr \rangle$ cases from regions of parameter space not seen during the training.

Based on the $\langle \snr \rangle > 10$ cases of the same dataset, Fig.~\ref{fig:fid-condbias} shows the multiplicative and additive bias terms (Equation \ref{eq:mc}) obtained by unweighted linear regression, in various bins of source galaxy parameters. 
Given the dependence on these galaxy parameters, we refer to the resulting biases as ``conditional''.
As for the training, the $\langle \snr \rangle$-cut of the validation set is arbitrary, but this simple threshold allows us to focus on the cases which matter most for a real shear measurement application.
In most bins of Fig.~\ref{fig:fid-condbias}, the conditional multiplicative biases $\mu$ have amplitudes close to $2 \cdot 10^{-3}$ (shown by the shaded horizontal band), while the additive biases $c$ are on the order of $10^{-4}$.
The most significant multiplicative biases are observed for galaxies with large \sersic indices (i.e., the most centrally concentrated light profiles).
This is easily explained, as the concentration of the light profile, related to the feature {\tt adamom\_rho4}, is particularly difficult to measure for small and faint galaxies.
In parts of parameter space in which features are unreliable ({\tt adamom\_rho4}, for example), the NNs will tend toward predictions satisfying an average galaxy.
In Fig.~\ref{fig:fid-condbias}, this directly leads to conditional bias trends, typically with a balance between positive and negative multiplicative biases.
We note that when considering average biases over an entire source galaxy population, these conditional biases might cancel out, at the price of some sensitivity to this source population.
It is also important to stress that the above analysis purposely avoids selection biases by disregarding cases for which $\langle \snr \rangle < 10$.
Such a selection cannot be made on individual realizations, leading us to the next training step.

We proceed by training the weight estimators, using the {\tt TW} dataset without any additional selection, which in a real application would mimic the observed data as closely as possible.
We again use committees with eight NNs, but reduce the capacity of the NNs to a single hidden layer of five nodes.
Figure~\ref{fig:fid-netvizw} shows an example of this small network structure.
For the present demonstration, larger NNs did not yield significantly better results.
The MSWB cost function, computing weighted averages of the shear point estimates per case, ceases to decrease after few hundred iterations.

We then run the predictions $\hat{g}_i$ and $w_i$ from both steps on the remaining {\tt VO} dataset.
Figure \ref{fig:fid-wbias} summarizes the results of this overall analysis of the shear estimates.
The two leftmost panels show shear residuals on each of the 200 cases, obtained by averaging the point estimates $\hat{g}_i$ alone accross each case, ignoring the weights.
Significant percent-level multiplicative shear biases can clearly be seen.
They can be attributed mainly to (1) noise and pixellation bias on small and low-\snr galaxies, and (2) the potentially shear-dependent selection function imposed by the feature measurement.
The central panels of this figure show residuals of the weighted average shear of each case, on the same data.
The use of weights reduces the overall multiplicative bias by an order of magnitude to the level of $10^{-3}$, with additive biases on the order of $10^{-4}$.
We conclude from this experiment that the weight estimators have successfully learned to down-weight ``difficult'' galaxies, and to mostly cancel out remaining biases of the point estimates.

The right panels of Fig. \ref{fig:fid-wbias} illustrate the distribution of weights for random individual galaxies.
One can observe that {\tt adamom\_flux} has a major influence on the weight value, but that other features in addition to {\tt adamom\_sigma} also contribute to the estimator. 
We stress that this reliance on other features is essential if we want the weights to counter biases introduced by any shear-sensitive selection function, which might have a complicated dependency on galaxy parameters.

As a side effect, the weights also introduce sensitivity to the true galaxy parameters. In Fig.~\ref{fig:fid-condbias-weights}, we present an analysis of the conditional biases similar to Fig.~\ref{fig:fid-condbias}, but now taking into account the weights, and for the full fiducial parameter space without any cut in $\langle \snr \rangle$.
The interpretation of these conditional biases of the weighted point estimates is not trivial.
The points in each bin show the multiplicative and additive biases one would obtain if all source galaxies would have their true properties within this particular bin, instead of following the distributions used for the weight training.
This is a very pessimistic point of view, as it assumes that we largely ignore the true galaxy properties.
Instead, for any practical application, the training set for the weights would be chosen to mimic the target galaxy population as accurately as possible.
The analysis shown in Fig.~\ref{fig:fid-condbias-weights} therefore gives a first idea about the sensitivity to the knowledge of source galaxies, in particular how important this knowledge is for training the weights.
It is reassuring to observe that the weights do not completely invalidate the very low sensitivity of the point estimates to the galaxy parameters. Impressively, the conditional multiplicative biases seen in Fig.~\ref{fig:fid-condbias-weights}, without any cut in \snr, are still sub-percent for the large majority of slices through true parameter space.
Nevertheless, comparing Figs.~\ref{fig:fid-condbias} and \ref{fig:fid-condbias-weights}, one might be led to wonder if the use of weights is not detrimental.
We stress again that a selection of galaxies based on $\langle \snr \rangle$, as done in Fig.~\ref{fig:fid-condbias}, is not possible for real data.
A substitute selection based on $\snr$ (or any other combination of observed features) will likely lead to selection biases, which can however be mitigated by the use of weights, as we will illustrate in Sect.~\ref{sec:euclid}.

\section{Correcting for a non-stationary PSF}
\label{sec:fiducial:varpsf-field}

\begin{figure*}[tbp]
\begin{center}
\includegraphics[width=0.7\linewidth]{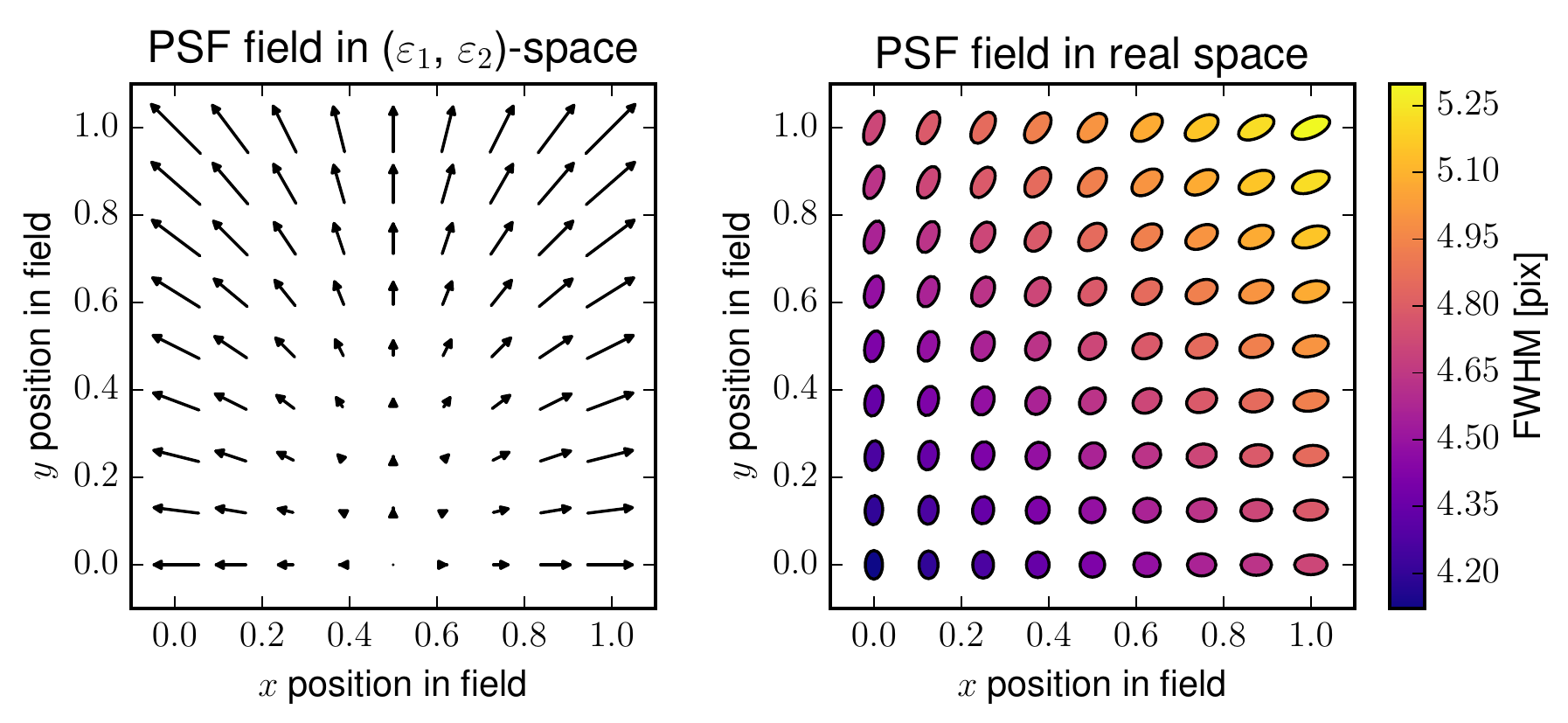} 
\caption{\label{fig:varpsf-field}
The simple spatially variable PSF field used for the fiducial experiments in Sect.~\ref{sec:fiducial:varpsf-field}.
In the \emph{left panel}, the PSF ellipticity across the field is shown by arrows representing the $(\varepsilon_1, \varepsilon_2)$-components. The maximum $\varepsilon_i$ component is 0.25. In the \emph{right panel}, the corresponding PSF shape is shown as ellipses across the field. The size of the arrows and ellipse symbols is not to scale with the field coordinates.
}
\end{center}
\end{figure*}

\begin{figure*}[tbp]
\begin{center}
\includegraphics[width=1.0\linewidth]{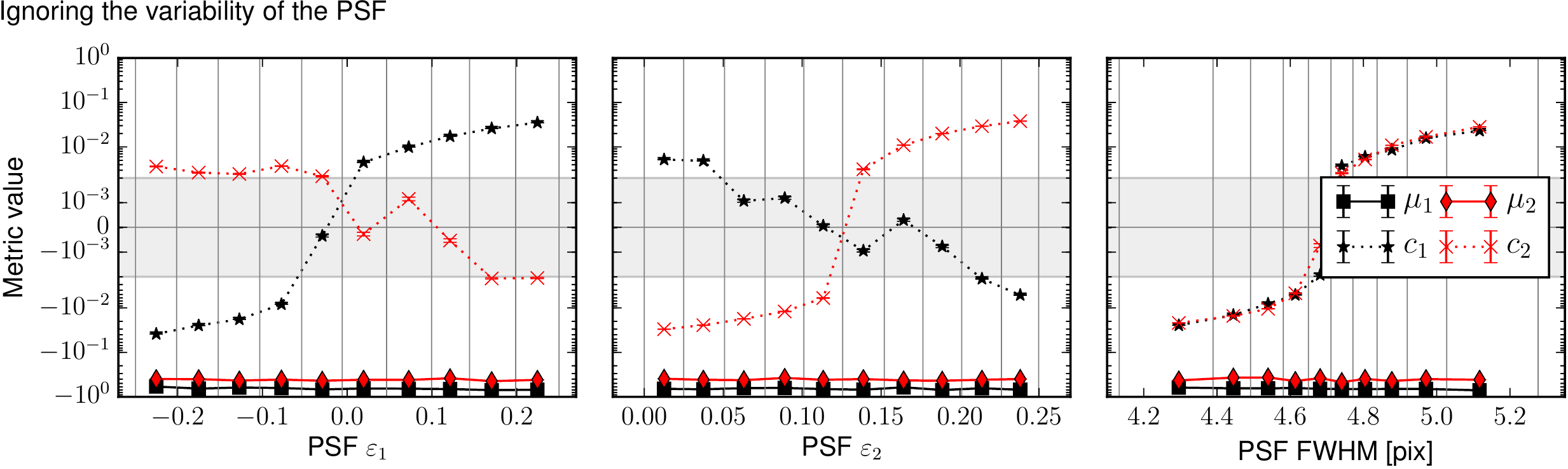} 
\includegraphics[width=1.0\linewidth]{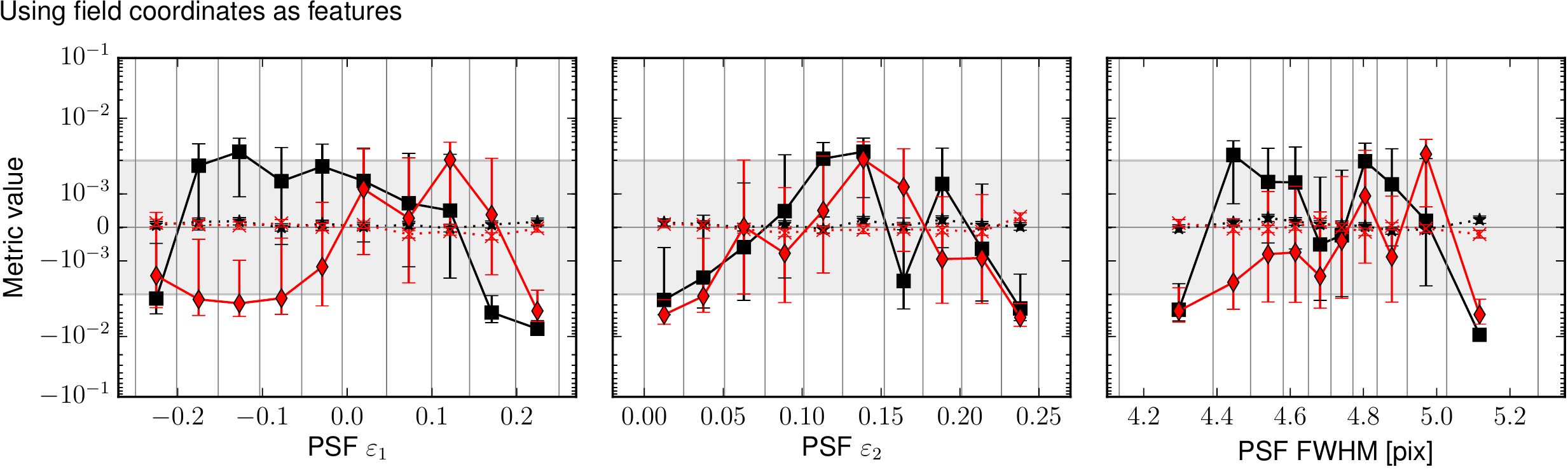} 
\includegraphics[width=1.0\linewidth]{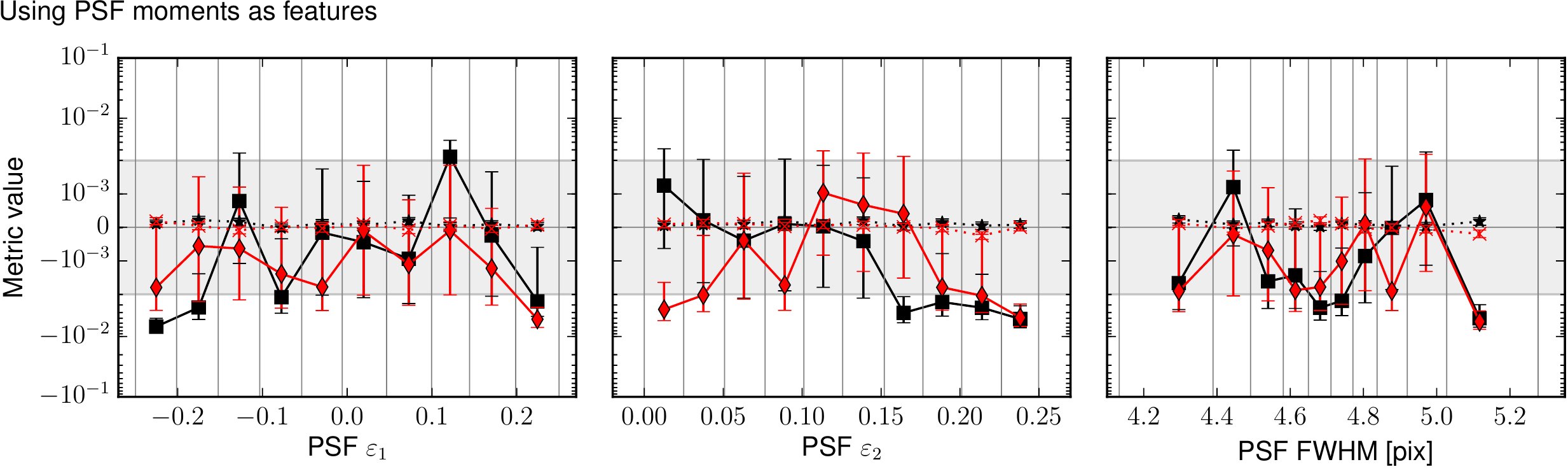} 
\caption{\label{fig:varpsf-biases}
Conditional biases of the shear point estimator $\hat{g}_i$ as function of true properties of the variable PSF, in equipopulous bins. The same training and validation datasets are used for all panels, but three different ML setups are shown. \emph{Top}: the ML is not informed about about the particular PSF of each galaxy. \emph{Middle}: the field coordinates ($x$, $y$) are used as input features. \emph{Bottom}: the shape parameters $\varepsilon_1$, $\varepsilon_2$, and $\sigma$ of the PSF are used as input features. As in Fig. \ref{fig:fid-condbias}, only cases with $\langle \snr \rangle \ge 10$ are considered for this unweighted analysis of the point estimates.
}
\end{center}
\end{figure*}

\begin{figure*}[tbp]
\begin{center}
\includegraphics[width=1.0\linewidth]{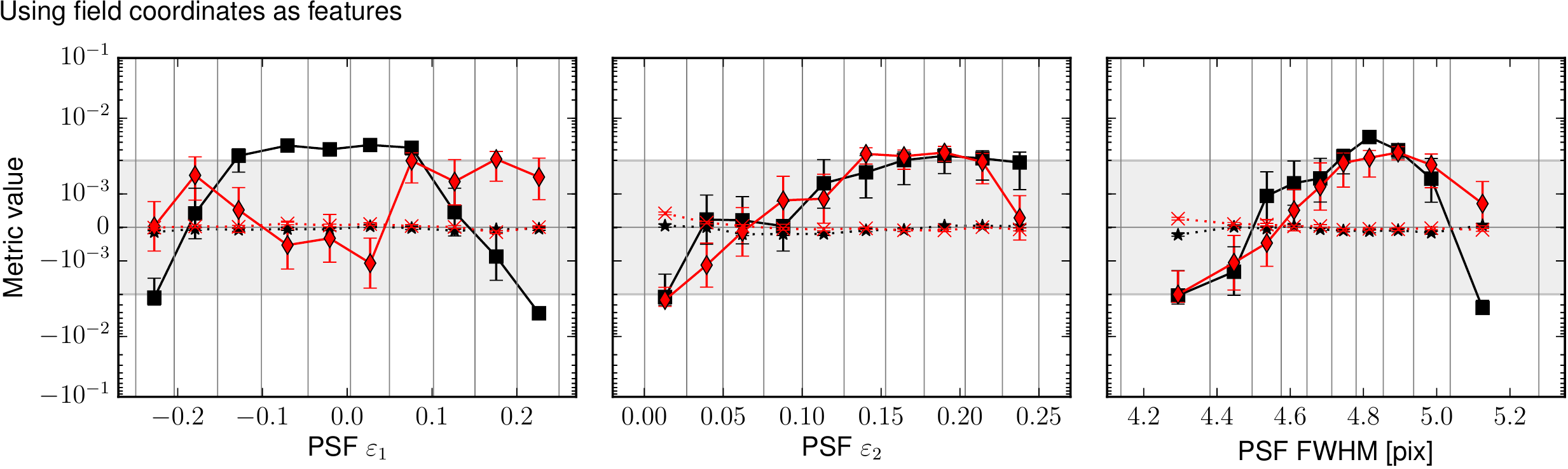} 
\includegraphics[width=1.0\linewidth]{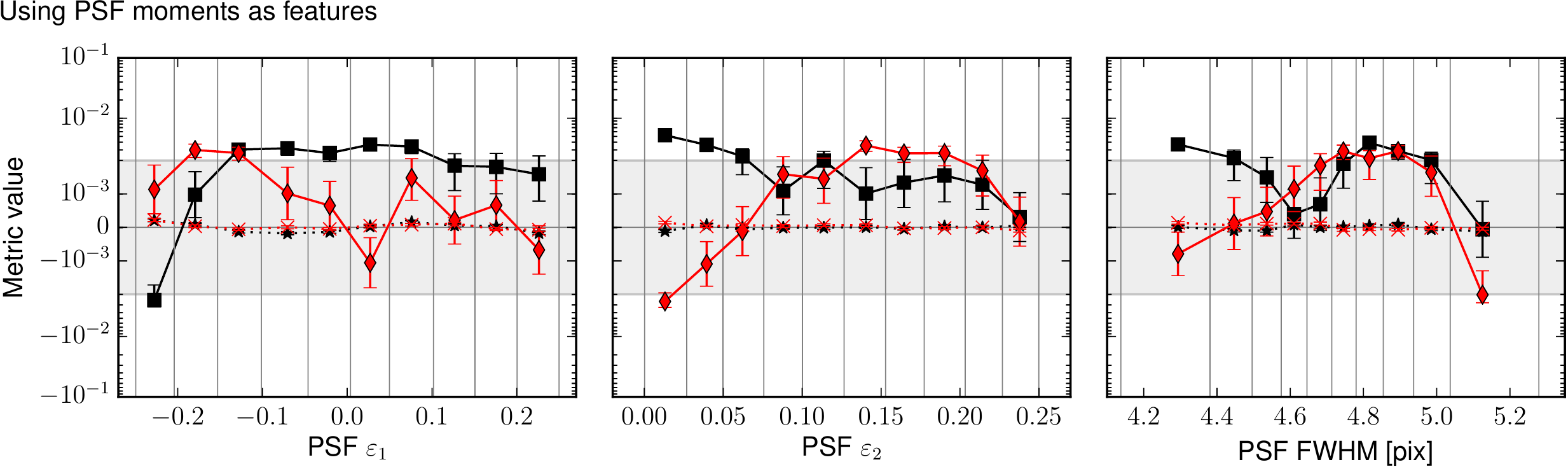} 
\caption{\label{fig:varpsf-biases-w}
Similar to Fig.~\ref{fig:varpsf-biases}, but including the predicted weights, and without any selection of $\langle \snr \rangle$. For the \emph{top panels}, the field coordinates ($x$, $y$) are used as input features to the ML, while for the \emph{bottom panels}, the PSF shape parameters $\varepsilon_1$, $\varepsilon_2$, and $\sigma$ are used.
To obtain $\mu_i$ and $c_i$ for each bin, an unweighted linear regression is performed on the weighted average $\hat{g}_i$ against the $g_i^{\mathrm{true}}$ per case, instead of performing a weighted least squares regression on all realizations of a bin. The later approach would result in an undesired weighting between different PSFs.
}
\end{center}
\end{figure*}

In the following, we demonstrate two alternative paths along which the ML method can correct for a variable PSF. For both situations, we assume that the PSF is exactly known, and regard the creation of a suitable PSF model to be a separate problem. 

The first approach is oriented toward dealing with a well-determined diffraction-limited PSF of a space-based telescope.
We assume that a model of the PSF across the field exists and is applicable to many exposures of a survey.
Given such a PSF model, one can directly use the field position of each galaxy as input features to an ML shear estimation.
The ML algorithm is trained on simulations that are generated using the same PSF model, and thereby directly learns how the PSF should be corrected for depending on the location of a source on the detector.
The advantage of using the source position as an input feature is that it optimally encodes, given the model, the knowledge about the PSF, and naturally provides a machine-learned interpolation of the PSF anywhere in the field. 
This approach can be generalized for a PSF model that depends, for example, on time or galaxy color.

The second approach is better suited for the stochastic atmospheric PSF of a ground-based survey.
For such data, a separate PSF model is usually constructed for each exposure, based on images of stars in the field.
It is intractable to train an ML algorithm to individually learn the spatial dependence of the PSF for each exposure.
But instead, one can train the ML to perform a PSF correction based on input features that directly describe the PSF shape.
For example, moment-based shape parameters describing the ellipticity and size of the PSF could be used, or parameters describing the PSF or its variability from a reference PSF in terms of an orthogonal basis.
The ML is then to be trained with a range of PSFs encountered in the whole survey, and naturally learns to interpolate the PSF shape parameters. 

\subsection{Implementation and simulations}
\label{sec:varpsfimpl}

To demonstrate these two approaches, we again use the fiducial galaxy parameters from Sect.~\ref{sec:fiducial:data}, but now randomly attribute a $(x, y)$ position in a virtual field of view to each case of the training and validation sets, with $x$ and $y$ uniformly drawn in the interval $[0, 1]$.
We generate the simulations with a different elliptical Gaussian PSF at each position, following a simple toy model shown in Fig.~\ref{fig:varpsf-field}.
The ellipticity component $\varepsilon_1$ varies linearly in the range $[-0.25, 0.25]$ along the $x$-axis of the field, while $\varepsilon_2$ evolves from $0.0$ to $0.25$ along $y$.
In addition, we vary the azimuthal average standard deviation $\sigma$ of the PSF according to $\sigma = 2.0 + 0.25 \cdot (x+y-1)$. 
The resulting field is qualitatively similar to the PSF-variation across the detector of a space telescope, although the variability is quantitatively exaggerated, somewhat closer to poor ground-based conditions.
We opt for this particular model to purposely include asymmetries both in $(\varepsilon_1, \varepsilon_2)$ and in $(x, y)$, so that PSF-related effects do not simply cancel out along slices of parameter space.

We implement the two described approaches by extending the five input features used in Sect.~\ref{sec:fiducial} in one case by the field coordinates $(x, y)$, and in the other case by the true PSF shape parameters $\varepsilon_1$, $\varepsilon_2$, and $\sigma$.
To allow for an acceptable correction, we increase the capacity of the involved NNs with respect to Sect.~\ref{sec:fiducial}.
We now use two hidden layers with ten (instead of five) nodes each for the $\hat{g}_i$-predicting networks, and one hidden layer with ten (instead of five) nodes for the weight-predicting NNs.

This increase in NN capacity, together with a potential increase of the simulation data size, would make a standalone training as realized in Sect.~\ref{sec:fiducial:data} unpractically costly with our simple implementation.
Instead of directly training a shear point estimator on noisy images, we therefore introduce a pretraining of the $\hat{g}_i$-predicting networks on a dedicated ``simpler'' dataset, targeting the true ellipticity of our simple \sersic galaxies (see Sect.~\ref{sec:ellipshortcut}).
This training data has a structure as shown in Fig.~\ref{fig:ellip}, and a pixel noise with a standard deviation of only 0.1 (instead of 1.0).
Owing to this simplification and pixel noise reduction, we can use significantly less data for this pretraining step, with 20\,000 cases and 50 realizations per case, amounting to $10^6$ stamps, compared to $10^7$ for {\tt TP} in the case of the stationary PSF.
The pretraining allows the ML to learn about the variable PSF-correction in low-noise conditions and with much faster iterations.
After convergence on the pretraining data (5000 iterations, one CPU-day per NN), we finish the training on data of the same structure and size as the set {\tt TP} of Sect.~\ref{sec:fiducial}, and targeting true shear.
In other words, the pretraining simply allows us to start the main training from a better starting point.
With this second step the NNs adapt to compensate for noise biases, and converge after a few hundred iterations corresponding to one more CPU-day.

To train the weights, a pretraining with the same simplifications is not helpful, as noise properties directly influence any optimal weighting solution.
Instead, we opt for introducing shape noise cancellation (SNC) into the training data, despite the negative effects described in Sect.~\ref{subsec:shearweights}.
Two realizations of each galaxy are drawn, with the intrinsic position angle (before the application of shear) differing by $90^{\circ}$.  
This allows to improve the precision of the analysis despite smaller training and validation data, and keeps our simple demonstration implementation tractable on widely available computers.
The aim of this section is to qualitatively demonstrate the correction of a non-stationary PSF.
While the use of SNC artificially shifts the weighting toward high-\snr galaxies, a good PSF correction is also important for those galaxies to yield low biases.
We therefore argue that the use of SNC in this section does not harm its conclusions. 
The use of more optimized ML implementations is likely required for any deeper analysis.

\subsection{Results}

Figure~\ref{fig:varpsf-biases} presents multiplicative and additive bias parameters of the point estimates $\hat{g}_i$ in bins of PSF shape, for the two alternative approaches.
For comparison, results from NNs of identical size but without PSF-related input features are shown in the top panels.
Following the procedure detailed in Sect.~\ref{sec:fiducial}, the bias analysis is performed on an independent validation set {\tt VP}.
In particular, the $(x, y)$ positions of the cases do not correspond to the positions probed in the training data.

For the PSF-agnostic NNs (top panels of Fig.~\ref{fig:varpsf-biases}), we observe the expected strong biases.
The components of the additive bias $c_i$ directly correlate with the corresponding ellipticity of the PSF.
The multiplicative biases are strongly negative in all bins, as these NNs were driven by the lack of information to blindly yield shear estimates close to the average true shear values for most of the galaxies, that is close to a zero shear.

On the middle and bottom panels of Fig.~\ref{fig:varpsf-biases}, we observe that both approaches to inform the ML about the variable PSF yield a very comparable quality in this first evaluation.
For most of the bins, the multiplicative biases stay close to the level of $10^{-3}$, with additive biases systematically far below this level.
Some significant multiplicative biases, although still sub-percent, are typically seen toward the boundaries of the PSF shape parameter space.
A simple example is the top-right corner of the field shown in Fig.~\ref{fig:varpsf-field}, which contains relatively few cases with the largest PSFs, leading to a poor training quality.
The resulting strong biases for these largest PSFs can be seen in the right-hand panels of Fig.~\ref{fig:varpsf-biases}.
For an application to a real survey, it could be beneficial to increase the number of training cases in regions of infrequent or extreme PSFs, until no significant biases remain.

In Fig.~\ref{fig:varpsf-biases-w}, we present a similar analysis, but including the predicted shear weights, and evaluated on all realizations of the fiducial parameter distributions without any selection apart from the successful feature measurement.
The residual multiplicative biases remain on the order of $2 \cdot 10^{-3}$ in most areas of the field, despite the strong variability of the ellipticity of the PSF.
For the small variability of a realistic space-based PSF, we can safely expect a further strong decrease of these biases. 
With these results, we have demonstrated the general feasibility of simple yet plausible approaches for ML algorithms to interpolate and correct for a variable PSF model across the field of view.

\section{Application to \euclid-like simulations} \label{sec:euclid}

We now turn to an application on simulations roughly mimicking realistic galaxy properties and \euclid-like imaging characteristics, including a narrow and undersampled PSF.
While still based on single \sersic profile galaxies on $64\times64$-pixel stamps without blends or artifacts, these simulations allow for some first best-case evaluation of the approach in a pixellation and noise regime comparable to the \euclid survey.
We use this application to further illustrate how the proposed method can be used to handle selection effects.

\subsection{Image simulations}

As in the previous sections, we simulate all training and validation images with {\tt GalSim}.
We follow \citet{Hoekstra:2017hg} in relying on the GEMS survey \citep{Rix:2004df} to obtain realistic distributions of source galaxy parameters.
We use a \euclid-like PSF, and a detector model combining the Poisson noise from the source and the zodiacal background with the read-out noise and pixellation of the VIS instrument.
In the following, we give a comprehensive description of these simulations.

\subsubsection{Galaxy properties from GEMS}
\label{sec:gems}

\begin{figure}[tbp]
\begin{center}
\includegraphics[width=0.8\linewidth]{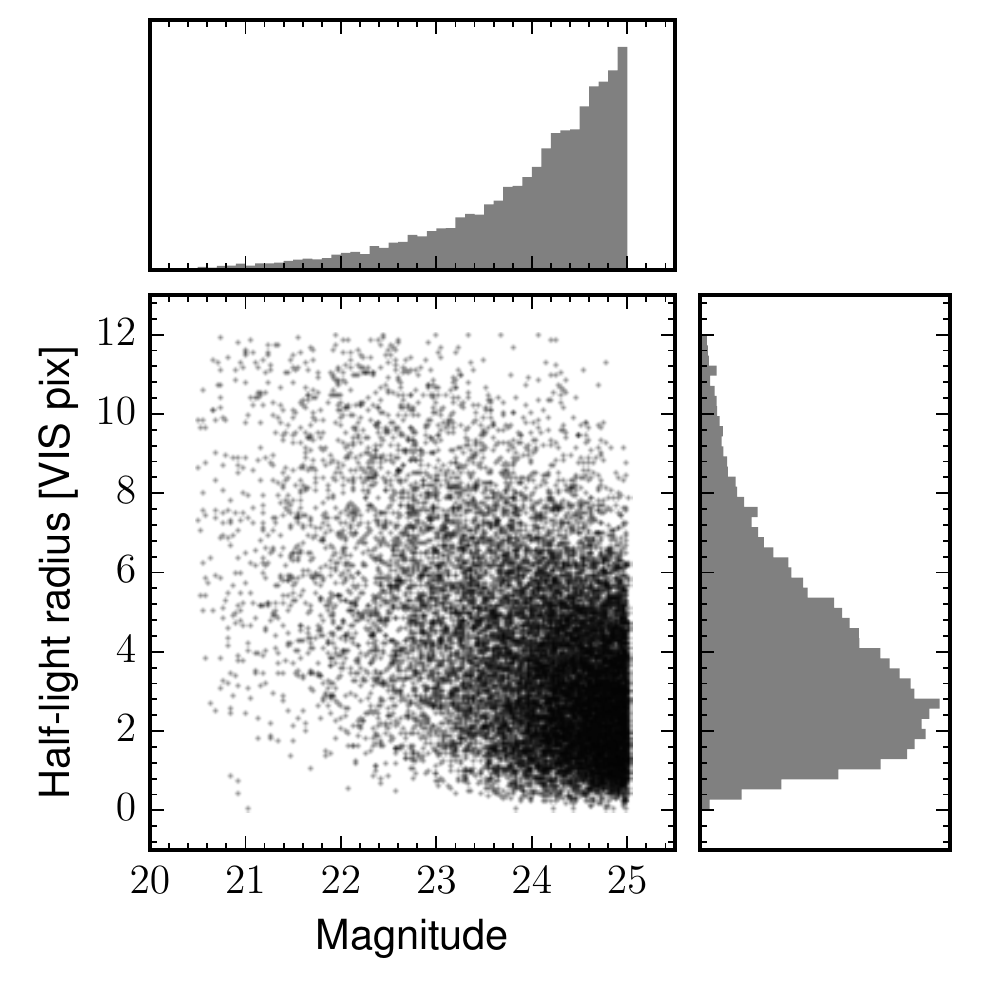} 
\includegraphics[width=0.8\linewidth]{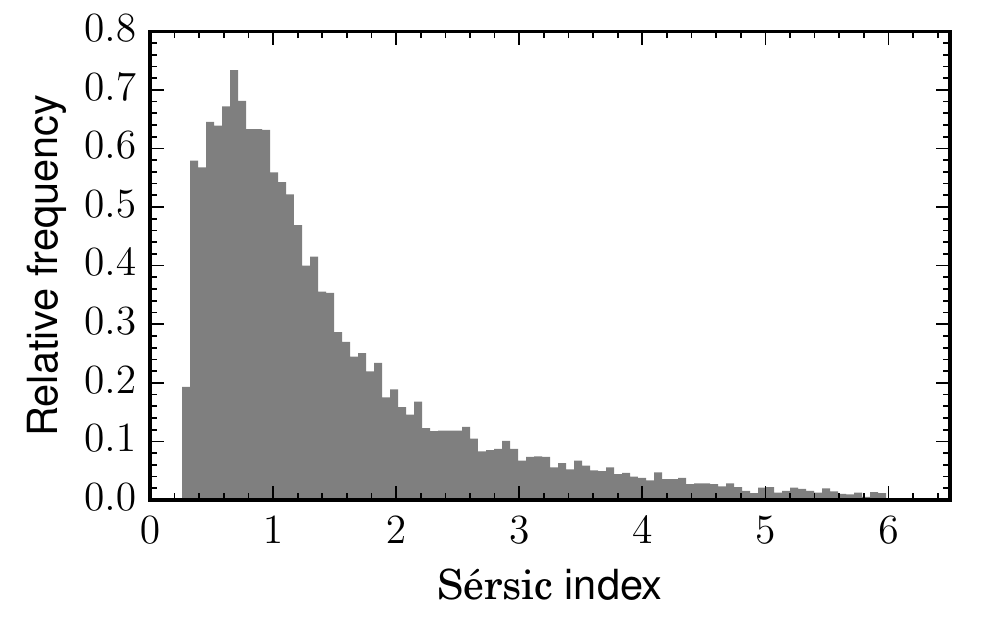} 
\caption{\label{fig:gems}
Distribution of half-light radii, magnitudes, and \sersic indices from the sources selected in GEMS to serve as ground truth for the \euclid-like simulations.
}
\end{center}
\end{figure}

The GEMS imaging survey \citep{Rix:2004df} was conducted with the Advanced Camera for Surveys on the Hubble Space Telescope (HST).
It covers 800 arcmin$^2$ in two bands, and reaches a 5-$\sigma$ depth of $28.3$ ($27.1$) in F606W (F850LP).
\citet{Rix:2004df} combine the images on a $0.03\arcsec$ per pixel grid, and provide a rich source catalog, notably containing morphological information based on the GALFIT  fitting software \citep{Peng:2002di}.

From the ultimate GEMS catalog with 130\,389 entries, we select 14\,661 $(11.2\%)$ detections to build our source catalog, satisfying all of the following criteria.

\begin{enumerate}
\item {\tt GEMS\_FLAG == 4} corresponds to sources classified as extended on HST images, with successful GALFIT measurement.
\item $|{\tt ST\_MAG\_BEST} - {\tt ST\_MAG\_GALFIT} | < 0.5$ rejects sources with discrepant photometric measurements. These sources could potentially suffer from contamination by blends.
\item $20.5 < {\tt ST\_MAG\_GALFIT} < 25.0$ selects sources up to 0.5 magnitude fainter than the nominal range of \euclid VIS weak-lensing source targets.
In the following, we simply identify the GEMS V (F606W) and \euclid VIS bands.
\item $0.3 < {\tt ST\_N\_GALFIT} < 6.0$ limits the range of measured \sersic indices {\tt ST\_N\_GALFIT} to match the simulation capabilities of {\tt GalSim}.
\item $0.0 < {\tt ST\_RE\_GALFIT} < 40.0$ limits the range of measured half-light radii {\tt ST\_RE\_GALFIT} to be within 0 and 12 VIS pixel.
\end{enumerate}
The last two criteria have a negligible impact on source number counts, and are included to simplify the simulations.
Figure \ref{fig:gems} summarizes the parameter distribution of the selected galaxies.

From this source catalog, we only use the magnitude {\tt ST\_MAG\_GALFIT}, \sersic index {\tt ST\_N\_GALFIT}, and half-light radius {\tt ST\_RE\_GALFIT} as joint input samples to some of the simulations of the present section.
Following \citet{Hoekstra:2017hg}, and to reduce the sparsity of our sample, we ignore the ellipticity measurements from GEMS.
Instead, we randomly draw the ellipticity $|\varepsilon^{\mathrm{true}}|$ of each galaxy from a truncated Rayleigh distribution with mode $\sigma = 0.25$ and a maximum value of $0.7$, and combine this modulus with a random uniformly distributed position angle.

\subsubsection{\euclid-like imaging conditions}
\label{sec:vis}

We adapt the technical parameters of our image simulations to mimic the planned design and specifications of \euclid's VIS instrument.
First of all, we build a simple yet roughly realistic PSF, using {\tt GalSim}'s {\tt OpticalPSF} functionality.
We assume an entrance pupil diameter ({\tt diam}) of 1.2 m, a linear {\tt obscuration} of $0.29$ (corresponding to a M2 diameter\footnote{A short description of the \euclid optical system can be found at \url{http://sci.esa.int/euclid}} of $0.35$ m), and six radial support struts ({\tt nstruts = 6}), leaving all other parameters of {\tt OpticalPSF} at default values.
To mimic the VIS bandpass, we generate monochromatic PSFs for wavelengths in steps of 50 nm between 550 and 900 nm \citep{Cropper:2016ig}.
We sample these PSFs on a grid of $0.02\arcsec$ per pixel, that is with a five times finer sampling than the native VIS pixel scale.
Finally, we stack these PSFs, weighting them using the spectrum of a G5V stellar template from the Pickles library \citep{Pickles:1998er} converted to photon counts per wavelength interval.
We chose to ignore any PSF variability and PSF color effects in this section.

\begin{figure}[tbp]
\begin{center}
\includegraphics[width=0.6\linewidth]{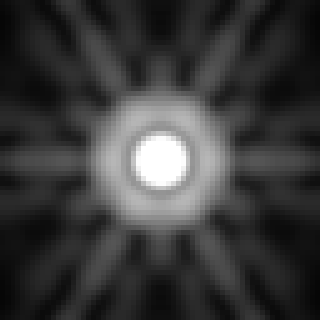} 
\caption{\label{fig:psfeuclid}
\texttt{GalSim}-generated PSF used for the \euclid-like simulations, on a grid of $0.02\arcsec$ per pixel, shown with a logarithmic grayscale.
}
\end{center}
\end{figure}

The resulting PSF is shown in Fig.~\ref{fig:psfeuclid}.
When highly oversampled, this PSF has a full width at half maximum (FWHM) of $0.1\arcsec$.
However, when rendered at the native VIS pixel scale of $0.1\arcsec$, the measured FHWM is typically $0.15\arcsec$, relatively close to the expected value of $0.17\arcsec$ for VIS \citep{Cropper:2016ig}.

Regarding the detector and background noise, we make the following assumptions for our \euclid-like simulations.
\begin{itemize}
\item Pixel size $l_\mathrm{pix}$: $0.1\arcsec$ \citep{Laureijs:2011wi}
\item Exposure time $t_\mathrm{exp}: 3 \times 565$ s \citep{Laureijs:2011wi}. Individual VIS exposures are planned to have an exposure time of 565~s, but here we directly simulate the depth of three exposures as a single acquisition. While \euclid takes four dithered exposures per step, only three exposures are available for a large fraction of the survey area, due to chip gaps. 
\item Gain $G$: $3.1$ electrons/ADU \citep{Niemi:2015kw}
\item Read-out noise: 4.2 electrons \citep{Cropper:2016ig}
\item Sky background $m_\mathrm{Sky}$: $22.35$ mag arcsec$^{-2}$ \citep{Refregier:2010uj}, which corresponds to a brightness of $2.5\cdot10^{-18}$ erg s$^{-1}$ cm$^{-2}$ \AA$^{-1}$ arcsec$^{-2}$, under the hypothesis of a tophat VIS spectral sensitivity. We assume that the sky level will be dominated by the zodiacal light.
\item Zero-point $Z_{\mathrm{p}}$: $24.6$ mag, as justified below.
\end{itemize} 
The true flux of a simulated galaxy of apparent magnitude $m$ is given by
\begin{equation}
F[\mathrm{ADU}] = \frac{t_\mathrm{exp}[\mathrm{s}]}{G} \cdot 10^{-0.4 (m - Z_{\mathrm{p}})}
\end{equation}
and the sky level by
\begin{equation}
F_{\mathrm{Sky}}[\mathrm{ADU}/\mathrm{pixel}] = (l_\mathrm{pix}[\arcsec])^{2} \cdot \frac{t_\mathrm{exp}[\mathrm{s}]}{G} \cdot 10^{-0.4 (m_\mathrm{Sky} - Z_{\mathrm{p}})}.
\end{equation}
The noise in our simulations is generated with {\tt GalSim}'s {\tt CCDNoise}, which encapsulates Poisson shot noise from the source and the sky level, and Gaussian read-out noise.
As in the previous sections, we do not consider the problem of background subtraction here, and directly simulate images with a perfectly subtracted sky level.

We empirically adjust the value of the instrumental zero-point $\mathrm{Z}_{\mathrm{p}}$ given above, so that our simulations mimic the expected \euclid depth performance as described in \citet{Cropper:2016ig}.
More precisely, we aim at obtaining an average \snr of 10 (following Eq. \ref{equ:sn}) on simulated $3\times565$-second exposures of elliptical 24.5-mag \sersic galaxies with parameters drawn from the GEMS distributions (Sect.~\ref{sec:gems}) and convolved with our \euclid-like PSF.
Clearly, the \snr of a source depends strongly on its extension.
We observe that limiting the observed size of sources to a half-light-diameter of $0.43\arcsec$, a value inspired by the description given in \citet{Cropper:2016ig}, would yield a similar zero-point within 0.1 mag.
Our estimate of the VIS zero-point of $Z_{\mathrm{p}} = 24.6$ is certainly a rough approximation, and we provide it here solely to make our simulations reproducible and comparable.

\begin{figure}[tbp]
\begin{center}
\includegraphics[width=1.0\linewidth]{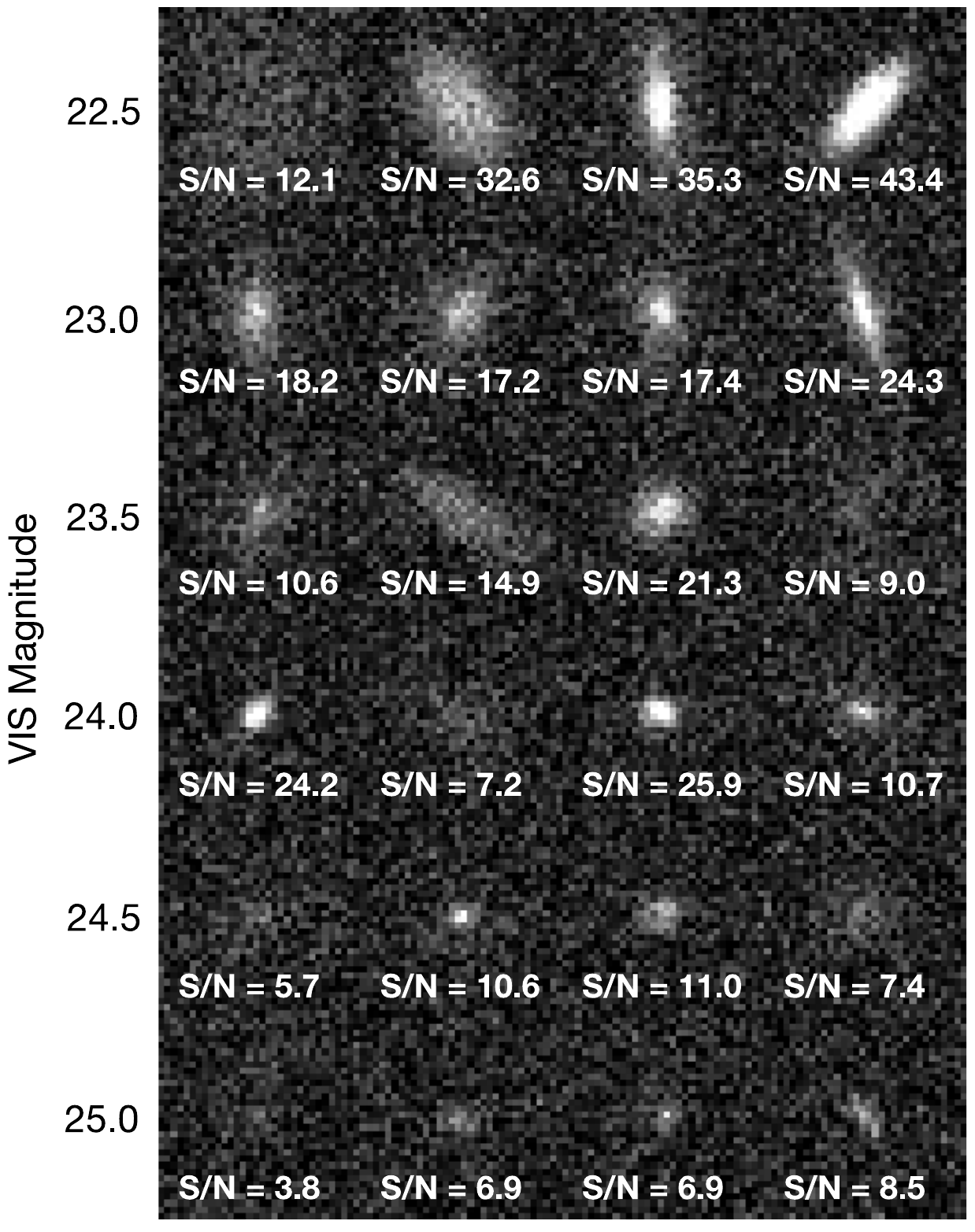} 
\caption{\label{fig:snrgems}
Cutouts of 32 $\times$ 32 pixels around galaxies from the \euclid-like simulations. Each line shows four galaxies randomly drawn from our GEMS source catalog described in the text, within a slice of $0.05$ mag around the true magnitude indicated on the left. The measured \snr (Equation \ref{equ:sn}) is given for each image.
}
\end{center}
\end{figure}

Figure \ref{fig:snrgems} shows single realizations of galaxies from the described simulations.
These galaxies are arranged according to their true magnitude, and labeled with the measured \snr, to illustrate the diversity of galaxy sizes, \sersic indices, and resulting \snr.
For a magnitude around 24.5, extended galaxies can be difficult to detect, while compact galaxies might be indistinguishable from foreground stars.
In practice, the selection of sources for shear analysis will unavoidably be based on noisy observed criteria potentially directly correlated with shear.
In the following application, we illustrate how our method can adapt to such a selection via training of the weight predictions $w_i$.

\subsection{Dataset structure and ML setup} \label{sec:euclid:nets}

\begin{table}[tbp]
\caption{Galaxy parameters to train the point estimator for the \euclid-like simulations.}
\label{table:euclidgalparams}
\centering
\begin{tabular}{l l} 
\hline\hline
Parameter & Distribution  \\ 
\hline
Shear components $g_1$, $g_2$ & $\mathcal{U}(-0.1, 0.1)$ \\
Intrinsic galaxy ellipticity $\varepsilon^{\mathrm{true}}$ & $\mathcal{R}(0.25)_{[0,\,0.7]}$ \\
S\'ersic index\tablefootmark{a} $n$ & $\mathcal{U}(0.3, 6.0)$  \\
Half-light radius $R$ [pix] & $\mathcal{U}(1.0, 10.0)$ \\
Magnitude & $\mathcal{U}(20.5, 25.0)$ \\
\hline
\end{tabular}
\tablefoot{
Notations are the same as in Table \ref{table:fidgalparams}. \\
\tablefoottext{a}{In practice, we grid the values for the S\'ersic index (see Table \ref{table:fidgalparams}).}
}
\end{table}

\begin{figure*}[htbp]
\begin{center}
\includegraphics[width=0.8\linewidth]{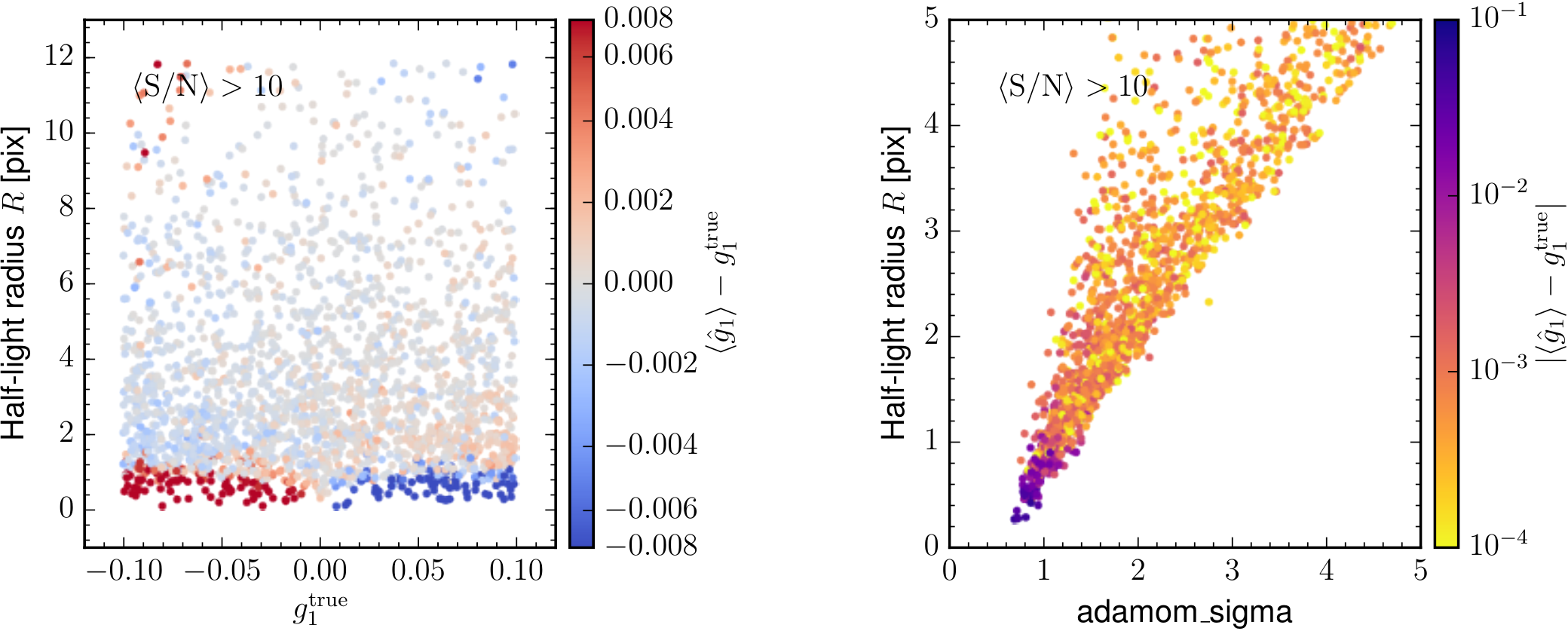} 
\caption{\label{fig:ec_radsigma}
Biases of the point estimates $\hat{g}_1$ on \euclid-like simulations of galaxies with shape parameters drawn from the GEMS catalog.
In both panels, each point represents a case consisting of one particular galaxy, and the shear bias is computed by averaging over 10\,000 rotated realizations of this galaxy (the data structure follows Fig.~\ref{fig:shear}).
For clarity, only cases with $\langle \snr \rangle > 10$ are shown.
In the \emph{right panel}, the amplitude of the shear bias is shown as function of half-light radius $R$ and the noisy measured {\tt adamom\_sigma} taken from a single realization of each case.
}
\end{center}
\end{figure*}

To apply our method to the \euclid-like simulations, we closely follow the procedure detailed in Sect.~\ref{sec:fiducial}. 
We generate a large overall validation dataset ({\tt VO}) composed of 200 cases of different shear with both components $g_i^{\mathrm{true}}$ uniformly drawn in $[-0.1; 0.1]$, and, for simplicity, no magnification.
Each of these cases contains 400\,000 galaxies drawn from the GEMS catalog defined in Sect.~\ref{sec:gems}, without any shape-noise cancellation scheme.
We note that this source catalog contains galaxies with half-light radii significantly smaller than the PSF.
The analysis of our method will use a subset of sources from this catalog, obtained by a selection based on observables.

To train the point estimators, we use simulations with galaxy parameters drawn from the simple distributions of Table \ref{table:euclidgalparams}.
These parameters roughly cover the range of galaxies present in the source catalog (see Fig.~\ref{fig:gems}).
We avoid very small galaxies when training the point estimators by introducing a lower bound on $R$ of 1.0 pixel.
Such a cut is motivated in Sect. \ref{sec:pracMLnotes}, and the particular threshold follows from an analysis which we describe further below.
We generate a training set {\tt TP} with ten times less galaxies than {\tt VO}, structured in 4000 cases of different galaxies and shears, and 2000 realizations per case.
As before, we only retain cases with $\langle \snr \rangle > 10$ for the training.
For increased efficiency, following Sect.~\ref{sec:varpsfimpl}, we also generate a special dataset without shear to pretrain the point estimators to predict the ellipticity of galaxy profiles, with 10\,000 galaxies and 100 realizations per galaxy.

To train the weights, we generate a dataset {\tt TW} very similar to {\tt VO}, with 200 cases of different shears and 200\,000 galaxies per case drawn from our GEMS source catalog.
Again, to demonstrate the feasibility of the method even on large datasets, we do not implement any shape-noise cancellation in this entire section.

As input to the ML, we use five features: \texttt{adamom\_g1}, \texttt{adamom\_g2}, \texttt{adamom\_sigma}, $\log(\texttt{adamom\_flux})$, and \texttt{adamom\_rho4}.
The NNs are set up in the same configuration as in Sect.~\ref{sec:fid:ml}, with two layers of five nodes for the $\hat{g}_i$ and a single layer of five nodes for the weights $w_i$, in committees of eight members.

\subsection{Analysis and results} \label{sec:euclidres}

\begin{figure*}[htbp]
\begin{center}
\includegraphics[width=1\linewidth]{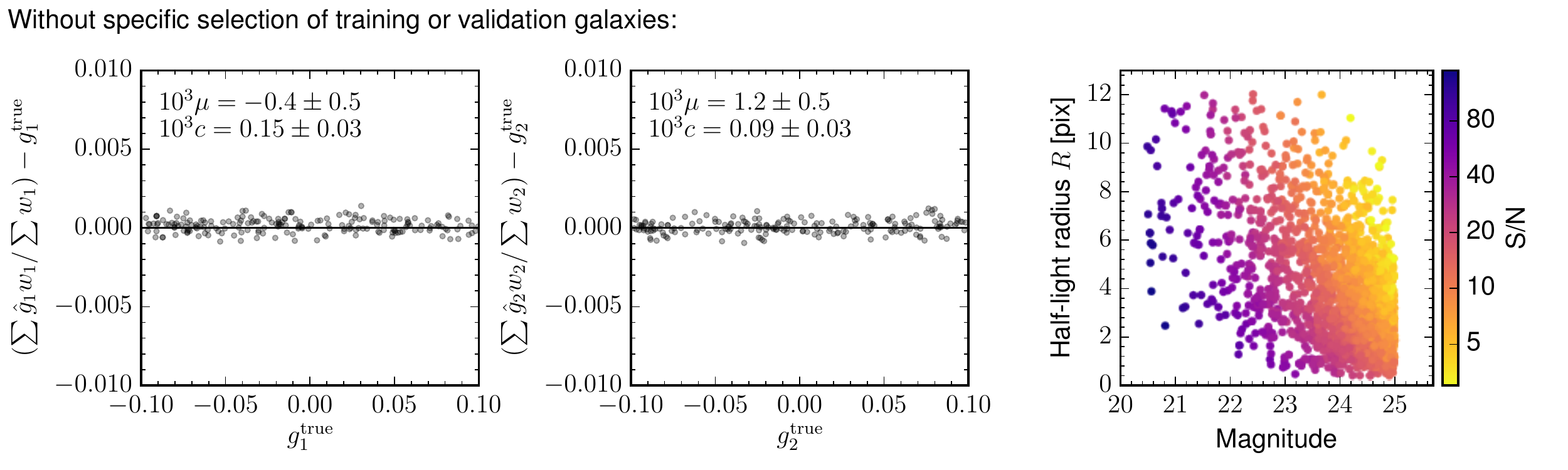} 
\includegraphics[width=1\linewidth]{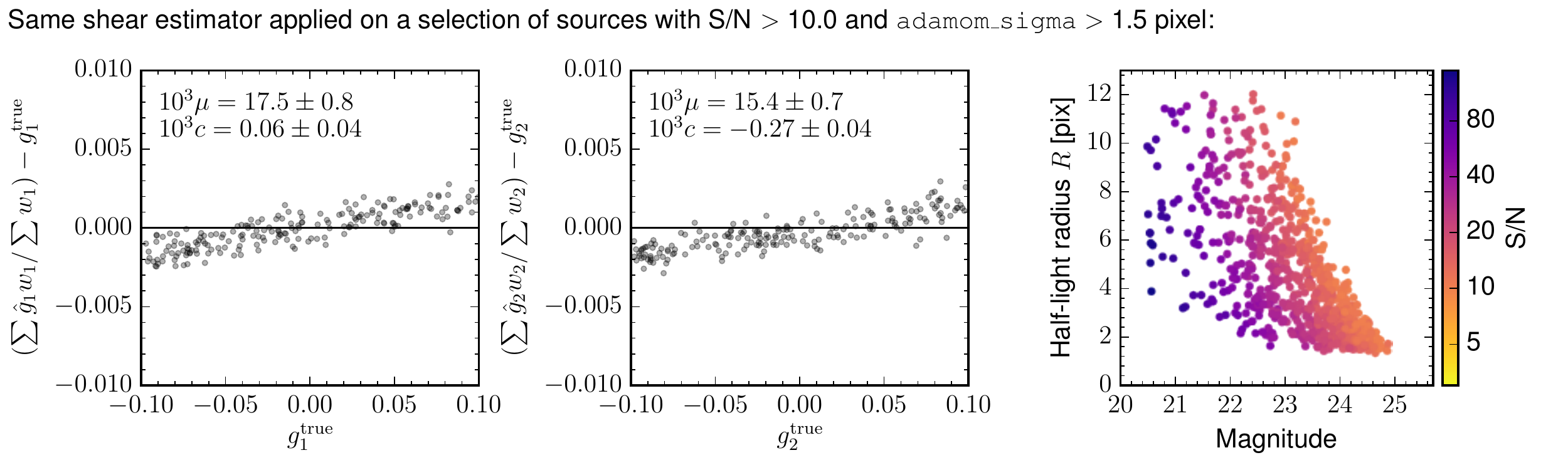} 
\includegraphics[width=1\linewidth]{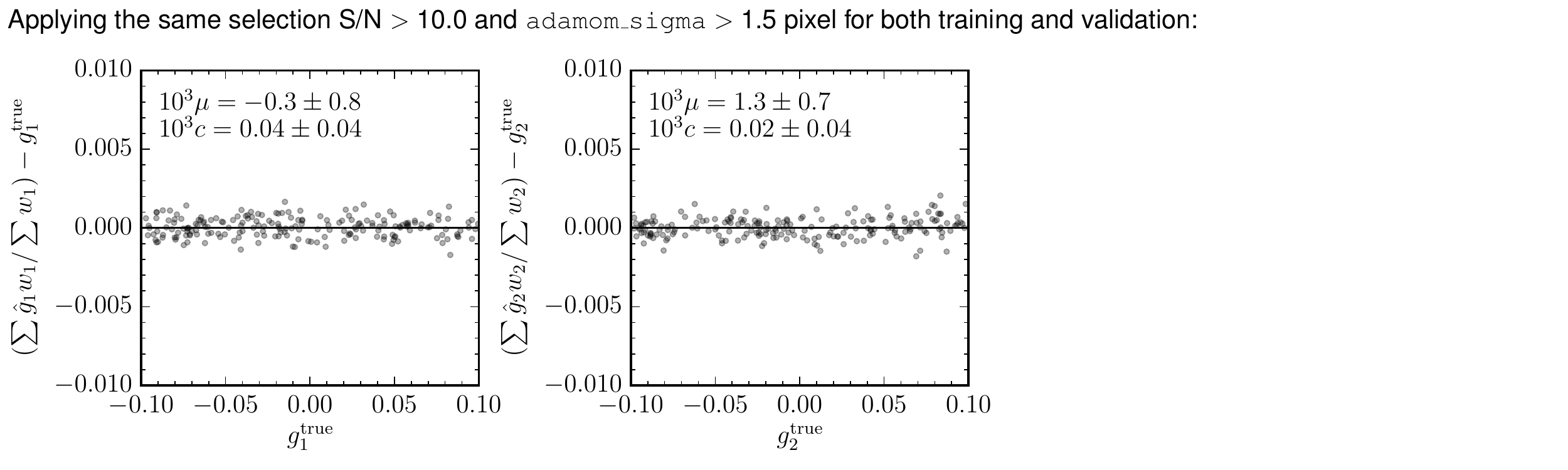} 
\caption{\label{fig:ec_selbias}
Illustration of the effect of a simple yet realistic selection $\snr > 10$ and ${\tt adamom\_sigma} > 1.5$ pixel on the shear bias obtained on \euclid-like simulations, and how the algorithm is trained to compensate for it. 
The \emph{top row} shows average shear estimation residuals on the validation dataset {\tt VO} (200 cases with 400\,000 galaxies per case) with weights trained on {\tt TW}, without specific selections based on observables, except for the success of the feature-measurement algorithm.
The \emph{right-hand panel} shows the measured \snr versus the true size and magnitude for a small random subset of galaxies from the validation set.
The \emph{middle row of panels} shows results from the same estimators, but with the selection function applied to the validation set, resulting in percent-level multiplicative biases $\mu_i$.
For the \emph{bottom panels}, the selection function has been applied both to the training set for the weights, and to the validation set.
The point estimators remain unchanged.
Only their weighting compensates for the selection, resulting in vanishing multiplicative and additive biases.
}
\end{center}
\end{figure*}

We start by examining the suitability of the training data {\tt TP} used for the point estimators $\hat{g}_i$, and in particular the choice for the lower bound on the true galaxy half-light radius $R$.
Figure~\ref{fig:ec_radsigma} shows residual biases of the point estimator $\hat{g}_1$, probed on galaxies drawn from the GEMS catalog.
As expected, the amplitude of the shear gets strongly underestimated for small unresolved galaxies.
This occurs with a relatively well defined threshold at $R \approx 1$ pixel.
For smaller galaxies, in absence of usable information, the ML tries to accommodate the cases at best by yielding a zero shear.
We tested that the occurrence of a threshold at this particular value is not dominated by our choice of the lower bound of $R = 1$ pixel for the training data.
An attempt to train the point estimators on galaxies with $R$ down to $0.5$ pixel does not improve the estimation quality on small galaxies from the GEMS catalog.
We therefore opt for a minimum $R$ of $1$ pixel as a simple choice for this first attempt to train the point estimators  $\hat{g}_i$ under \euclid-like conditions.

One can also observe on the left panel of Fig.~\ref{fig:ec_radsigma} that slightly larger galaxies, around $R = 2$ pixel, show an opposite bias.
We interpret this ``overshooting'' as consequence of the difficulty the ML has with the almost unresolved galaxies around $R = 1$ pixel and the noise in the size estimation feature available to the ML.
An analogous effect is seen with the fiducial simulations in Fig.~\ref{fig:fid-condbias}, and discussed in Sect.~\ref{sec:fid:ml}.
Recall that a point estimator with no conditional biases at all cannot be expected given noisy or uninformative data.
The proposed ML algorithm will use the weights $w_i$ to minimize overall biases by counterbalancing those residual conditional biases.

On the right panel of Fig.~\ref{fig:ec_radsigma}, we observe that our shear point estimation for galaxies with $\langle \snr \rangle > 10$ is not accurate for measured sizes ${\tt adamom\_sigma} < 1.5$ pixel.
This motivates the choice of a somewhat realistic selection function: for the following analysis, we consider the rejection of galaxies with individually measured $\snr < 10$ and ${\tt adamom\_sigma} < 1.5$ pixel.
This selection rejects about $60\%$ of the simulated sources from the datasets used in this section.
The cut in measured size is a first plausible action to reject stars when applying the method to a real survey.
It also contributes to maintaining a low sensitivity of our shear estimator to the distribution of true galaxy sizes.

Considering the shear point estimators to be of satisfactory quality, we proceed by training the weights.
This is the step at which the ML method learns about the selection function of a survey.
The same selection function is applied to the dataset used to train the weights, for the algorithm to adapt to the selected population of source galaxies.
For illustration purposes, we perform this $w_i$-training twice, on the same simulations {\tt TW}: once with the selection $\snr > 10$ and ${\tt adamom\_sigma} > 1.5$ pixel, and once without.

Figure~\ref{fig:ec_selbias} summarizes the resulting shear biases of the weighted estimates.
For the upper panels, no selection is performed. The method achieves low overall biases using all measurable galaxies of the source catalog, despite the significant conditional biases (i.e., sensitivity) shown in Fig.~\ref{fig:ec_radsigma}.
It is therefore not unexpected that significant overall biases appear when the described selection is applied to the validation simulations, as shown in the middle row panels.
Indeed, by changing for example the distribution of true half-light radii $R$ of the galaxy population, the selection alters the cancelling balance achieved by the weighting.
This is a first source of the bias observed in this situation.
In addition, the measured quantities \snr and {\tt adamom\_sigma} on which the selection is made unavoidably correlate with ellipticity and shear at some level.
This second source of bias comes entirely from the selection itself.
Experiments with selections performed on the true galaxy parameters, which are uncorrelated to the shear, suggest that the first bias source is dominating in this particular situation.
A more comprehensive analysis of these effects is however beyond the scope of this paper.
Finally, the bottom panels of Fig.~\ref{fig:ec_selbias} demonstrate that excellent accuracy is achieved when the weights are trained with the same selection as the one applied to the validation set.
The overall multiplicative bias is now at a level of $10^{-3}$ for both components, and consistent with zero given the finite precision.
The overall additive bias $c$ is below $10^{-4}$ in amplitude.
We stress that the same point estimators $\hat{g}_i$ are used in all panels of this figure.
Only their weighting changes.
The NNs predicting the weights $w_i$ successfully learn to balance out the biases resulting from the considered selection.

While we have used a very simple distribution of galaxies to train the point estimates, we have assumed in the above an ideal knowledge of the population of source galaxies to train the weights.
We lift this assumption in the next section.

\section{Application to GREAT3} \label{sec:great3}

In this section, we describe the application of our ML-method to simulations from the GREAT3 challenge.
This allows us to quantify its performance on a reference dataset for which results from other shape measurement methods are available.
Through this experiment, we continue to use only simple single \sersic profiles to draw the training data.
The performance on both the parametric bulge + disk models and the real galaxies of GREAT3 therefore gives a first handle on how sensitive our machine learning method is to the fidelity of the galaxies used for the training.

\subsection{GREAT3 data}

GREAT3 is the latest data analysis challenge for weak-lensing shear measurement algorithms \citep{Mandelbaum:2014dq, Mandelbaum:2015gc}.
While the blind phase of GREAT3 ended in April 2014, the challenge simulations still provide a welcome benchmark for the evaluation of algorithm, and are publicly available\footnote{\url{http://great3challenge.info}}.
Some of us participated in the blind phase with an earlier version of this approach, under the name \emph{MegaLUT}. We summarize the key differences between this early attempt and the present work in Appendix \ref{changesG3}.

The GREAT3 data is structured in several branches of somewhat increasing complexity.
In this work, we focus exclusively on branches for which the PSF of each galaxy is provided (in the form of pixellated images), and which mimic single epoch observations.
To ease the bias analysis, we also limit ourselves to the so-called ``constant-shear'' branches, containing 200 ``subfields'' with stationary shears, and 10\,000 galaxy stamps per subfield.
We stress that our ML approach draws no advantage from this stationary shear situation, and that the same trained algorithm could equally well be applied to variable-shear branches.
These choices leave us with four branches, corresponding to the combinations of either ground- or space-based observations of either ``control'' or ``real'' galaxies.
The ``control'' galaxies were drawn for GREAT3 as combinations of two S\'ersic profiles -- one for the bulge, one for the disk -- based on model fits to Hubble Space Telescope (HST) observations, while the ``real'' galaxies use actual HST images.
Following the standard GREAT3 nomenclature, the four branches we analyze are control-ground-constant (CGC), control-space-constant (CSC), real-ground-constant (RGC), and real-space-constant (RSC).

Within each GREAT3 branch, the PSF model varies widely between subfields. For example, in space-based branches, subfields have PSFs of different telescopes, with different number of spikes.
Attempting to train single NNs to correct for this large diversity of PSFs within a branch has no motivation from any potential application of the method to real survey data, and would require a large number of features characterizing the PSF shape.
We therefore opt for an individual ML training for each subfield, with dedicated training simulations using the PSF of this same subfield.
With this choice, the algorithm uses no NN input features describing the PSF, as the latter is identical for every galaxy in a subfield.
The disadvantage is that we require 200 sets of simulations and ML trainings for each branch. Given the related large computational cost, we keep the algorithm as simple as possible for this section.

\subsection{Simulation parameters, features, ML configuration, and training}
\label{sec:great3params}

We place ourselves in the position of challenge participants, and train and apply our algorithm without using any information that was hidden during the competition.
In addition, we perform no iterative adjustments based on bias evaluations on the GREAT3 data.
We only employ our own internal validation datasets to test our ML setup, as we detail below.

\begin{figure*}[htbp]
\begin{center}
\includegraphics[width=0.9\linewidth]{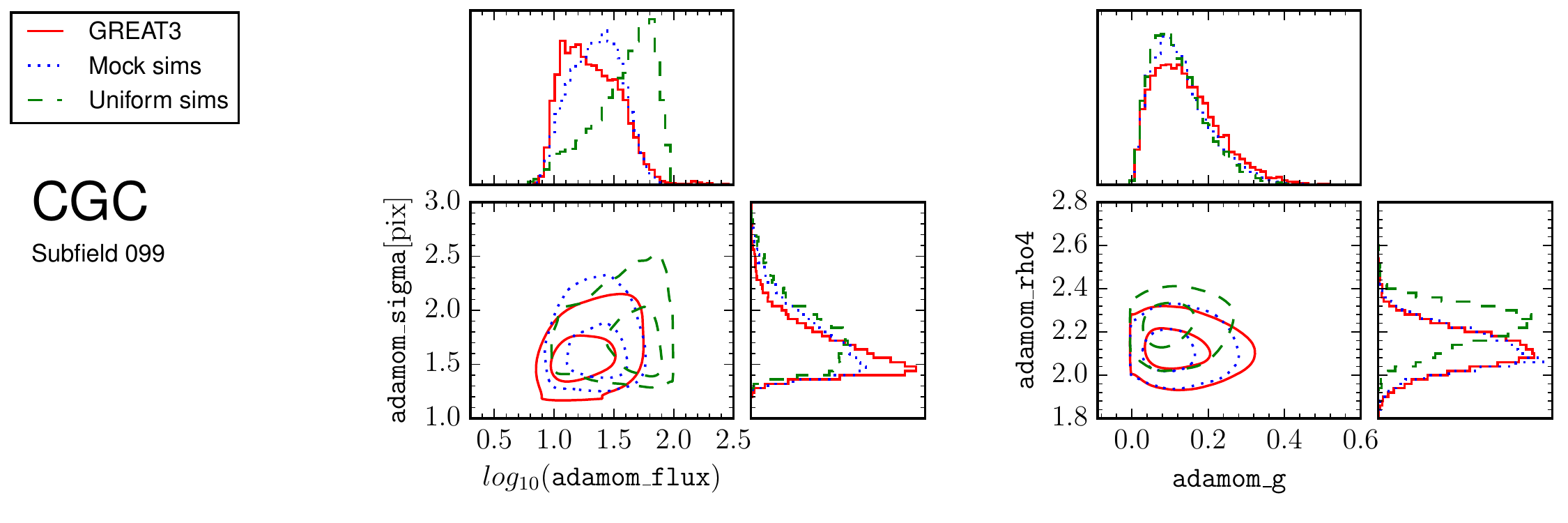} 
\includegraphics[width=0.9\linewidth]{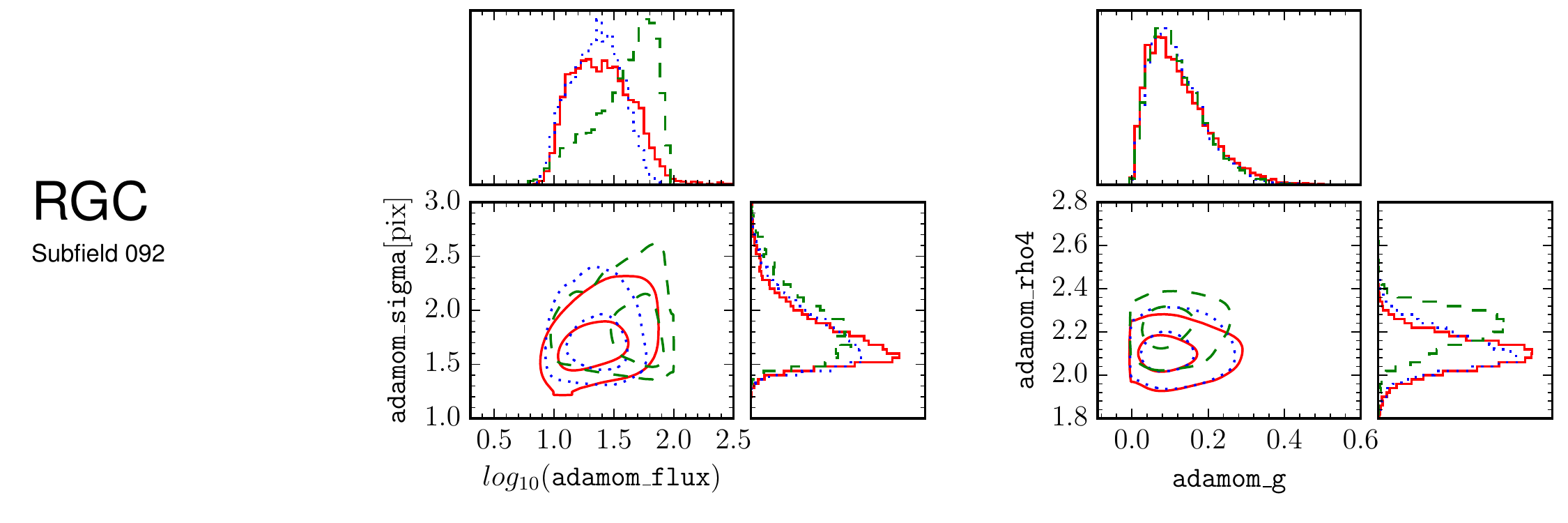} 
\includegraphics[width=0.9\linewidth]{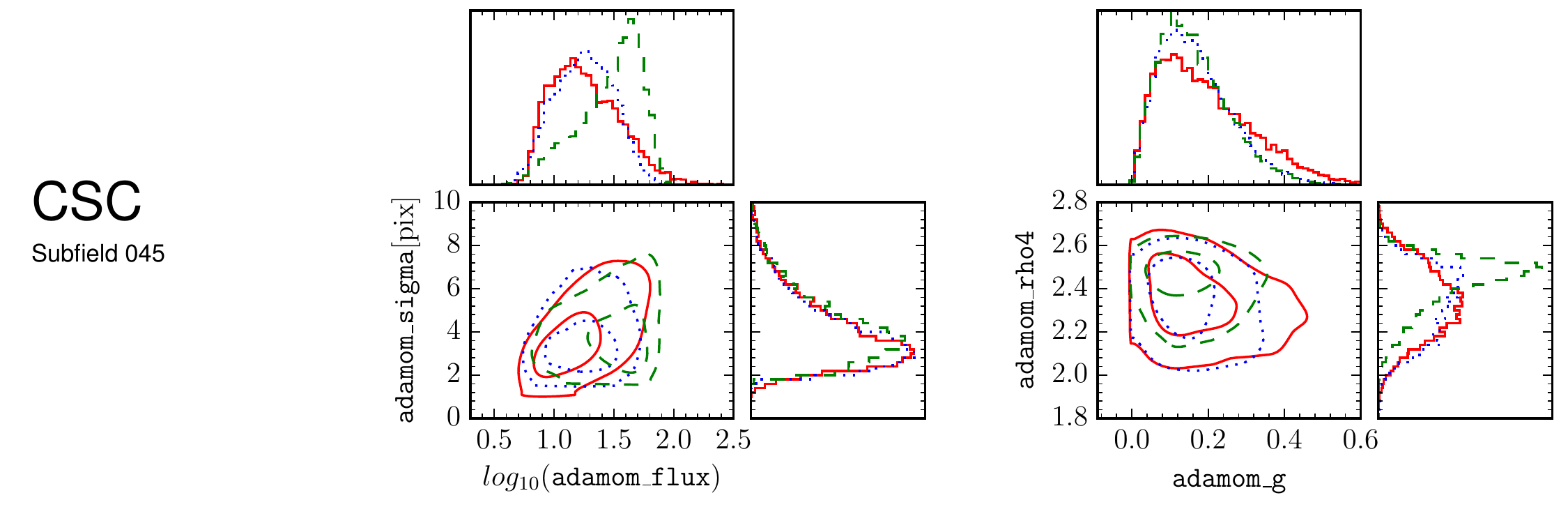} 
\includegraphics[width=0.9\linewidth]{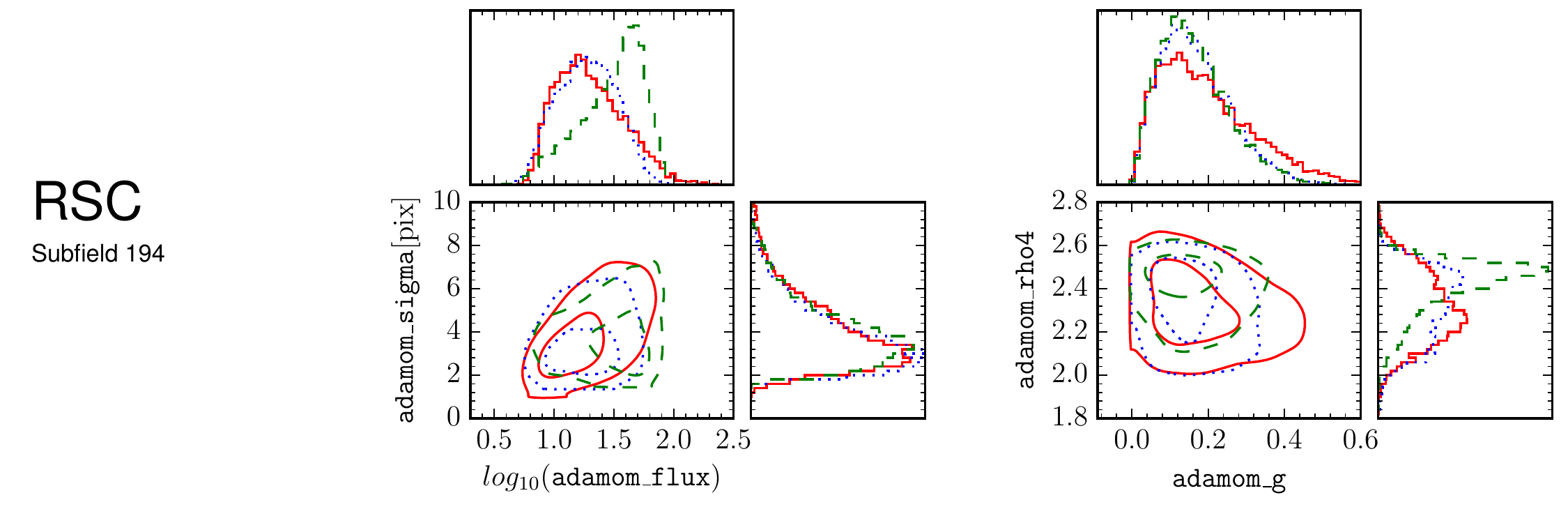} 
\caption{\label{fig:g3_fig_1}
Measured distributions of galaxy features in the GREAT3 data and in ``mock'' and ``uniform'' simulations, for the subfield with the sharpest PSF in each of the four considered branches. $\texttt{adamom\_g}$ denotes the modulus of $\texttt{adamom\_g1} + \mathrm{i}\cdot\texttt{adamom\_g2}$ (as described in Sect.~\ref{sec:adaptativemoments}). Measurements on the GREAT3 data are shown with solid red lines, measurements on the ``mock'' simulations (using the PSF and noise-level of the corresponding subfield) in dotted blue lines, and measurements on the ``uniform'' simulations in dashed green.
}
\end{center}
\end{figure*}

Let us first discuss the parameters of the training and validation simulations, which we draw with intentionally simple single \sersic profiles.
To adapt our simulations to the GREAT3 data, we inspect the distributions of measured features using the subfield with the sharpest PSF of each branch. Figure~\ref{fig:g3_fig_1} shows these distributions, and compares them to two simulation types: one which covers the range of measured sizes, fluxes, and galaxy ellipticities using very simple uniform distributions for the true \sersic parameters (denoted ``uniform''), and one which roughly mimics the GREAT3 data with slightly more flexibility (denoted ``mock'').
The true parameter distributions behind these two simulation types are given in Table~\ref{table:g3simparams}.
We use them for both the ``control'' and ``real'' branches.
Following the same approach as in the previous sections, the uniform simulation type is designed to generate training sets for the point estimates, while the more representative mock simulations are used to train weights, and generate validation sets.
It can be seen on Fig.~\ref{fig:g3_fig_1} that even for the mock type, the match between GREAT3 and our simulations is only very approximate.
A better correspondence could certainly be achieved by adopting joint parameter distributions, if required.
We argue however that these simple simulations are sufficient for the present analysis, given the desired low sensitivity of the algorithm to the galaxy parameter distribution.

\begin{table*}[htbp]
\caption{Parameter distributions of the two types of simulations used to generate the training data for each GREAT3 subfield.}
\label{table:g3simparams}
\centering
\begin{tabular}{l c c c c}
\hline\hline
Branch type: & \multicolumn{2}{c}{Ground-based} & \multicolumn{2}{c}{Space-based} \\
Training set to learn the prediction of: & Point estimate & Weight & Point estimate & Weight \\
Simulation type: & Uniform & Mock & Uniform & Mock \\
\hline
Shear components $g_1$, $g_2$ & 0 & $\mathcal{U}(-0.1, 0.1)$ & 0 & $\mathcal{U}(-0.1, 0.1)$ \\
Galaxy ellipticity modulus $\varepsilon^{\mathrm{true}}$ & $\mathcal{R}(0.2)_{[0,\,0.7]}$ & $\mathcal{R}(0.2)_{[0,\,0.7]}$ & $\mathcal{R}(0.2)_{[0,\,0.7]}$ & $\mathcal{R}(0.2)_{[0,\,0.7]}$ \\
S\'ersic index\tablefootmark{a} $n$ & $\mathcal{U}(0.5, 4)$ & $\mathcal{U}(0.5, 2.5)$ & $\mathcal{U}(0.5, 4)$ & $\mathcal{U}(0.5, 4)$ \\
Flux $F$ [counts] & $\mathcal{U}(10, 100)$ & $\mathcal{N}(15, 20)_{[10,\,200]}$ & $\mathcal{U}(10, 100)$ & $\mathcal{N}(0, 30)_{[10,\,200]}$ \\
Half-light radius $R$ [pix] & $\mathcal{U}(0.75, 3.0)$ & $\mathcal{N}(1.0, 0.8)_{[0.75,\,3.0]}$ & $\mathcal{U}(1.25, 10.0)$ & $\mathcal{N}(2.5, 3.5)_{[1.25,\,10.0]}$\\
\hline
\end{tabular}
\tablefoot{
We use $\mathcal{U}(a, b)$ to denote the uniform distribution between $a$ and $b$, $\mathcal{N}(\mu, \sigma)$ to denote a normal distribution with mean $\mu$ and variance $\sigma^2$, and $\mathcal{R}(\sigma)$ for a Rayleigh distribution with mode $\sigma$. Intervals in subscript denote the range to which we clip a distribution, so that no sample falls outside of the given interval.\\
\tablefoottext{a}{In practice, we grid the values for the S\'ersic index instead of drawing them randomly. This significantly speeds up the galaxy stamp generation, as \texttt{GalSim} can reuse cached S\'ersic profiles.}
}
\end{table*}

Both for the prediction of the point estimate $\hat{g}_i$ and for the weights $w_i$, we use the four input features \texttt{adamom\_g}$i$, \texttt{adamom\_sigma}, \texttt{adamom\_flux}, and \texttt{adamom\_rho4}.
Indeed, given the constant PSF and noise properties within each subfield, the ML algorithm does not require distinct input about these at the location of each galaxy.
In terms of a real-life application, this configuration corresponds to the assumption that the ML is optimally informed about the PSF and background noise.
We tested that adding the feature \texttt{adamom\_g2} (\texttt{adamom\_g1}) when predicting $\hat{g}_1$ and $w_1$ ($\hat{g}_2$ and $w_2$) does not improve the performance for this particular analysis.

The use of plain \sersic profiles with well-defined ellipticities allows us to simplify the first step of the training algorithm, by training a point estimator for ellipticity instead of shear, as described in Sect.~\ref{sec:ellipshortcut}, and done in previous sections as a from of pretraining.
For each subfield of the branches CGC, RGC, CSC, and RSC, we generate a training set with a structure as illustrated in Fig. \ref{fig:ellip}, with $1000$ cases, and drawing from the ``uniform'' parameter distributions of Table \ref{table:g3simparams}.
Interestingly, we observe that we obtain good results when training the estimator $\hat{g}_i$ without adding noise to the generated stamps.
Validations show that, in absence of pixel noise, we can opt for a small number of only $10$ realizations per case.
These realizations still differ in the sub-pixel positioning of the \sersic profiles on their stamps, which is uniformly distributed.
Individually for each case, we randomly select one of the nine provided ``star'' images and use it as kernel to convolve the galaxies by the PSF.

Training the weights requires simulations with the full noise level, and therefore significantly larger training sets.
To keep the computational cost tractable, we chose to include shape noise cancellation (SNC) into these simulations.
We follow the scheme introduced in Sect.~\ref{sec:fiducial:varpsf-field}, with two realizations of each galaxy rotated by $90^{\circ}$ before shearing and PSF convolution.
The same SNC scheme is implemented in the GREAT3 simulations \citep{Mandelbaum:2014dq}.
We acknowledge that our ML method will adjust to optimally exploit this feature of GREAT3, by exaggeratedly down-weighting low-\snr galaxies (see the discussion in Sect.~\ref{subsec:shearweights}).
However, we motivate our choice in favor of SNC by the higher precision achievable on the GREAT3 data.
This allows in particular for a better demonstration of residual biases related to the difference in galaxy models between training and GREAT3 data, on which we focus in this section.
We generate training sets for the weights with 200 cases and 1000 realizations per case (i.e., 500 different galaxies per case), drawing from the ``mock'' parameter distributions.
We add simple stationary Gaussian noise in the stamps, with a variance measured on the GREAT3 subfields via Equation \ref{equ:skymad}.

With the same mock parameter distributions we also generate internal validation sets for the space- and ground-based branches.
These sets mimic the GREAT3 data by using the same PSFs and noise properties as the CGC and CSC branches, as well as a structure of 200 subfields with 10\,000 stamps each.
The difference to GREAT3 is that galaxies are drawn with the same single-\sersic parametrization used for the training.
The resulting validation sets, which we call the fiducial branches fGC and fSC, provide a reference for our first analysis of the sensitivity to the realism of the training data.

For all estimators, we use committees of eight members with two hidden layers and five nodes per layer, as described in Sect.~\ref{subsec:nndetails}.
The feature measurement fails on typically less than $0.1\%$ of the galaxies (internal or GREAT3), and no other selections on the datasets are performed.

\subsection{Results and discussion}

\begin{figure*}[htbp]
\begin{center}
\includegraphics[width=1.0\linewidth]{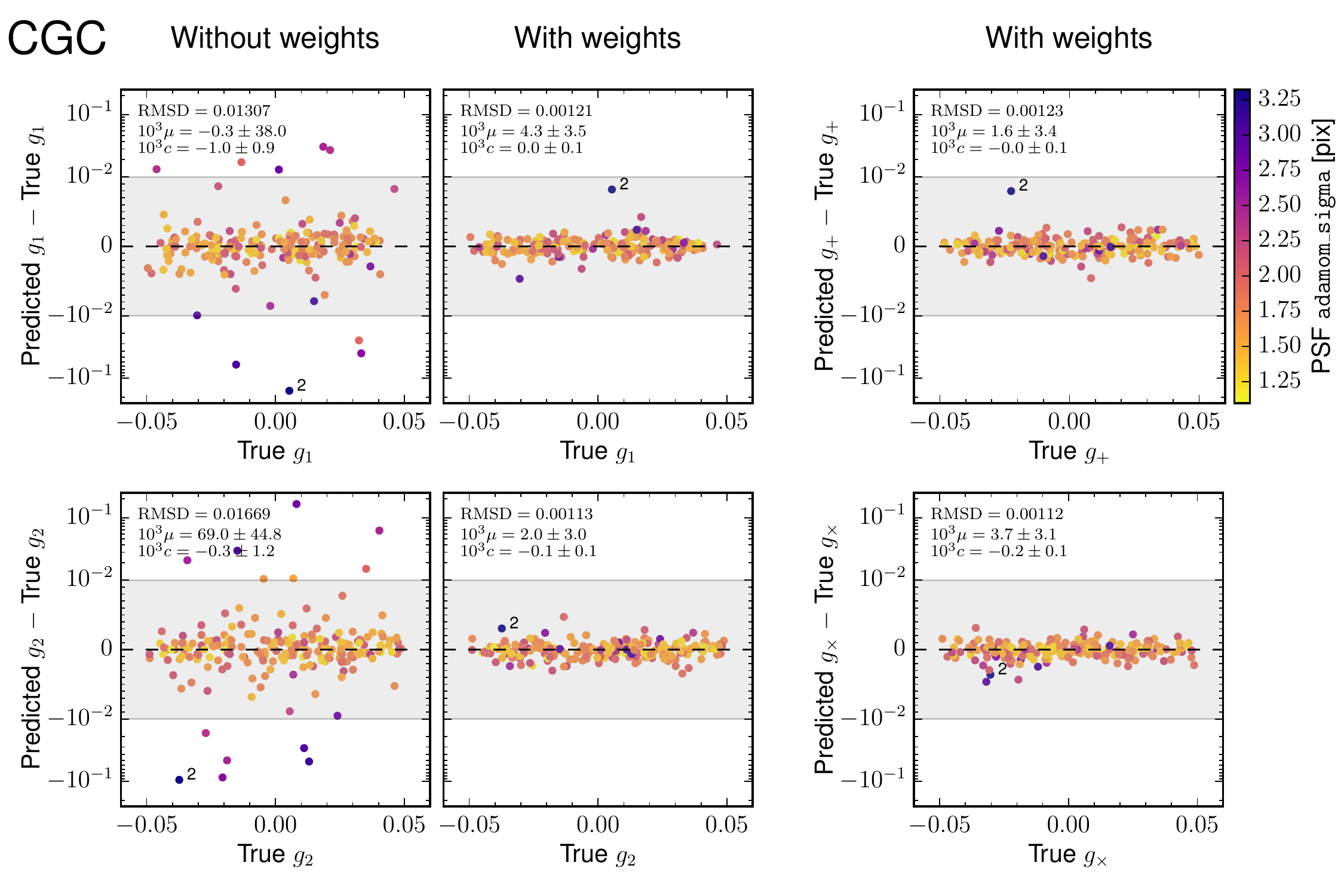} 
\caption{\label{fig:g3_fig_2_cgc}
Analysis of the shear estimation errors on the CGC branch of GREAT3.
In each panel, points show the residuals of the average estimated shears against the true shears of the 200 subfields.
The different panels show different components of the shear, in the frame of the pixel grid ($g_1$ and $g_2$) and in a frame rotated to be aligned with the PSF anisotropy ($g_+$ and $g_{\times}$), as defined in Sect.~\ref{sec:biases}.
The residuals are shown on a linear scale within the shaded area, and using a logarithmic scale outside of this region.
The \emph{leftmost panels} show residuals of unweighted average shears, i.e., ignoring the weights predicted by the second step of the algorithm.
In all panels, colors of the datapoints encode the size of the PSF, with darker colors corresponding to broader PSFs. The infamous subfield labeled ``2'' has the worst PSF of this GREAT3 branch, with a strong defocus.
}
\end{center}
\end{figure*}

\begin{figure*}[htbp]
\begin{center}
\includegraphics[width=1.0\linewidth]{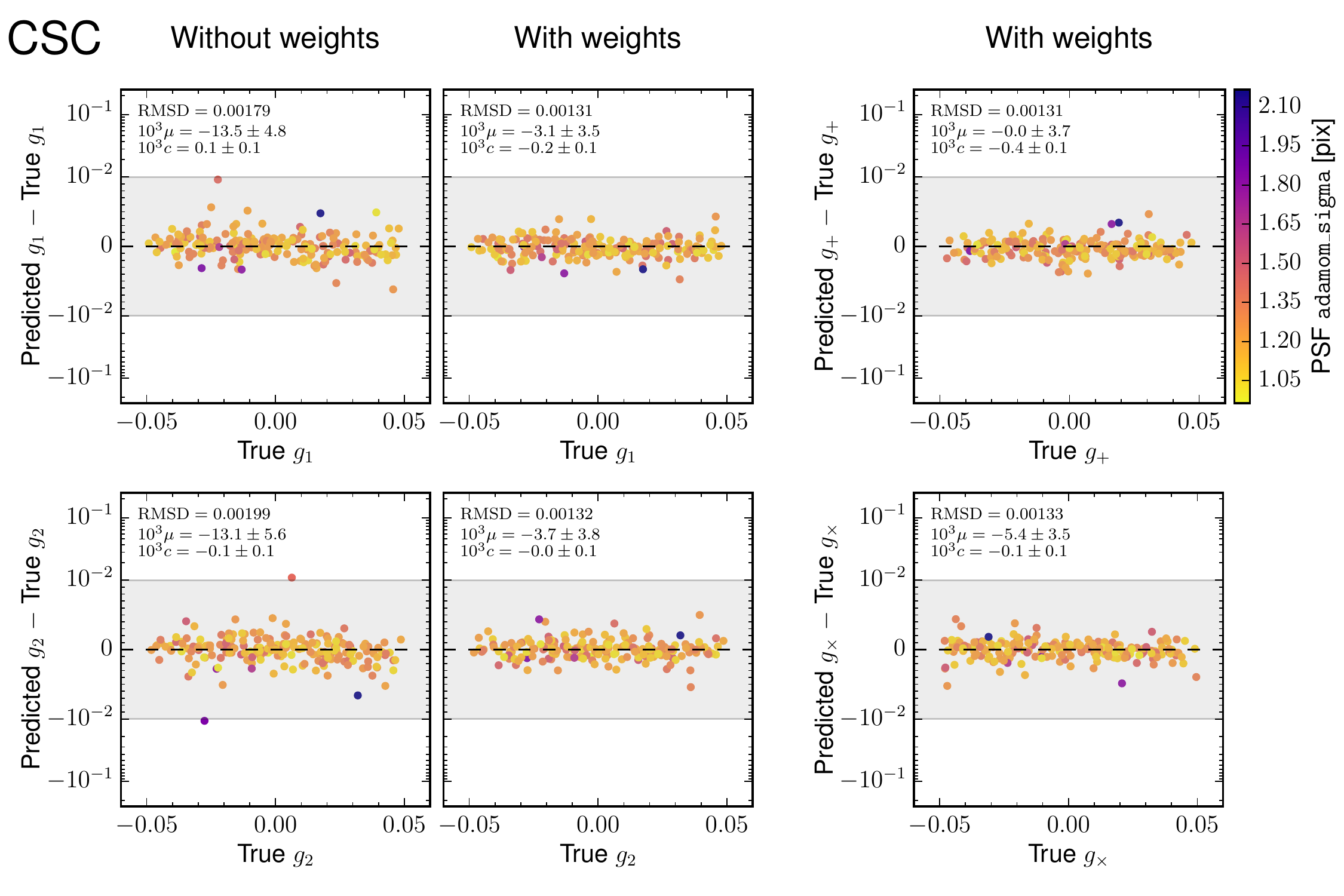} 
\caption{\label{fig:g3_fig_2_csc}
Shear estimation errors on the CSC branch, otherwise similar to Fig.~\ref{fig:g3_fig_2_cgc}.
}
\end{center}
\end{figure*}

\begin{table*}
\caption{Multiplicative and additive biases measured on GREAT3 branches, and ``fiducial'' validation sets (fGC and fSC) mimicking the GREAT3 control branches but using simple S\'ersic galaxy profiles. See Fig.~\ref{fig:g3_cross_bias} for a visualization.}
\label{table:g3results}
\centering
\newlength{\defaulttabcolsep} 
\setlength{\defaulttabcolsep}{\tabcolsep} 
\setlength{\tabcolsep}{0.42em} 
\begin{tabular}{c c c c c c c c c}
\hline\hline
Branch & $\mu_1\cdot10^3$ & $\mu_2\cdot10^3$ & $\mu_{+}\cdot10^3$ & $\mu_{\times}\cdot10^3$ & $c_1\cdot10^3$ & $c_2\cdot10^3$ & $c_{+}\cdot10^3$ & $c_{\times}\cdot10^3$\\
\hline
fGC & $-1.50\pm2.78$ & $+3.18\pm2.04$ & $+1.97\pm2.28$ & $-0.30\pm2.62$ & $+0.02\pm0.08$ & $+0.01\pm0.06$ & $+0.08\pm0.07$ & $-0.02\pm0.07$ \\
CGC & $+4.34\pm3.52$ & $+1.99\pm3.03$ & $+1.58\pm3.43$ & $+3.70\pm3.06$ & $+0.02\pm0.09$ & $-0.11\pm0.08$ & $-0.03\pm0.09$ & $-0.16\pm0.08$ \\
RGC & $-5.42\pm3.82$ & $-4.72\pm3.42$ & $-3.84\pm3.34$ & $-7.98\pm3.79$ & $+0.12\pm0.09$ & $-0.11\pm0.08$ & $+0.29\pm0.08$ & $-0.02\pm0.09$ \\
\hline
fSC & $-0.97\pm1.49$ & $-1.87\pm1.50$ & $-1.55\pm1.41$ & $-1.18\pm1.57$ & $-0.04\pm0.04$ & $-0.00\pm0.05$ & $+0.07\pm0.04$ & $-0.02\pm0.04$ \\
CSC & $-3.11\pm3.55$ & $-3.73\pm3.78$ & $-0.04\pm3.73$ & $-5.37\pm3.52$ & $-0.19\pm0.09$ & $-0.00\pm0.09$ & $-0.35\pm0.09$ & $-0.09\pm0.09$ \\
RSC & $-5.49\pm3.15$ & $-6.40\pm3.15$ & $-3.77\pm3.08$ & $-7.70\pm3.25$ & $+0.04\pm0.08$ & $+0.04\pm0.08$ & $-0.01\pm0.08$ & $-0.08\pm0.08$ \\
\hline
\end{tabular}
\setlength{\tabcolsep}{\defaulttabcolsep} 
\end{table*}

\begin{figure}[htbp]
\begin{center}
\includegraphics[width=1.\linewidth]{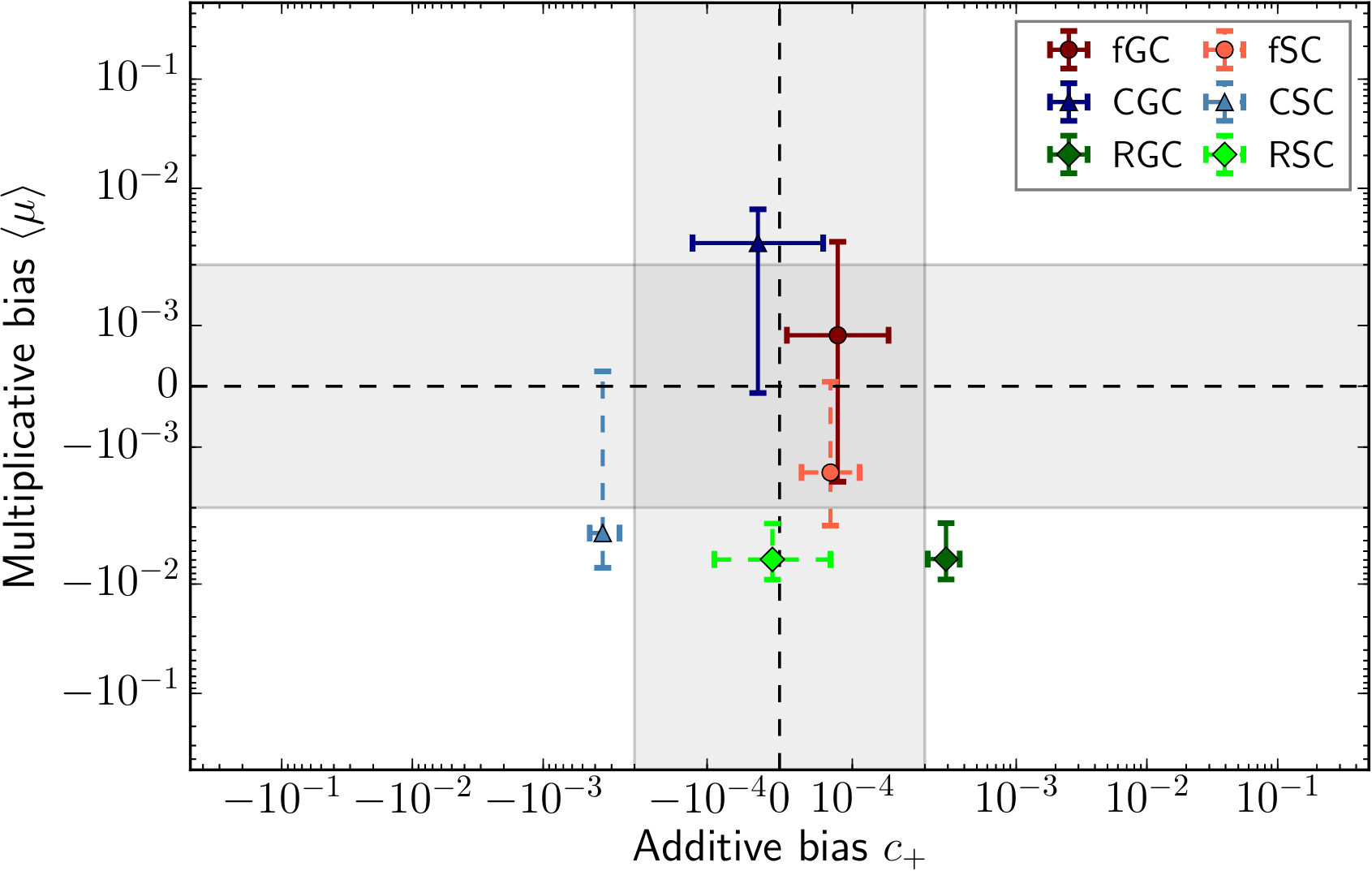} 
\caption{\label{fig:g3_cross_bias}
Multiplicative bias $\mu$ averaged over the components 1 and 2, against the additive bias $c_+$ defined in the coordinate system of the PSF anisotropy.
This figure can be directly compared with Fig.~17 of the GREAT3 result paper \citep{Mandelbaum:2015gc}. 
Space branches are shown with dashed error bars.
The axes are linear within the the gray-shaded region, and logarithmic outside. 
}
\end{center}
\end{figure}

\begin{figure*}[htbp]
\begin{center}
\includegraphics[width=0.8\linewidth]{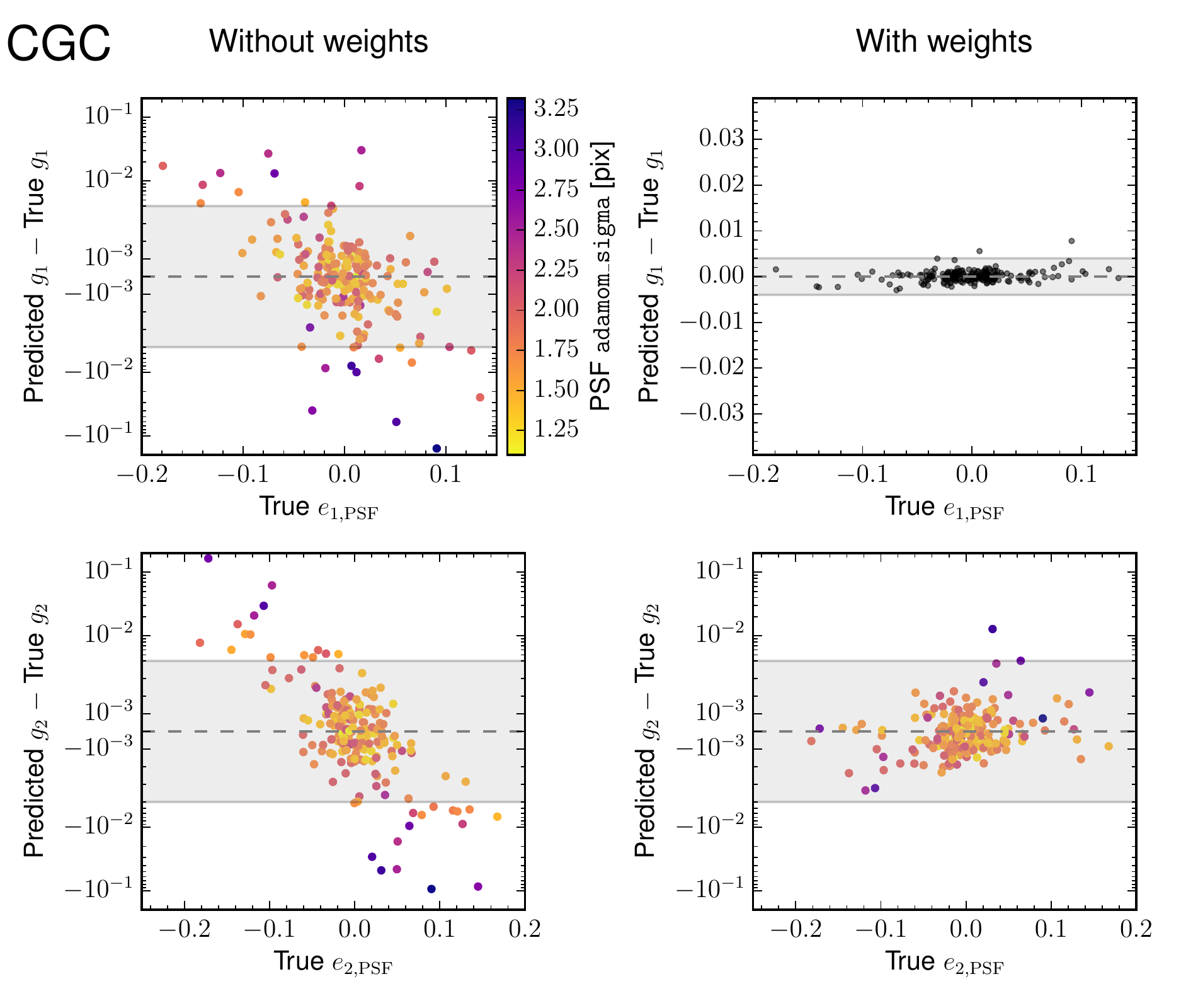} 
\caption{\label{fig:g3_fig_3_cgc}
Shear estimation residuals as a function of PSF anisotropy, for the CGC branch of GREAT3.
Points show the residuals of the average estimated shear components of the 200 subfields, against the anisotropy of the PSF for the same component.
The \emph{left panels} show residuals of unweighted average shears, and the \emph{right panels} show residuals of weighted average shears.
For this figure only, we use the distortion $e$ to quantify the anisotropy of the PSF, following Eq. (6) of \citet{Mandelbaum:2014dq}.
The shaded region covers the same range of residuals in all panels.
Its extension, and the axes range of the \emph{top right panel} is chosen to be directly comparable with Fig. 3 from \citet{Huff:2017ww}.
In the three other panels, the residuals are shown on a linear scale within the shaded area, and using a logarithmic scale outside of this region.
}
\end{center}
\end{figure*}

Figures \ref{fig:g3_fig_2_cgc} and \ref{fig:g3_fig_2_csc} show the residuals of these average predicted shears with respect to the true shears on the 200 subfields of the CGC and CSC branches, respectively. 
In particular for the ground-based branch CGC, the use of weights drastically improves the precision on these average $g_i$, reducing the RMS deviation of the residuals by a factor $\sim$$10$.
This can easily be understood, as the PSFs of CGC have a large range of quality (seeing), shown by the color scale in Fig.~\ref{fig:g3_fig_2_cgc}.
For the subfields with very wide PSF (dark purple points), many of the galaxies with small intrinsic extension carry little shear information.
The weight training successfully adapts to these PSFs.
For the space-based branch CSC (Fig.~\ref{fig:g3_fig_2_csc}), the increase in precision due to the ML-learned weighting of galaxies is smaller, but here the weights bring a significant improvement in multiplicative bias as well.
The situation for the RGC and RSC branches is qualitatively and quantitatively very similar.

Multiplicative and additive biases are computed from linear least-squares fits to the residuals shown in these figures.
Table~\ref{table:g3results} presents the complete set of bias measurements for all considered branches.
The results labeled with fGC and fSC are obtained from the internal validation sets, which use the same single-\sersic galaxy parametrization as in the training datasets.
Results from this table are visualized in Fig.~\ref{fig:g3_cross_bias}.

We observe that no significant biases can be detected on our fiducial sets, and that the results for the ``control'' branches CGC and CSC are equally good except for a small $c_{+}$ of $3.5\cdot10^{-4}$ for CSC.
Arguably, the tendency toward negative $\mu$ for the real galaxy branches indicates a moderate ``model bias'', that is a bias resulting from the training with overly simplistic \sersic profiles.
We note that these biases are still sub-percent.

Figure~\ref{fig:g3_fig_3_cgc} gives a different view on the quality of the PSF correction achieved for the CGC branch.
The same unweighted and weighted average shear residuals are shown against the anisotropy of the PSF, to allow a direct comparison with the recent work by \citet{Huff:2017ww}.
All subfields are represented within this figure, including the widest PSFs.
By down-weighting galaxies for which no accurate shear predictions can be obtained, the ML achieves a PSF-correction with almost no significant residuals.

We stress again that the precision and accuracy achieved by our method is improved by the training against simulations with SNC.
This allows the method to exploit the given GREAT3 data beyond the level that a training for real sky conditions would achieve.

An important conclusion from the application of our ML method to the GREAT3 benchmark is that a rather small set of ML input features seems to extract sufficient shear information to be highly competitive with other state-of-the art shape measurement methods.
While the input features can likely be further improved, we take this observation as an indication that a much larger set of features, or a deep learning approach, might not be required. 

Let us recall some potential reasons for residual biases in the present GREAT3 analysis:
\begin{enumerate}
\item For both the ``real-galaxy'' and ``control'' branches, the galaxy models used for the training (simple S\'ersic profiles) differ significantly from the galaxies in the GREAT3 data. 
\item The PSF information used for the training is taken from the pixellated images of the PSF model, which have the same coarse sampling as the galaxy images.
\item The adjustment of galaxy property distributions of the training data was kept very simple.
\end{enumerate}
As we summarize in the next section, these key aspects can all be addressed in a survey pipeline by generating training data with actual galaxies (e.g., from HST observations), and using a finely sampled model of the PSF.

\section{Toward a comprehensive shear pipeline}
\label{sec:outlooks}

A number of observational conditions and astrophysical effects were not included in the tests and demonstrations conducted in this paper.
We list some of the most important ones as well as further challenges in the following, and briefly give our point of view on how they can be addressed in future work.

\begin{enumerate}

\item \emph{Galaxy colors}, if not accounted for, lead to intolerable shear estimation biases for broadband space-based imaging data.
These biases arise from the strong wavelength dependence of the diffraction-limited PSF, and the large variety of galaxy SEDs.
Furthermore, spatial variations of color across galaxy profiles -- the so-called color gradients -- introduce biases which cannot be neglected for surveys like \euclid \citep{Voigt:2012fi, Semboloni:2013ee, Cropper:2013gc, Er:2018ek}.
These color gradients are not observable on individual galaxies without multiband space-based imaging.
However, they do correlate with galaxy colors (or types), accessible from ground-based photometry.
Since these galaxy colors vary statistically with the environment, a correction for color-related effects must be performed at the level of individual galaxies to avoid spatial variations in the shear bias.
For a machine learning method, the natural way to perform this correction is by including representative galaxy colors, color gradients, and wavelength-dependent PSFs in the training simulations, and by using observed galaxy colors as additional input features.

\item \emph{Blended sources} are frequent in the crowded images of real surveys.
Galaxies can be interacting in close physical proximity, or blend with other unrelated galaxies or stars by projection.
These blends affect the shear measurement by changing the effective source ellipticity distribution which leads to non-negligible biases (see, e.g., 
\citealt{Dawson:2016db} and \citealt{Samuroff:2018hr} for a recent discussion and the impact on DES).
Simply rejecting all recognisable blends is harmful as it would (1) remove a significant fraction of source galaxies and (2) lead to entangled biases in the shear correlation function \citep{Hartlap:2011ef}.
A potential way to mitigate these problems is to obtain the ML input features from a multi-object fitting of simple elliptical profiles to the observed images instead of the moment measurement used in this paper.
This fitting could be done without forward-modeling any PSF convolution, and therefore be of comparable computational cost than the adaptive moment measurement.

Galaxies also blend with sources below the detection limit, which result in a form of correlated pixel noise.
This leads to significant biases, with a dependence on the local source density \citep[][and Martinet et al. in prep.]{Hoekstra:2017hg}.
To correct for this effect in line with the algorithm discussed in this paper, one could include faint galaxies in the training simulations, and add features probing the density of these sources to the ML input.

\item \emph{Image artifacts and cosmic rays} require masking to mitigate their impact.
A fitting approach to feature measurement allows to easily ignore masked pixels. We note that the \texttt{GalSim} implementation of adaptive moments, which can be seen as a fit of an elliptical Gaussian profile, does handle masks \citep{Rowe:2015ema}.

\item \emph{Multiple exposures} of the same fields are usually acquired in any survey.
However, in this study, we assumed that each source galaxy was characterized by only one set of measured features.
Two possibilities to deal with multiple exposures are either to co-add the images ahead of the feature measurement, or to combine results from the multiple exposures a posteriori, at the catalog level.
\citet{Bernstein:2002gq} discuss the advantages and disadvantages of the two approaches, and recommend combining the information at the catalog level rather than at the image level.
This notably avoids problems related to the differing PSFs of the exposures and the interpolation of pixels for registration.
Whatever solution is explored, it might be beneficial to include the combination in a training process, and thereby use ML to correct for any resulting biases.

\item \emph{Astrometric distortions} of the images are currently not handled.
To take these distortions into account, features could be directly measured in a world coordinate system (WCS) rather than in pixel space, or the shear predictions could be converted from pixel space to WCS a posteriori.

\item \emph{The realism of the morphology of training galaxies}, that is their fidelity to the real source galaxies from the survey, is important to avoid ``model bias''.
In Fig.~\ref{fig:g3_cross_bias}, we have observed a possible indication for such a model bias when training our ML algorithm with simple \sersic profiles and applying it to the real galaxy branches of the GREAT3 challenge.
However, the described approach can well be trained with more sophisticated analytical simulations, or simulations based on high-\snr observations of real galaxies.
We have shown that a definition of the ellipticity of a galaxy image is not required.
We expect the realism to become more important with additional ML features, such as galaxy colors.
Data augmentation techniques might be necessary to generate large enough training sets.

\item \emph{Ensemble properties of the population of source galaxies} have to be estimated, so that the simulated training sets are sufficiently representative of the survey data given some requirements.
This paper illustrates how ML methods can be designed to favor low conditional biases, thereby minimizing the sensitivity to the distribution of galaxy properties (see, e.g., Fig.~\ref{fig:fid-condbias-weights}).
But some sensitivity to the training population inevitably remains.
While results from our simple application to GREAT3 (Sect.~\ref{sec:great3params}) suggest that this sensitivity might be manageable, it will be important to quantify this aspect in future work, with realistic assumptions.
The estimation of these ensemble properties will likely make use of observations that are deeper than the cosmic shear survey, and it also requires knowledge of the selection function.

\item \emph{The choice and design of input features} was intentionally kept simple in this paper, to demonstrate feasibility and to obtain a benchmark solution.
It remains to be tested if other measurements can extract more information from the galaxy images.
Some preprocessing of features to perform an approximative analytical correction for the PSF \citep[e.g., with DEIMOS,][]{Melchior:2011je} could reduce the required capacity of the ML algorithm and the size of the training data.

\item \emph{The shear estimation formalism} of point estimates and weights is a simple but potentially insufficient description of the shear information extracted from galaxy images.
One of the next development stages of the proposed ML approach could be to predict shear probability distributions represented for example by Gaussian mixture models as done for photometric redshifts in \citet{DIsanto:2018jd}.

\end{enumerate}

\section{Summary and conclusions}
\label{sec:conclusion}

This paper explores the use of supervised machine learning to address the problem of weak-lensing shear measurement.
We propose and analyze an algorithm based on artificial neural networks, which regress a point estimate and predict a weight for each shear component of each source galaxy.
The algorithm is trained on image simulations with known shear, and uses measured shape parameters of the observed galaxy images as input features.
This training is divided into two distinct steps and makes use of unconventional cost functions, as summarized below.

First, the point estimator is optimized to minimize its bias, as far as possible, individually in all regions of the ``nuisance parameter space'' corresponding to the different galaxies encountered in the survey.
This aims at obtaining a lowest possible sensitivity to the distributions of these parameters.

Then, in a second step, the weight prediction is trained to optimize the accuracy of the weighted average point estimator over an assumed distribution of these nuisance parameters.
A particularity of this training is that there are no ``true'' target weights.
The learned weights accommodate for the variable \snr of the different galaxies, and counterbalance residual biases of the point estimators.
Furthermore, the weights can straightforwardly compensate for biases originating from the observational selection of the galaxies, if the same selection function is applied to the training simulations.
This feature is interesting on its own, as it can also be employed to address selection biases encountered by other shape measurement methods.

We demonstrate the potential of our method by applying it to a range of different simulations.
The first application confirms the convergence of the training algorithm, and shows that a low sensitivity of the point estimator to the true half-light radius, \sersic index, ellipticity modulus and average \snr of simulated galaxies is achieved.
The introduction of trained weights successfully eliminates overall biases on a population of galaxies whose properties are assumed, at this stage, to be perfectly known.
We then show how the method is able to correct for a PSF model with highly variable ellipticity, while still using neural networks of modest size with two hidden layers of ten nodes.
More realistic simulations, based on galaxy parameters from the GEMS survey and mimicking the \euclid VIS instrument allow for a first quantitative best-case assessment.
We introduce a simple yet plausible observational selection function, which rejects low-\snr and unresolved sources. Training the weights with the same selection function, we obtain multiplicative and additive biases consistent with zero with an uncertainty below $10^{-3}$ and $10^{-4}$, respectively. 
Lastly, we use the GREAT3 dataset to illustrate the robustness of the method to discrepancies between the distributions of galaxy parameters in the data to be analyzed and in the training simulations.
On this GREAT3 data we also observe that a multiplicative model bias, originating from the measurement on real galaxies with an algorithm trained with simple \sersic profiles, remains sub-percent.
We note that this is a pessimistic point of view. 
Given that the proposed approach does not require the definition of a true ellipticity, it can well be trained with simulations based on high-\snr observations of real galaxies.

We do not foresee major obstacles in the further development of machine learning shear measurement methods.
The low computational cost at runtime (few ms per galaxy) and the ease with which intricate effects can be integrated into the training simulations are strong advantages of these approaches.
To address effects relevant for upcoming surveys, the complexity of the training sets likely has to be increased.
The current capabilities of the {\tt GalSim} software package already allow analyses with much higher fidelity to the real sky compared to our present demonstrations.
We stress that the aim of this paper is to demonstrate the feasibility of a shape measurement by neural networks rather than to describe a finalized pipeline.
Obtaining good results on an idealized challenge such as GREAT3 does not allow us to conclude on the readiness for real data.
A further step in the realism of image simulations (in particular, using crowded stamps) is required.
However, a conclusion of the present work is that machine learning offers a promising route to address the interwoven biases affecting weak-lensing measurements.

\begin{acknowledgements}

We wish to thank our referee for pertinent and helpful comments, 
Riccardo de Lutio and Bal\'azs K\'egl for useful discussions,
and the organizers of the GREAT3 challenge, as well as the developers of \galsim, for their thorough work.
Parts of this work were discussed during an international team meeting on ``Cosmology with size and flux magnification'' led by Alan Heavens and HH at the International Space Science Institute (ISSI).
MT acknowledges support by a fellowship of the Alexander von Humboldt Foundation.
TK and FC are supported by the Swiss National Science Foundation (SNSF).
RN and TS acknowledge support from the German Federal Ministry for Economic Affairs and Energy (BMWi) provided via DLR under project no. 50QE1103.
HH and MT are supported by the DFG Emmy Noether grant Hi 1495/2-1.
This work greatly benefits from the community-developed \texttt{scipy} \citep{numpyscipy}, \texttt{matplotlib} \citep{matplotlib}, and \texttt{astropy} \citep{Robitaille:2013cd} packages.

\end{acknowledgements}

\bibliographystyle{aa}
\bibliography{papers}

\begin{appendix}

\section{Illustration of an inverse regression}\label{appendix:invreg}

\begin{figure*}[htbp]
\begin{center}
\includegraphics[width=0.9\linewidth]{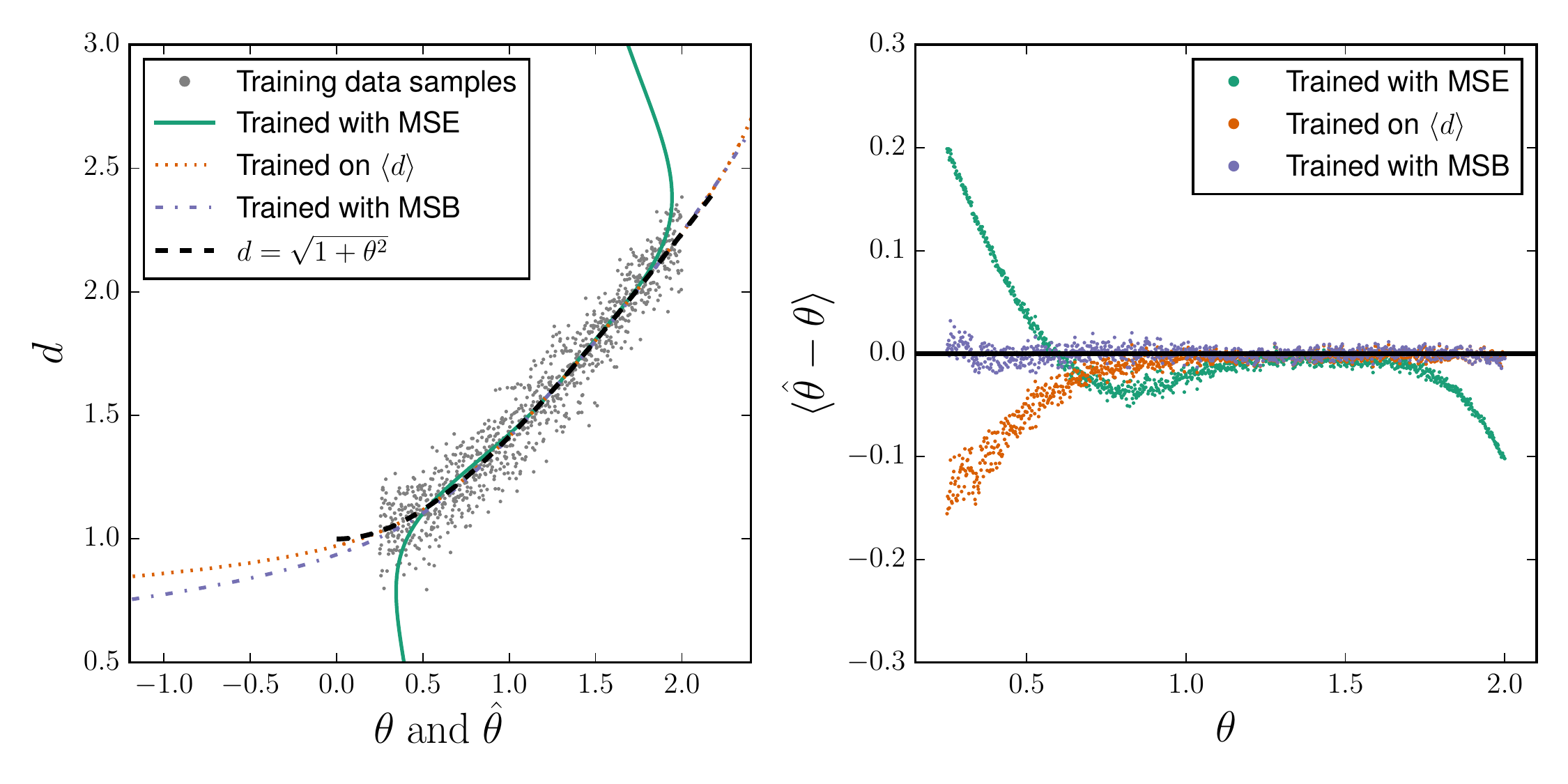} 
\caption{\label{fig:MSB_demo}
One-dimensional example of inverse regression $\hat{\theta}$ of an explanatory variable $\theta$ from noisy observations of the dependent variable $d$.
The \emph{left panel} shows a small number of samples of the training set as datapoints, and three different estimators corresponding to the same NN trained with three different error functions, as discussed in the text.
The \emph{right panel} shows the bias of each estimator as function of true $\theta$.
}
\end{center}
\end{figure*}

In this appendix, we illustrate the effect of NN error functions on inverse regressions in the case of a simple one-dimensional example.
Let $d$ be a noisy observed variable, which depends on some explanatory variable $\theta \ge 0$ following the relation
\begin{equation}
d = \sqrt{1 + \theta^2} + n,
\end{equation}
with $n \sim \mathcal{N}(0, 0.1)$.
Datapoints in the left panel of Fig.~\ref{fig:MSB_demo} illustrate a few samples drawn from this relation.

Our objective is to obtain an accurate statistical point estimator $\hat{\theta}$ of the explanatory variable, as a function of the observable $d$.
More precisely, by inverse regression, we want to obtain this estimator empirically, using a training set consisting of simulated observations of $d$ and the corresponding true values of the explanatory variable.

Following the approach presented in Sect.~\ref{sec:nomenclature}, we generate a structured training set with 1000 cases of different $\theta$ uniformly drawn from $[0.25, 2.0]$, and 1000 realizations of $d$ per case.
We build a training set of such an exaggeratedly large size to avoid stochastic behavior in this illustrative example.
For all experiments of this appendix, we use a NN with a single input of the observation $d$, one hidden layer of five nodes, and one output node for the point estimate $\hat{\theta}$.

To start, we train the network with a conventional MSE error function.
This training ignores the structure of cases and realizations, and plainly minimizes the mean square error between the output $\hat{\theta}$ and the true $\theta$ across all samples of the training set.
The resulting estimator is shown with a green line in Fig.~\ref{fig:MSB_demo}, to be read as a function of $d$.
The function follows the locus one would obtain by averaging $\theta$ in fine bins of $d$ for the whole training set, as the arithmetic average of a sample is the point with respect to which the mean square deviation is minimal.

In the right panel of this same figure, the green points show the bias of this MSE-trained estimator, evaluated on a validation set of similar structure to the training set, as a function of true $\theta$.
A strong bias can be seen over the full domain, even in the quasi-linear part of the relation close to the boundary $\theta = 2$. This underlines that even for simple Gaussian noise and a linear relation, the MSE error function does not yield in an accurate estimator, given any real-life distribution of explanatory variables.

As next alternative, let us consider a training aiming at the average of realizations $\left<d\right>$ per case, or equivalently per fine bin of true $\theta$ (vertical slices in the left panel of Fig.~\ref{fig:MSB_demo}).
For the present model, this corresponds to training on noiseless data. 
The resulting estimator is shown in orange, and indeed it follows closely the relation $d = \sqrt{1 + \theta^2}$ over the considered domain.
As the right panel shows, this training leads to a strongly biased estimator in the nonlinear part where $\theta < 1$. As described in Sect.~\ref{invregprob}, a NN which is not aware of the noise in its inputs will be affected by noise bias.

Finally, we train the same NN to minimize the mean square bias (MSB) of its predictions, exploiting the training data structure of cases and realizations. The resulting estimator is shown in purple on Fig.~\ref{fig:MSB_demo}.
For observations $d \lesssim 1.0$, this $\hat{\theta}$ returns slightly higher values than the estimator trained on noiseless data, effectively compensating for the noise bias of the latter.
The right panel shows the dramatic improvement in accuracy achieved by the MSB training.
We note that this behavior does not depend on the Gaussian nature of the noise in the observations. The NN would equally learn to compensate for other noise distributions.

\section{Python implementation}
\label{Code}

We make publicly available the code that we developed in the scope of this work, including scripts and configuration files generating the presented results and figures, at \url{https://astro.uni-bonn.de/~mtewes/ml-shear-meas/}.
A frozen copy of the code is also available at the CDS\footnote{\url{http://cdsarc.u-strasbg.fr/viz-bin/qcat?J/A+A/621/A36}}.

The core code comes in the form of two separate python packages:
\begin{description}
\item{{\tt tenbilac}} is an artificial neural network library for noisy features, implementing the peculiar distinction between training cases and realizations.
It is implemented in python and numpy, and agnostic about the particular application to galaxy shape measurement.
\item{{\tt momentsml}} is a library providing a toolbox for experimenting with shear and shape estimators, build around GalSim\footnote{\url{https://github.com/GalSim-developers/GalSim}} and astropy\footnote{\url{http://www.astropy.org}}.
It includes a simple wrapper to process GREAT3 data, and an interface to {\tt tenbilac}.

\end{description}

\section{Evolution of the method since the GREAT3 challenge}
\label{changesG3} 

We participated in the GREAT3 challenge in 2014 with an earlier version of the approach, under the name \emph{MegaLUT}.
Results presented in this paper are based on a substantial evolution of the algorithm since these original submissions to the GREAT3 challenge.
In the following, we briefly summarize the main differences between the present work, and the code used in \citet{Mandelbaum:2015gc}.

For the GREAT3 challenge participation, we optimized an MSE cost function (with no distinction between realizations and cases), targeting the ellipticity of a galaxy.
To improve the performance despite this flawed cost function, we made use of training sets with high \snr compared to the challenge data.
Furthermore, we did not train any weight prediction, but instead rejected faint and small galaxies using hard thresholds on input features.
These aspects contributed both to a suboptimal performance and to a much higher sensitivity of the method to its training data.

\end{appendix}

\end{document}